\documentclass[aps,pre,floats,floatfix,showpacs,superscriptaddress,nofootinbib,twocolumn]{revtex4}
\usepackage{amsmath,amssymb,multirow}
\usepackage[pdftex]{graphicx}
\usepackage[tight]{subfigure}

\newcommand{\SM}{\mathbb{S}^1}
\newcommand{\HM}{\mathbb{H}^2}

\renewcommand{\leq}{\leqslant}
\renewcommand{\geq}{\geqslant}

\begin{document}

\title{Hyperbolic Geometry of Complex Networks}

\author{Dmitri Krioukov}
\affiliation{Cooperative Association for Internet Data Analysis
(CAIDA), University of California, San Diego (UCSD), La Jolla, CA
92093, USA}

\author{Fragkiskos Papadopoulos}
\affiliation{Department of Electrical and Computer Engineering,
University of Cyprus, Kallipoleos 75, Nicosia 1678, Cyprus}

\author{Maksim Kitsak}
\affiliation{Cooperative Association for Internet Data Analysis
(CAIDA), University of California, San Diego (UCSD), La Jolla, CA
92093, USA}

\author{Amin Vahdat}
\affiliation{Department of Computer Science and Engineering,
University of California, San Diego (UCSD), La Jolla, CA 92093, USA}

\author{Mari{\'a}n Bogu{\~n}{\'a}}
\affiliation{Departament de F{\'\i}sica Fonamental, Universitat de
Barcelona, Mart\'{\i} i Franqu\`es 1, 08028 Barcelona, Spain}

\begin{abstract}

We develop a geometric framework to study the structure and function
of complex networks. We assume that hyperbolic geometry underlies
these networks, and we show that with this assumption, heterogeneous
degree distributions and strong clustering in complex networks
emerge naturally as simple reflections of the negative curvature and
metric property of the underlying hyperbolic geometry. Conversely,
we show that if a network has some metric structure, and if the
network degree distribution is heterogeneous, then the network has
an effective hyperbolic geometry underneath. We then establish a
mapping between our geometric framework and statistical mechanics of
complex networks. This mapping interprets edges in a network as
non-interacting fermions whose energies are hyperbolic distances
between nodes, while the auxiliary fields coupled to edges are
linear functions of these energies or distances. The geometric network
ensemble subsumes the standard configuration model and classical
random graphs as two limiting cases with degenerate geometric
structures. Finally, we show that targeted transport processes
without global topology knowledge, made possible by our geometric
framework, are maximally efficient, according to all efficiency
measures, in networks with strongest heterogeneity and clustering,
and that this efficiency is remarkably robust with respect to even
catastrophic disturbances and damages to the network structure.

\end{abstract}


\maketitle

\begin{figure*}
    \centerline{
        \subfigure[]
            {\includegraphics[width=2.3in]{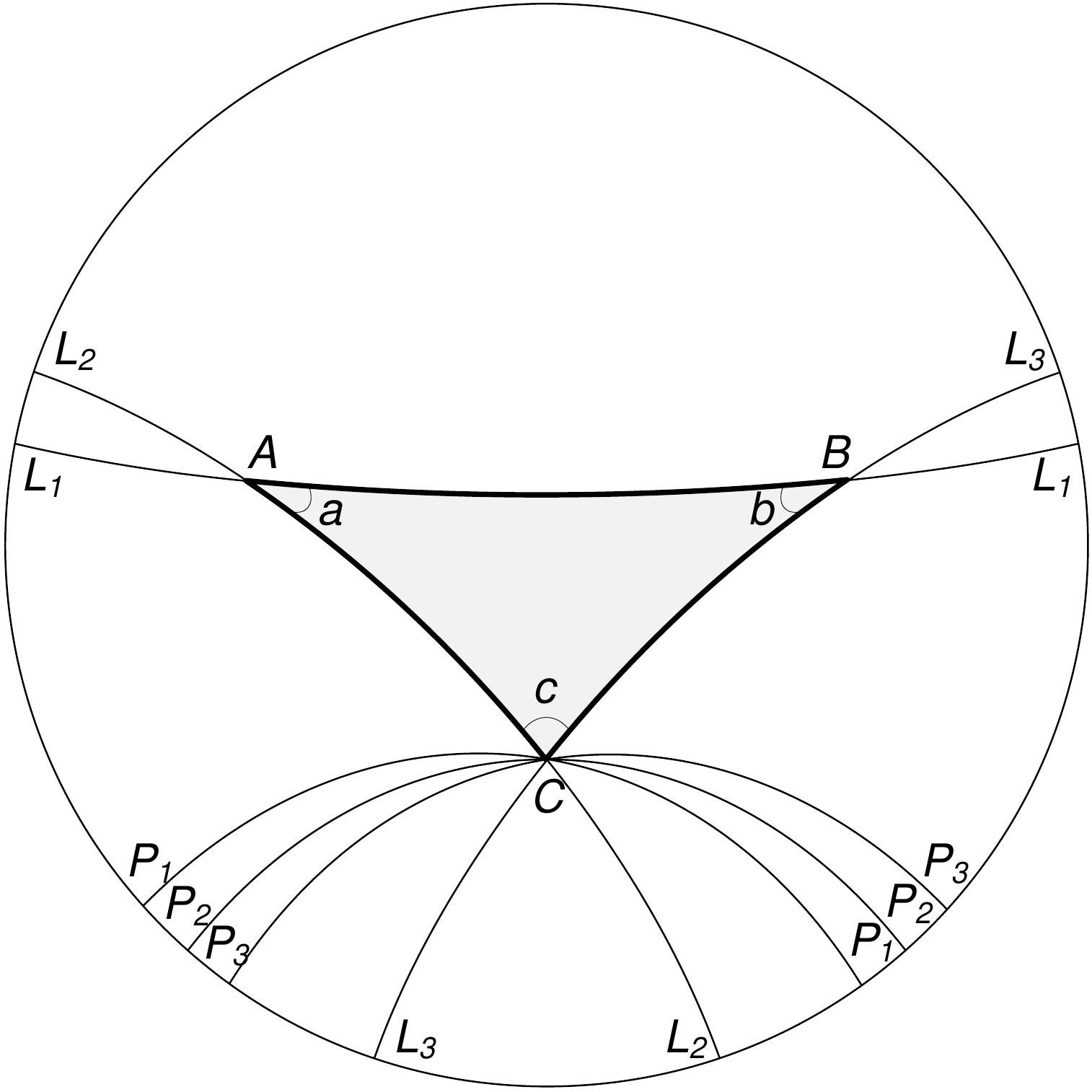}}
        \hfill
        \subfigure[]
            {\includegraphics[width=2.3in]{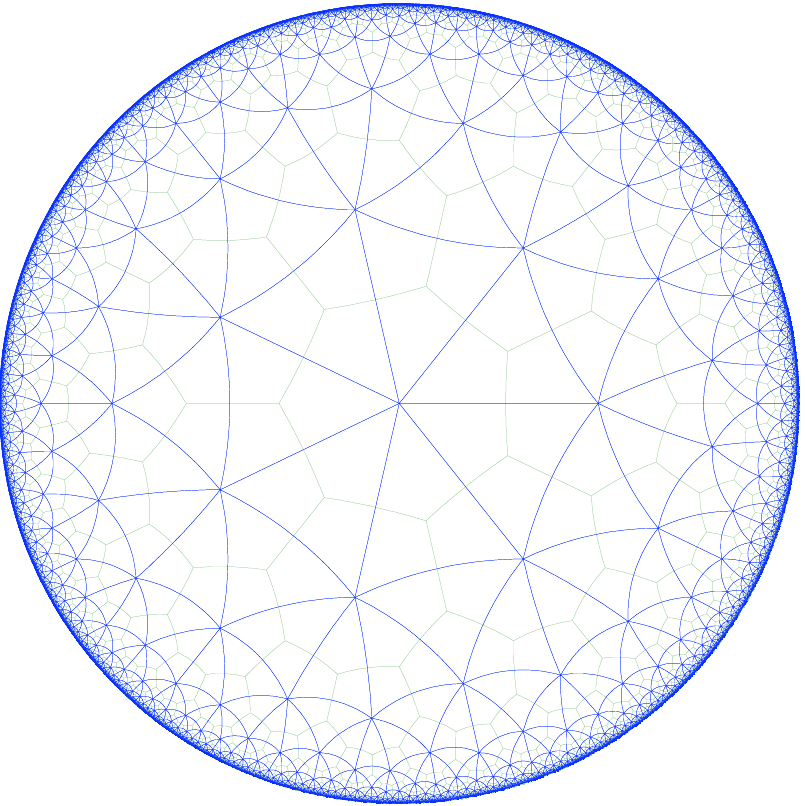}}
        \hfill
        \subfigure[]
            {\includegraphics[width=2.3in]{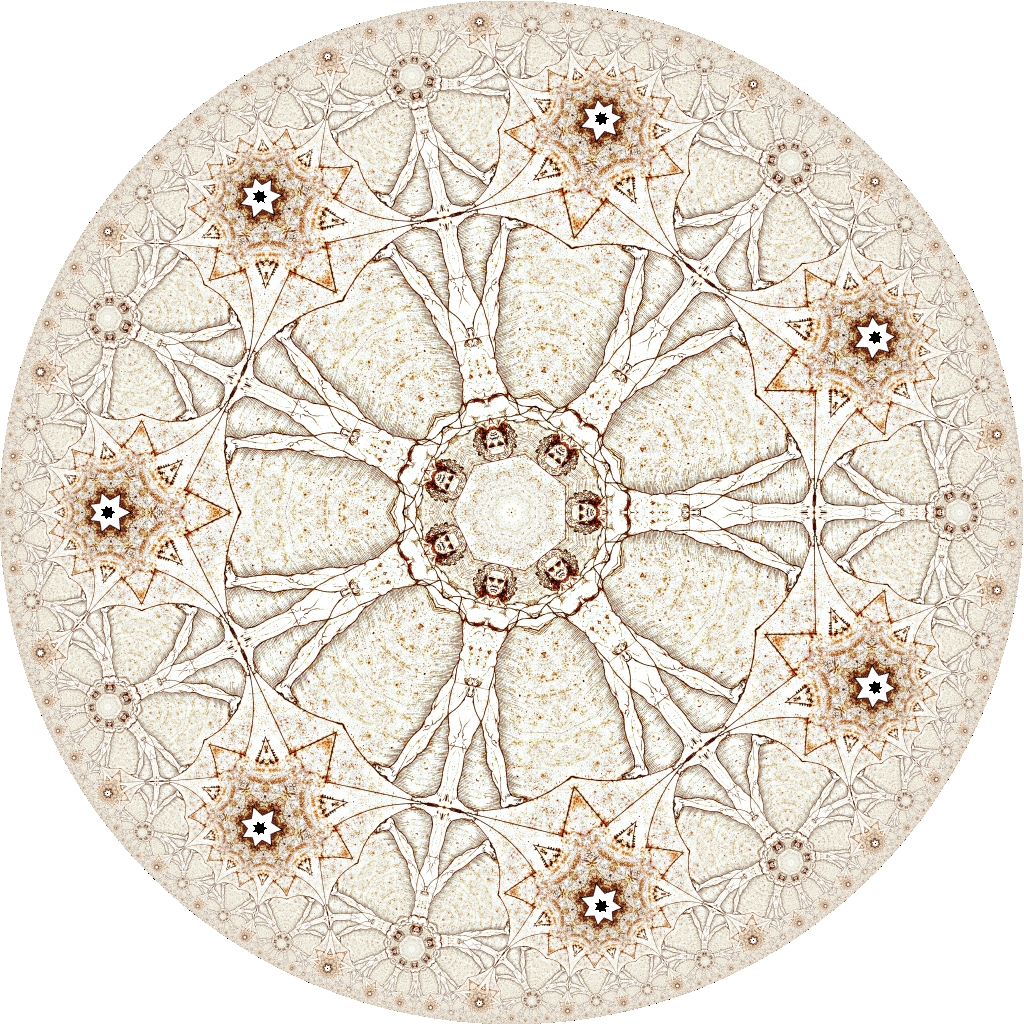}}
    }
    \caption{(Color online) Poincar\'e disk model. In {\bf(a)}, $L_{1,2,3}$ and $P_{1,2,3}$
    are examples of hyperbolic lines. Lines
    $L_{1,2,3}$ intersect to form triangle $ABC$. The sum of its
    angles $a+b+c<\pi$. As opposed to Euclidean geometry,
    there are infinitely many lines (examples
    are $P_{1,2,3}$) that are parallel to line $L_1$ and go through
    a point $C$ that does not belong to $L_1$. In~{\bf(b)}, a
    $\{7,3\}$-tessellation of the hyperbolic plane by equilateral triangles,
    and the dual $\{3,7\}$-tessellation by regular heptagons are shown.
    All triangles and heptagons
    are of the same hyperbolic size but the size of their Euclidean
    representations exponentially decreases as a function of the
    distance from the center, while their number exponentially
    increases. In~{\bf(c)}, the exponentially increasing number
    of men illustrates the exponential expansion of hyperbolic
    space. The {\tt Poincar\'e} tool~\cite{poincare} is used to construct
    a $\{7,7\}$-tessellation of the hyperbolic plane, rendering
    a fragment of {\it The Vitruvian Man\/} by Leonardo da Vinci.
    \label{fig:poincare}
    }
\end{figure*}

\section{Introduction}

Geometry has a proven history of success, helping to make impressive
advances in diverse fields of science, when a geometric fabric
underlying a complex problem or phenomenon is identified. Examples
can be found everywhere. Perhaps the most famous one is general
relativity, interpreting gravitation as a curved geometry. Quite a
contrasting example comes from the complexity theory in computer
science, where apparently intractable computational problems
suddenly find near optimal solutions as soon as a geometric
underpinning of the problem is discovered~\cite{GoWi95}, leading to
viable practical applications~\cite{DiKrHuClRi05}. Yet another
example is the recent conjecture by Palmer~\cite{Palmer09}
suggesting that many ``mysteries'' of quantum mechanics can be
resolved by the assumption that a hidden fractal geometry underlies
the universe.

Inspired by these observations, and following~\cite{KrPa09}, we
develop here a geometric framework to study the structure and
function of complex
networks~\cite{Newman10-book,Dorogovtsev10-book}. We begin with the
assumption that hyperbolic geometry underlies these networks.
Although difficult to visualize, hyperbolic geometry, briefly
reviewed in Section~\ref{sec:primer}, is by no means anything
exotic. In fact it is the geometry of the world we live in. Indeed,
the relativistic Minkowski spacetime is hyperbolic, and so is the
anti-de Sitter space~\cite{Maldacena99,GuKl98,Witten98}. On the
other hand, hyperbolic spaces can be thought of as smooth versions
of trees abstracting the hierarchical organization of complex
networks~\cite{ClMo08}, a key observation providing a high-level
rationale, Section~\ref{sec:motivations}, for our hyperbolic hidden
space assumption. In Section~\ref{sec:models} we show that from this
assumption, two common properties of complex network topologies
emerge naturally. Namely, heterogeneous degree distributions and
strong clustering appear, in the simplest possible settings, as
natural reflections of the basic properties of underlying hyperbolic
geometry. The exponent of the power-law degree distribution, for
example, turns out to be a function of the hyperbolic space
curvature. Fortunately, unlike in~\cite{Palmer09}, for instance, we
can directly verify our assumption. In
Section~\ref{sec:model_equivalence} we consider the converse
problem, and show that if a network has some metric
structure---tests for its presence are described
in~\cite{SeKrBo08}---and if the network's degree distribution is
heterogeneous, then the network does have an effective hyperbolic
geometry underneath.

Many different pieces start coming together in
Section~\ref{sec:fermi}, where we show that the ensembles of
networks in our framework can be analyzed using standard tools in
statistical mechanics. Hyperbolic distances between nodes appear as
energies of corresponding edges distributed according to Fermi-Dirac
statistics. In this interpretation, auxiliary fields, which have
been considered as opaque variables in the standard exponential
graph formalism~\cite{DoMeSa03b,PaNe04,GaLo09,AnBi09,Bianconi09},
turn out to be linear functions of underlying distances between
nodes. The chemical potential, Boltzmann constant, etc., also find
their lucid geometric interpretations, while temperature appears as
a natural parameter controlling clustering in the network. The
network ensemble exhibits a phase transition at a specific value of
temperature, caused---as usual---by a non-analyticity of the partition
function. This phase transition separates two regimes in the
ensemble, cold and hot. Complex networks belong to the cold regime,
while in the hot regime, the standard configuration
model~\cite{ChLu02b} and classical random graphs~\cite{ErRe59} turn
out to be two limiting cases with degenerate geometric structures,
Section~\ref{sec:random_graphs}.
Sections~\ref{sec:degree_distribution} and~\ref{sec:clustering}
analyze the degree distribution and clustering as functions of
temperature in the two regimes.

Finally, in Section~\ref{sec:simulations}, we shift our attention to
network function. Specifically, we analyze the network efficiency
with respect to targeted communication or transport processes
without global topology knowledge, made possible by our geometric
approach. We find that such processes in networks with strong
heterogeneity and clustering, guided by the underlying hyperbolic
space, achieve the best possible efficiency according to all
measures, and that this efficiency is remarkably robust with respect
to even catastrophic levels of network damage. This finding
demonstrates that complex networks have the optimal structure,
allowing for routing with minimal overhead approaching its
theoretical lower bounds, a notoriously difficult longstanding
problem in routing theory, proven unsolvable for general
graphs~\cite{KoPe06}.

\section{Hyperbolic geometry}\label{sec:primer}

In this section we review the basic facts about hyperbolic geometry.
More detailed accounts can be found
in~\cite{Anderson05-book,CaFlo97,BuBuIv01-book,Ratcliffe06-book,BridsonHaefliger99-book,BuSch07-book,Gromov07-book}.

There are only three types of isotropic spaces: Euclidean (flat),
spherical (positively curved), and hyperbolic (negatively curved).
Hyperbolic spaces of constant curvature are difficult to envisage
because they cannot be isometrically embedded into any Euclidean
space. The reason is, informally, that the former are ``larger'' and
have more ``space'' than the latter.

Because of the fundamental difficulties in representing spaces of
constant negative curvature as subsets of Euclidean spaces, there
are not one but many equivalent models of hyperbolic spaces. Each
model emphasizes different aspects of hyperbolic geometry, but no
model simultaneously represents all of its properties. In special
relativity, for example, the hyperboloid model is commonly used,
where the hyperbolic space is represented by a hyperboloid. Its two
different projections to disks orthogonal to the main axis of the
hyperboloid yield the Klein and Poincar\'e unit disk models. In the
latter model, the whole infinite hyperbolic plane $\mathbb{H}^2$,
i.e., the two-dimensional hyperbolic space of constant curvature
$-1$, is represented by the interior of the Euclidean disk of
radius~$1$, see Fig.~\ref{fig:poincare}. The boundary of the disk,
i.e., the circle $\SM$, is not a part of the hyperbolic plane, but
represents its infinitely remote points, called boundary at infinity
$\partial\HM$. Any symmetry transformation on $\HM$ translates to a
symmetry on $\partial\HM$, and {\it vice versa}, a cornerstone of
the anti-de Sitter space/conformal field theory
correspondence~\cite{Maldacena99,GuKl98,Witten98}, where quantum
gravity on an anti-de Sitter space is equivalent to a quantum field
theory without gravity on the conformal boundary of the space.
Hyperbolic geodesic lines in the Poincar\'e model, i.e., shortest
paths between two points at the boundary, are disk diameters and
arcs of Euclidean circles intersecting the boundary perpendicularly.
The model is conformal, meaning that Euclidean angles between
hyperbolic lines in the model are equal to their hyperbolic values,
which is not true with respect to distances or areas. Euclidean and
hyperbolic distances, $r_e$ and $r_h$, from the disk center, or the
origin of the hyperbolic plane, are related by
\begin{equation}\label{eq:re.vs.rh}
r_e=\tanh\frac{r_h}{2}.
\end{equation}
The model is generalizable for any dimension $d\geq2$, in which case
$\mathbb{H}^d$ is represented by the interior of the unit ball whose
boundary $\mathbb{S}^{d-1}$ is the boundary at infinity
$\partial\mathbb{H}^d$. The model is related via the stereographic
projection to another popular model---the upper half-space
model---where $\mathbb{H}^d$ is represented by a ``half'' of
$\mathbb{R}^d$ span by vectors $\mathbf{x}=(x_1,x_2,\ldots,x_d)$
with $x_d>0$. The boundary at infinity $\partial\mathbb{H}^d$ in
this case is the hyperplane $x_d=0$ instead of $\mathbb{S}^{d-1}$.
Essentially any $d$-dimensional space $X$ with a $(d-1)$-dimensional
boundary can be equipped with a hyperbolic metric structure, with the
$X$'s boundary playing the role of the boundary at infinity
$\partial X$.

Given the abundance of hyperbolic space representations, we are free
to choose any of those, e.g., the one most convenient for our
purposes. Unless mentioned otherwise, we use the {\em native\/}
representation in the rest of the paper. In this representation, all
distance variables have their true hyperbolic values. In polar
coordinates, for example, the radial coordinate $r$ of a point is
equal to its hyperbolic distance from the origin. That is, instead
of~(\ref{eq:re.vs.rh}), we have
\begin{equation}\label{eq:re.vs.rh-native}
r \equiv r_h = r_e.
\end{equation}

A key property of hyperbolic spaces is that they expand faster than
Euclidean spaces. Specifically, while Euclidean spaces expand
polynomially, hyperbolic spaces expand exponentially. In the
two-dimensional hyperbolic space $\mathbb{H}^2_\zeta$ of constant
curvature $K=-\zeta^2<0$, $\zeta>0$, for example, the length of the
circle and the area of the disk of hyperbolic radius $r$ are
\begin{eqnarray}
L(r) &=& 2\pi\sinh\zeta r,\label{eq:hyperbolic-circle-length}\\
A(r) &=& 2\pi(\cosh\zeta r - 1),\label{eq:hyperbolic-disk-area}
\end{eqnarray}
both growing as $e^{\zeta r}$ with $r$. The hyperbolic distance $x$
between two points at polar coordinates $(r,\theta)$ and
$(r',\theta')$ is given by the hyperbolic law of cosines
\begin{equation}\label{eq:x-zeta}
\cosh\zeta x=\cosh{\zeta r}\cosh{\zeta r'}-\sinh{\zeta r}\sinh{\zeta r'}\cos{\Delta\theta},
\end{equation}
where $\Delta\theta=\pi-|\pi-|\theta-\theta'||$ is the angle between
the points. Equations~(\ref{eq:hyperbolic-circle-length}-\ref{eq:x-zeta})
converge to their familiar Euclidean analogs at $\zeta\to0$. For
sufficiently large $\zeta r$, $\zeta r'$, and
$\Delta\theta>2\sqrt{e^{-2\zeta r}+e^{-2\zeta r'}}$, the hyperbolic
distance $x$ is closely approximated by
\begin{equation}\label{eq:x-approx}
x = r + r' + \frac{2}{\zeta}\ln\sin\frac{\Delta\theta}{2} \approx
r + r' + \frac{2}{\zeta}\ln\frac{\Delta\theta}{2}.
\end{equation}
That is, the distance between two points is approximately the sum of
their radial coordinates, minus some $\Delta\theta$-dependent
correction, which goes to zero at $\zeta\to\infty$.

Hyperbolic spaces are similar to trees. In a $b$-ary tree (a tree
with branching factor $b$), the analogies of the circle length or
disk area are the number of nodes at distance exactly $r$ or not
more than $r$ hops from the root. These numbers are $(b+1)b^{r-1}$
and $[(b+1)b^r-2]/(b-1)$, both growing as $b^r$ with $r$. We thus
see that the metric structures of $\mathbb{H}^2_\zeta$ and $b$-ary
trees are the same if $\zeta=\ln b$: in both cases circle lengths
and disk areas grow as $e^{\zeta r}$. In other words, from the
purely metric perspective, $\mathbb{H}^2_{\ln b}$ and $b$-ary trees
are equivalent. Informally, trees can therefore be thought of as
``discrete hyperbolic spaces.'' Formally, trees, even infinite ones,
allow nearly isometric embeddings into hyperbolic spaces. For
example, any tessellation of the hyperbolic plane (see
Fig.~\ref{fig:poincare}) naturally defines isometric embeddings for
a class of trees formed by certain subsets of polygon sides. For
comparison, trees do not generally embed into Euclidean spaces.
Informally, trees need an exponential amount of space for branching,
and only hyperbolic geometry has it.

\begin{table}
\begin{centering}
\caption{Characteristic properties of Euclidean, spherical, and
hyperbolic geometries. {\it Parallel lines} is the number of lines
that are parallel to a line and that go through a point not
belonging to this line, and $\zeta=\sqrt{|K|}$. \label{table:geometries}}
\begin{tabular}{|l|l|l|l|}\hline
Property & Euclidean & Spherical & Hyperbolic \\ \hline
Curvature $K$& $0$ & $>0$ & $<0$ \\ \hline
Parallel lines & $1$ & $0$ & $\infty$ \\ \hline
Triangles are & normal & thick & thin \\ \hline
Shape of triangles &
\includegraphics[width=.3in]{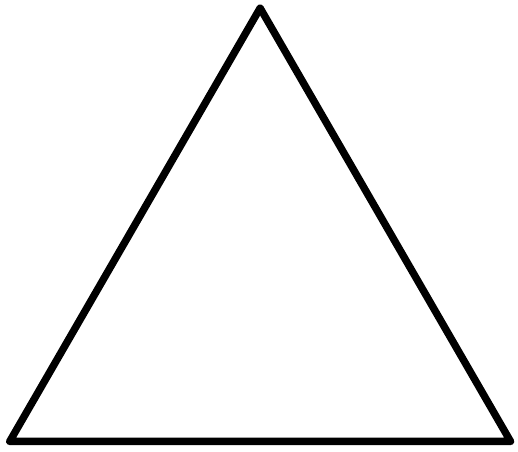} &
\includegraphics[width=.3in]{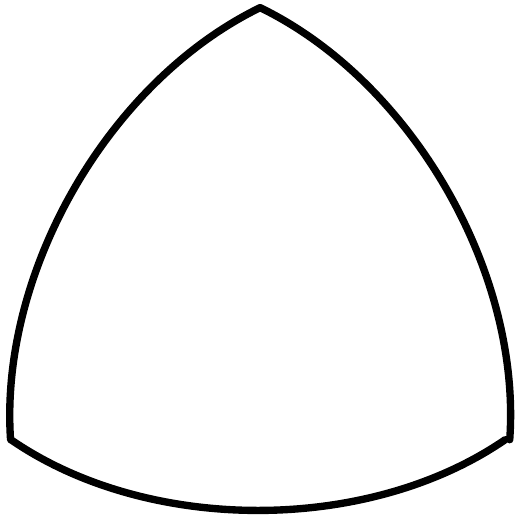} &
\includegraphics[width=.3in]{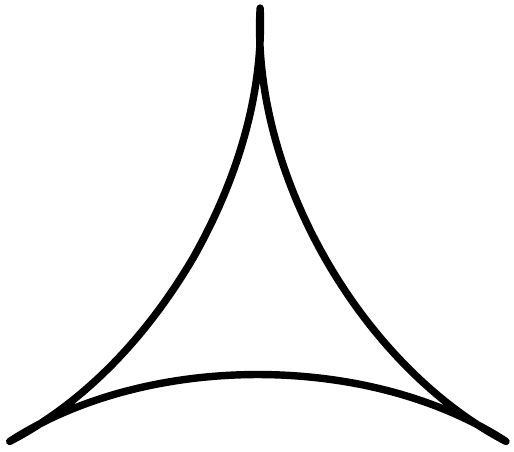}
\\ \hline
Sum of $\triangle$ angles & $\pi$ & $>\pi$ & $<\pi$ \\ \hline
Circle length & $2\pi r$ & $2\pi\sin\zeta r$ & $2\pi\sinh\zeta r$ \\ \hline
Disk area & $2\pi r^2/2$ & $2\pi(1-\cos\zeta r)$ & $2\pi(\cosh\zeta r - 1)$ \\
\hline
\end{tabular}
\end{centering}
\end{table}

Table~\ref{table:geometries} collects these and other characteristic
properties of hyperbolic geometry and juxtaposes them against the
corresponding properties of Euclidean and spherical geometries.

\section{Topological heterogeneity versus geometrical hyperbolicity}\label{sec:motivations}

In this section we make high-level observations suggesting the
existence of intrinsic connections between hyperbolic geometry and
the topology of complex networks.

Complex networks connect distinguishable, heterogeneous elements
abstracted as nodes. Understood broadly, this heterogeneity implies
that there is at least some taxonomy of elements, meaning that all
nodes can be somehow classified. In most general settings, this
classification implies that nodes can be split in large groups
consisting of smaller subgroups, which in turn consist of even
smaller subsubgroups, and so on. The relationships between such
groups and subgroups can be approximated by tree-like structures,
sometimes called {\it dendrograms}, which represent hidden
hierarchies in networks~\cite{ClMo08}. But as discussed in the
previous section, the metric structures of trees and hyperbolic
spaces are the same. We emphasize that we do not assume that the node
classification hierarchy among a particular dimension is strictly a
tree, but that it is approximately a tree. As soon as it is at least
approximately a tree, it is negatively curved~\cite{Gromov07-book}.
This argument obviously applies only to a snapshot of a network
taken at some moment of time. A logical question is how these
taxonomies emerge. Clearly, when a network begins to form, the node
classification is degenerate, but as more and more nodes join the
network and evolve in it, they tend to diversify and specialize,
thus deepening their classification hierarchy. The distance between
nodes in such hierarchies is then a rough approximation of node
similarity, and the more similar a pair of nodes, the more likely
they are connected.

We consider several examples suggesting that these general
observations apply to different real networks. Social networks form
the most straightforward class of examples, where network community
structures~\cite{GiNe02,BoPa04} represent hidden
hierarchies~\cite{WatDoNew02}. More concretely, in paper citation
networks, the underlying geometries can approximately be the
relationships between scientific subject categories, and the closer
the subjects of two papers, the more similar they are, and the more
likely they cite each other~\cite{Redner98,BoMaGo04-pnas}.
Classifications of web pages (or more specifically, of the Wikipedia
pages~\cite{MuIt07,CraCo08}) also show the same effect: the more
similar a pair of web pages, the more likely that there is a
hyperlink between them~\cite{menczer02-pnas}. In biology, the
distance between two species on the phylogenetic tree is a widely
used measure of similarity between the
species~\cite{Phylogenetic-book00}. Note that this example
emphasizes both the existing taxonomy of elements and their
evolution. The evolution of the Internet is yet another paradigmatic
example. In the beginning, there were only a couple of computers
connected to each other, but then the network
grew~\cite{internet-history} splitting into a collection of
independently administered networks, called autonomous systems
(ASs), whose number and diversity have been growing
fast~\cite{DhDo08}. Currently, ASs can be classified based on their
geographic position and coverage, size, number and type of
customers, business role, and many other
factors~\cite{DhDo08,DiKrRi06}.

\begin{figure}
\centerline{\includegraphics[width=3in]{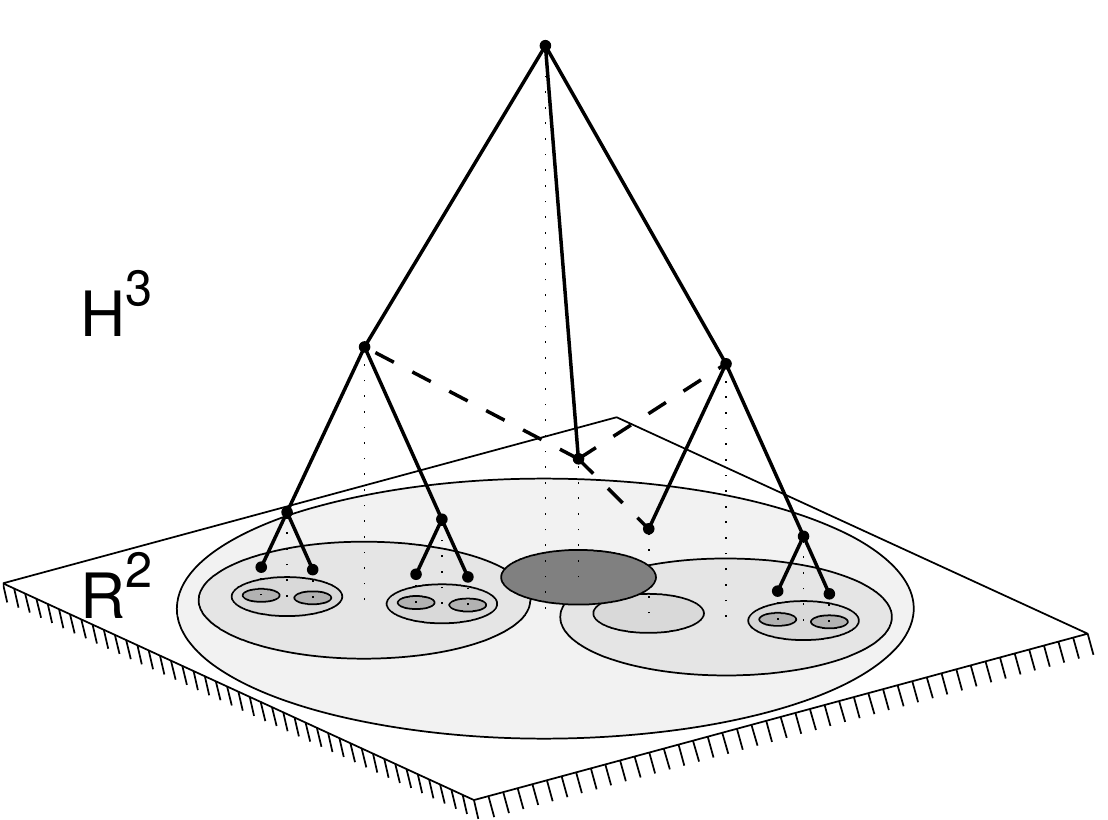}}
\caption{Mapping between disks in the Euclidean plane $\mathbb{R}^2$
and points in the Poincar\'e half-space model of the three-dimensional
hyperbolic space $\mathbb{H}^3$~\cite{Anderson05-book}. The
$x,y$-coordinates of disks in $\mathbb{R}^2$ are the
$x,y$-coordinates of the corresponding points in $\mathbb{H}^3$. The
$z$-coordinates of these points in $\mathbb{H}^3$ are the radii of
the corresponding disks. This mapping represents the tree-like
hierarchy among the disks. Two points in $\mathbb{H}^3$ are
connected by a solid link if one of the corresponding disks is the
minimum-size disk that fully contains the other disk. This hierarchy
is not perfect; thus, the tree structure is approximate. The darkest
disk in the middle partially overlaps with three other disks at
different levels of the hierarchy. Two points in $\mathbb{H}^3$ are
connected by a dashed link if the corresponding disks partially
overlap. These links add cycles to the tree. The shown structure is
thus not strictly a tree, but it is hyperbolic~\cite{Gromov07-book}.
\label{fig:R2H3}}
\end{figure}

The general observation that the metric structure of node similarity
distances is hyperbolic follows from the mathematical fact
illustrated in Fig.~\ref{fig:R2H3}. We assume there that a point in
$\mathbb{R}^2$ represents an abstract node attribute or
characteristic, while a Euclidean disk in $\mathbb{R}^2$ represents
a collection of all the attributes for a given node in the network.
The network itself is not shown. Instead we visualize a hidden
hierarchy arising from the overlapping disks. The more two disks
overlap, the more similar the sets of characteristics of the two
corresponding nodes, that is, the more similar the nodes themselves.
But the mapping between disks $\mathbb{R}^2$ and nodes in
$\mathbb{H}^3$ in Fig.~\ref{fig:R2H3} is such that the more the two
disks overlap, the hyperbolically closer are the corresponding two nodes.
Formally, if the ratio of the disks' radii $r,r'$ is bounded by
a constant $C$, $1/C \leq r/r' \leq C$, and the Euclidean distance
between their centers is bounded by $Cr$, then the hyperbolic
distance between the corresponding nodes in $\mathbb{H}^3$ is
bounded by some constant $C'$, which depends only on $C$, and not on
the disk radii or center locations~\cite{Gromov07-book}. The
converse is also true. Therefore, the distances between nodes based
on similarity of their attributes can be mapped to distances in a
hyperbolic space, assuming that node attributes possess some metric
structure ($\mathbb{R}^2$ in the above example) in the first place.

\section{Hyperbolic geometry yields heterogeneous topology}\label{sec:models}

We now put the intuitive considerations in the previous section to
qualitative grounds. We want to see what network topologies emerge
in the simplest possible settings involving hyperbolic geometry.

\subsection{Uniform node density at curvature $K=-1$}\label{sec:uniform-density}

Since the one-dimensional hyperbolic space $\mathbb{H}^1$ does not
exist, the simplest hyperbolic space is the hyperbolic plane
$\mathbb{H}^2$ discussed in Section~\ref{sec:primer}. The simplest
way to place $N\gg1$ nodes on the hyperbolic plane is to distribute
them uniformly over a disk of radius $R\gg1$, where $R$ abstracts
the depth of the hidden tree-like hierarchy. We will see below that
$R$ is a growing function of $N$, reflecting the intuition in
Section~\ref{sec:motivations} that the network hierarchy deepens
with network growth. The hyperbolically uniform node density implies
that we assign the angular coordinates $\theta\in[0,2\pi]$ to nodes
with the uniform density $\rho(\theta)=1/(2\pi)$, while according to
Eqs.~(\ref{eq:hyperbolic-circle-length},\ref{eq:hyperbolic-disk-area})
with $\zeta=1$, the density for the radial coordinate $r\in[0,R]$ is
exponential
\begin{equation}\label{eq:rho(r)-uniform}
\rho(r)=\frac{\sinh r}{\cosh R - 1} \approx e^{r-R} \sim e^r.
\end{equation}
To form a network, we need to connect each pair of nodes with some
probability, which can depend only on hyperbolic distances $x$
between nodes. The simplest connection probability function is
\begin{equation}\label{eq:p(x)-step}
p(x)=\Theta(R-x),
\end{equation}
where $\Theta(x)$ is the Heaviside step function. We will justify
and relax this choice in Section~\ref{sec:fermi}. This connection
probability means that we connect a pair of nodes by a link only if
the hyperbolic distance~(\ref{eq:x-zeta}) between them is $x \leq
R$.

\begin{figure}
\centerline{\includegraphics[width=2.5in]{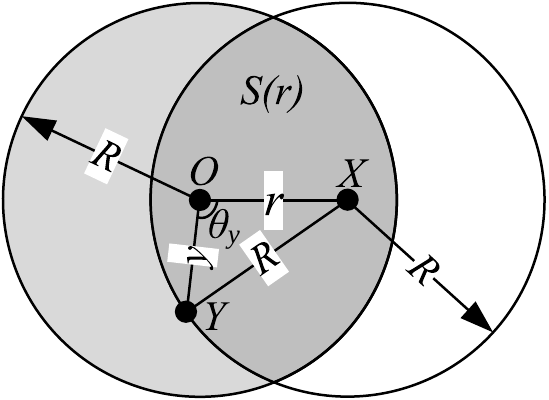}}
\caption{The expected degree of a node at point $X$ located at
distance $r$ from the origin $O$ is proportional to the area of the
dark-shaded intersection $S(r)$ of the two disks of radius $R$. The
first disk, centered at $O$, contains all the nodes, distributed
within it with a uniform density. The second disk, centered at $X$,
is defined by the connection probability $p(x)$, which is either $1$
or $0$ depending on whether the distance $x$ from $X$ is less or greater
than $R$. The node at $X$ is connected to all the nodes lying in the
dark-shaded intersection area~$S(r)$. \label{fig:S(r)}}
\end{figure}

\begin{figure}
\centerline{\includegraphics[width=2.5in]{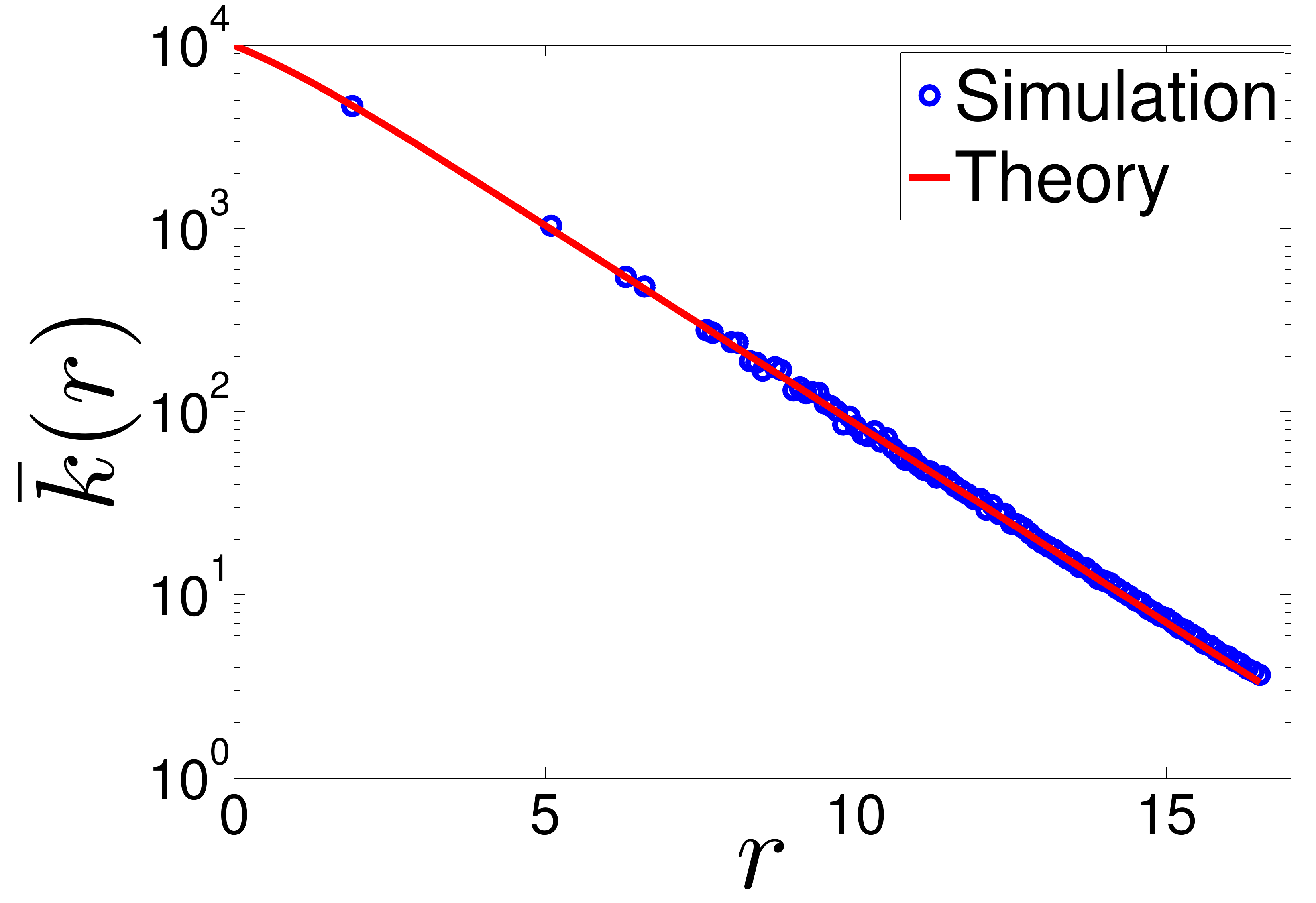}}
\caption{(Color online) Average degree at distance $r$ from the origin for a
network with $N=10000$ and $R=16.55$.
} \label{fig:bar_k_r}
\end{figure}

The network is now formed, and we can analyze its topological
properties. We are first interested in the most basic one, the
degree distribution $P(k)$, to compute which we have to calculate
the average degree $\bar{k}(r)$ of nodes located at distance $r$
from the origin. Since the node density is uniform, $\bar{k}(r)$ is
proportional to the area $A(r)$ of the intersection $S(r)$ of the
two disks shown in Fig.~\ref{fig:S(r)}. Specifically,
$\bar{k}(r)=\delta A(r)$ with node density $\delta=N/[2\pi(\cosh R -
1)]$. The area element $dA$ in polar coordinates $(y,\theta)$ is $dA
= \sinh y\,dyd\theta$, cf.\
Eqs.~(\ref{eq:hyperbolic-circle-length},\ref{eq:hyperbolic-disk-area})
with $\zeta=1$. Therefore, the intersection area
$A(r)=\iint_{S(r)}dA$ is given by the following integration
illustrated in Fig.~\ref{fig:S(r)}
\begin{eqnarray}\label{eq:S(r)-def}
A(r) &=& 2\int_0^{R-r} \sinh y\,dy \int_0^\pi d\theta  +
2\int_{R-r}^R \sinh y\,dy \int_0^{\theta_y} d\theta \nonumber \\
&=& 2\pi[\cosh(R-r)-1] + 2\int_{R-r}^R  \theta_y \sinh y\,dy,
\end{eqnarray}
where $\theta_y\in[0,\pi]$ is given by the hyperbolic law of
cosines~(\ref{eq:x-zeta}) for the triangle $\triangle OXY$ in
Fig.~\ref{fig:S(r)}
\begin{equation}\label{eq:theta_y}
\cosh R = \cosh r \cosh y - \sinh r \sinh y \cos \theta_y.
\end{equation}
Solving the last equation for $\theta_y$ and substituting the result
into~(\ref{eq:S(r)-def}) yields the exact expression for $A(r)$ and
consequently for the average degree $\bar{k}(r)$
\begin{widetext}
\begin{eqnarray}\label{eq:step-function-final}
\bar{k}(r) &=& \frac{N}{2\pi(\cosh R - 1)} \left\{2\pi(\cosh R - 1) -
2 \cosh R \left( \arcsin \frac{\tanh(r/2)}{\tanh R}
+ \arctan \frac {\cosh R \sinh(r/2)}
{\sqrt{\sinh(R+r/2)\sinh(R-r/2)}} \right) \right. \nonumber\\
&+&  \left.
 \arctan \frac {(\cosh R + \cosh r)\sqrt{\cosh2R-\cosh r}}
{\sqrt{2}(\sinh^2R - \cosh R - \cosh r)\sinh(r/2)}
-\arctan \frac {(\cosh R - \cosh r)\sqrt{\cosh2R-\cosh r}}
{\sqrt{2}(\sinh^2R + \cosh R - \cosh r)\sinh(r/2)}
\right\},
\end{eqnarray}
\end{widetext}
which perfectly matches simulations in Fig.~\ref{fig:bar_k_r}. For
large $R$ this terse exact expression is closely approximated by
\begin{equation}\label{eq:k(r)-uniform}
\bar{k}(r)=
N\left\{\frac{4}{\pi}e^{-r/2}-\left(\frac{4}{\pi}-1\right)e^{-r}\right\}\approx\frac{4}{\pi}Ne^{-r/2},
\end{equation}
where the last approximation holds for large $r$.

The average degree in the network is then
\begin{equation}
\bar{k}=\int_0^R\rho(r)\bar{k}(r)dr \approx\frac{8}{\pi}Ne^{-R/2},
\end{equation}
from which we conclude that if we want to generate an $N$-node
network with a target average degree $\bar{k}$ we have to select the
disk radius $R=2\ln[8N/(\pi\bar{k})]$. We see that $R$ scales with
$N$ as $R\sim\ln N$, i.e., the same way as the depth of a balanced
tree with its size. We also observe that by fixing
\begin{equation}\label{eq:N.vs.R}
N = \nu\,e^{R/2},
\end{equation}
we gain control over the average degree in a network via parameter
$\nu=\pi\bar{k}/8$, using which we rewrite~(\ref{eq:k(r)-uniform})
as
\begin{equation}\label{eq:k(r)-uniform-no-N}
\bar{k}(r)=\frac{\bar{k}}{2}e^{(R-r)/2}\sim e^{-r/2}.
\end{equation}

To finish computing the degree distribution $P(k)$ we treat the
radial coordinate $r$ as a hidden variable in the terminology
of~\cite{BoPa03}, yielding $P(k) = \int_0^R g(k|r)\rho(r)\,dr$,
where the propagator $g(k|r)$ is the conditional probability that a
node with hidden variable $r$ has degree $k$. For sparse networks
this propagator is Poissonian~\cite{BoPa03},
$g(k|r) = e^{-\bar{k}(r)}\bar{k}(r)^k/k!$,
using which we finally obtain
\begin{equation}
P(k)=
2\left(\frac{\bar{k}}{2}\right)^2\frac{\Gamma(k-2,\bar{k}/2)}{k!}
\sim k^{-3}.
\end{equation}
That is, the node degree distribution is a power law.

This result is remarkable as we have done nothing to enforce this
power law. Network heterogeneity has naturally emerged as a direct
consequence of the basic properties of hyperbolic geometry
underlying the network. Indeed, the observed power law is a
combination of two exponentials~\cite{newman05}, node density
$\rho(r)$ in~(\ref{eq:rho(r)-uniform}) and average degree
$\bar{k}(r)$ in~(\ref{eq:k(r)-uniform-no-N}), both reflecting
the exponential expansion of space in hyperbolic geometry discussed
in Section~\ref{sec:primer}.

\subsection{Quasi-uniform node density at arbitrary negative curvature}

We next relax two constraints in the model. The first constraint is
that the node density is exactly uniform. We let it be
quasi-uniform,
\begin{equation}\label{eq:rho(r)-alpha}
\rho(r) = \alpha\frac{\sinh\alpha r}{\cosh\alpha R-1} \approx
\alpha\,e^{\alpha(r-R)}\sim e^{\alpha r},
\end{equation}
that is, exponential with exponent $\alpha>0$. In terms of the
analogy with trees in
Sections~\ref{sec:primer},\ref{sec:motivations}, this relaxation is
equivalent to assuming that the hidden tree-like hierarchy has the
average branching factor $b=e^\alpha$. Second, we let the curvature
of the hyperbolic space be any $K=-\zeta^2$ with $\zeta>0$. The node
density is exactly uniform now only if $\alpha=\zeta$.

The exact expression for the average degree $\bar{k}(r)$ of nodes at
distance $r$ from the origin is the same as before,
\begin{eqnarray}\label{eq:k(r)-alpha-def}
\bar{k}(r) &=& \frac{N}{2\pi}\iint_{S(r)}\rho(y)\,dyd\theta\\
&=&
N\left\{\int_0^{R-r}\rho(y)\,dy+\frac{1}{\pi}\int_{R-r}^R\rho(y)\theta_y\,dy\right\},\nonumber
\end{eqnarray}
but we cannot compute it exactly to yield an answer analogous
to~(\ref{eq:step-function-final}). However, approximations are easy.
The main approximation deals with the angle $\theta_y$ in
Fig.~\ref{fig:S(r)}. Instead of~(\ref{eq:theta_y}), we now have
according to~(\ref{eq:x-zeta})
\begin{equation}
\cosh\zeta R = \cosh\zeta r \cosh\zeta y - \sinh\zeta r\sinh\zeta y\cos\theta_y,
\end{equation}
which for large $R$, $r$, and $y$ yields
$\theta_y=2e^{\zeta(R-r-y)/2}$. Substituting this $\theta_y$ in the
integral for $\bar{k}(r)$~(\ref{eq:k(r)-alpha-def}), using there the
approximate expression for $\rho(y)$ in~(\ref{eq:rho(r)-alpha}), and
introducing notation $\xi=(\alpha/\zeta)/(\alpha/\zeta-1/2)$, we
obtain
\begin{eqnarray}
\bar{k}(r) &=& N\left\{ \frac{2}{\pi}\xi e^{-\zeta r/2} -
\left(\frac{2}{\pi}\xi-1\right)e^{-\alpha r}\right\}\label{eq:k(r)-alpha-zeta-to-start-with}\\
&=&
\begin{cases}
N(2\xi/\pi)e^{-\zeta r/2}&\text{if $\alpha>\zeta/2$,}\\
N(1+\zeta r/\pi)e^{-\zeta r/2}&\text{if $\alpha\to\zeta/2$,}\\
N(1-2\xi/\pi)e^{-\alpha r}&\text{if $\alpha<\zeta/2$.}
\end{cases}
\end{eqnarray}
The average degree $\bar{k}$ in the whole network is now
\begin{widetext}
\begin{equation}\label{eq:bar-k-alpha-zeta}
\bar{k} = \frac{2}{\pi}\xi^2N
\left\{e^{-\zeta R/2}+e^{-\alpha R}\left[\alpha\frac{R}{2}\left(
\frac{\pi}{4}\left(\frac{\zeta}{\alpha}\right)^2-(\pi-1)\frac{\zeta}{\alpha}+(\pi-2)\right)
-1\right]\right\},
\end{equation}
\end{widetext}
and its limit at $\alpha\to\zeta/2$ is well defined,
\begin{equation}\label{eq:bar-k-alpha=zeta/2}
\bar{k}\xrightarrow[\alpha\to\zeta/2]{}N\frac{\zeta}{2}R\left(1+\frac{\zeta}{2\pi}R\right)e^{-\zeta R/2}.
\end{equation}
If $\alpha/\zeta>1/2$, we can neglect the second term in~(\ref{eq:bar-k-alpha-zeta}),
leading to
\begin{equation}
\bar{k}=\frac{2}{\pi}\xi^2Ne^{-\zeta R/2}.
\end{equation}
That is, the condition controlling the average degree in the
network changes from~(\ref{eq:N.vs.R}) to
\begin{equation}\label{eq:N.vs.R,zeta}
N = \nu\,e^{\zeta R/2},
\end{equation}
where the control parameter $\nu=\pi\bar{k}/(2\xi^2)$. This control is the less accurate,
the closer the $\alpha$ to $\zeta/2$. Indeed, as $\alpha$ approaches
$\zeta/2$, the relative contribution to the total average degree coming
from the second term in~(\ref{eq:bar-k-alpha-zeta}) increases.
In particular, if $\alpha/\zeta=1/2$, then $\xi$ is undefined,
meaning that $\nu$ can no longer be $\pi\bar{k}/(2\xi^2)$.
If instead of solving Eq.~(\ref{eq:bar-k-alpha=zeta/2}) to
find radius $R$ for given $N$ and $\bar{k}$,
we fix $R$ according to~(\ref{eq:N.vs.R,zeta}) with some $\nu\equiv\nu_0$,
then similar to~\cite{KraRe05}, the average degree in~(\ref{eq:bar-k-alpha=zeta/2})
will grow polylogarithmically with the networks size,
\begin{equation}
\bar{k}=\nu_0\ln\frac{N}{\nu_0}\left(1+\frac{1}{\pi}\ln\frac{N}{\nu_0}\right).
\end{equation}

If we neglect the second terms in~(\ref{eq:bar-k-alpha-zeta},\ref{eq:k(r)-alpha-zeta-to-start-with})
at $\alpha/\zeta>1/2$, then using~(\ref{eq:N.vs.R,zeta}), we rewrite (\ref{eq:k(r)-alpha-zeta-to-start-with}) as
\begin{equation}\label{eq:k(r)-alpha-zeta}
\bar{k}(r)=\frac{\bar{k}}{\xi}e^{\zeta(R-r)/2}\sim
e^{-\zeta r/2}.
\end{equation}
That is, somewhat surprisingly, the scaling of the average degree
$\bar{k}(r)$ with radius $r$ does not depend on the exponent
$\alpha>\zeta/2$ of the node density.
Proceeding as in Section~\ref{sec:uniform-density}, the degree
distribution $P(k)$ for $\alpha>\zeta/2$ is then
\begin{equation}\label{eq:P(k)-H2-final-exact}
P(k)=
2\frac{\alpha}{\zeta}\left(\frac{\bar{k}}{\xi}\right)^{2\alpha/\zeta}\frac{\Gamma(k-2\alpha/\zeta,\bar{k}/\xi)}{k!}\sim
k^{-(2\alpha/\zeta+1)}.
\end{equation}
For arbitrary values of $\alpha/\zeta>0$ the degree distribution
scales as
\begin{equation}\label{eq:P(k)-H2-final}
P(k) \sim k^{-\gamma}, \quad \text{with}\; \gamma =
\begin{cases}
2\frac{\alpha}{\zeta}+1&\text{if $\frac{\alpha}{\zeta} \geq \frac{1}{2}$},\\
2&\text{if $\frac{\alpha}{\zeta} \leq \frac{1}{2}$}.
\end{cases}
\end{equation}

We observe that the node density exponent $\alpha$ and the space
curvature $\zeta$ affect the heterogeneity of network topology,
parameterized by $\gamma$, only via their ratio $\alpha/\zeta$. This
result is intuitively expected in view of the analogy to trees
discussed in Sections~\ref{sec:primer},\ref{sec:motivations}, since
a tree with branching factor $b=e^\alpha$ is metrically equivalent
to the two-dimensional hyperbolic space with curvature
$K=-\alpha^2$. In other words, the branching factor of a tree and
the curvature of a hyperbolic space are two different measures of
the same metric property---how fast the space expands.
Result~(\ref{eq:P(k)-H2-final}) states then that the topology of
networks built on top of these metric structures depends only on the
appropriate normalization, $\alpha/\zeta$, between the two measures.

\begin{figure}
    \centerline{\includegraphics[width=3.5in]{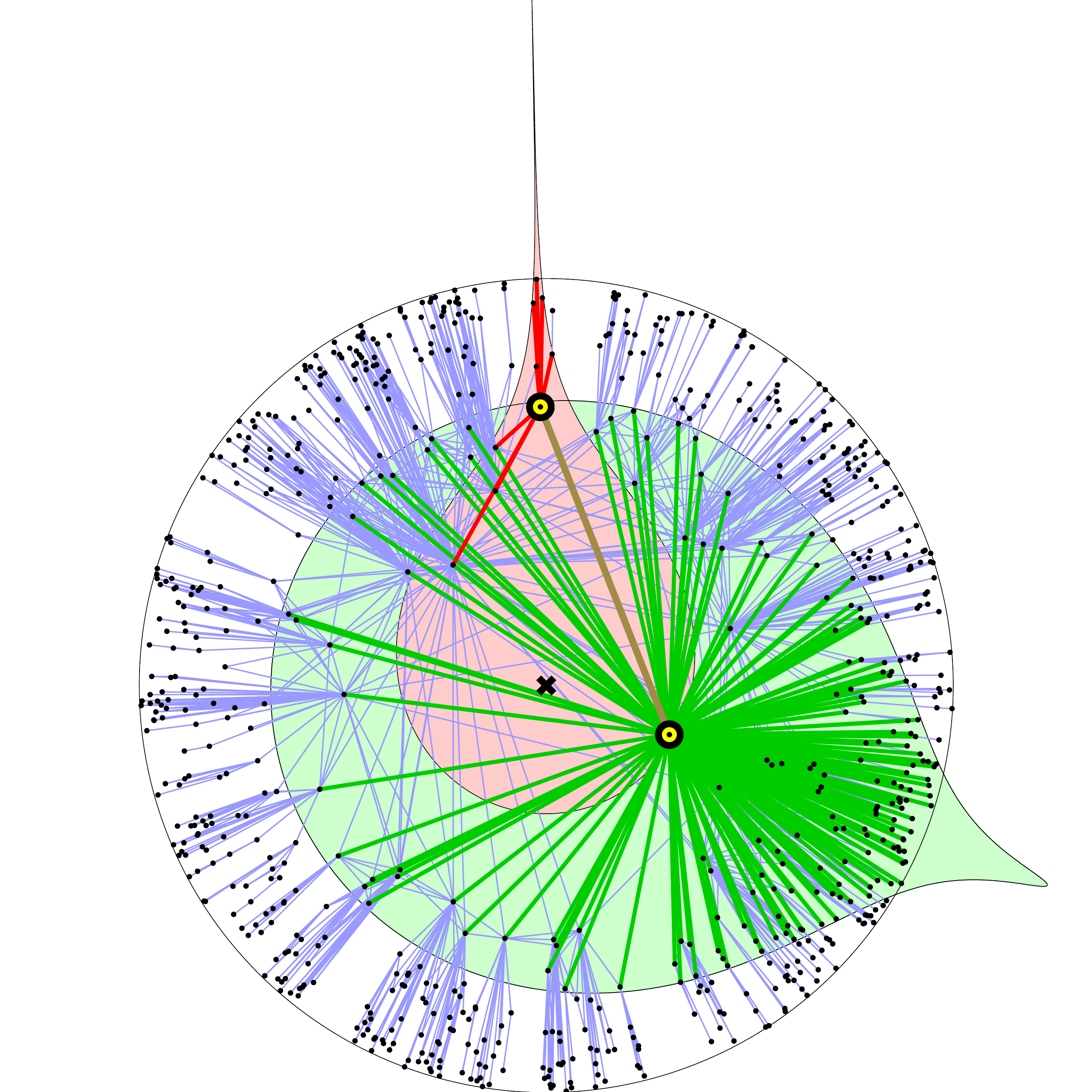}}
    \caption{(Color online)
    A modeled network with $N=740$ nodes,
    power-law exponent $\gamma=2.2$,
    and average degree $\bar{k}=5$
    embedded in the hyperbolic disk of curvature $K=-1$ and radius
    $R=15.5$ centered at the origin shown by the cross.
    For visualization purposes, we use the native hyperbolic space
    representation~(\ref{eq:re.vs.rh-native}).
    Therefore, the shown network occupies a small part of the
    whole hyperbolic plane in Fig.~\ref{fig:poincare}.
    The shaded areas show two hyperbolic disks
    of radius $R$ centered at the circled nodes located
    at distances $r=10.6$ (upper node) and $r=5.0$ (lower node) from the origin.
    The shapes of these disks
    are defined by~(\ref{eq:x-zeta}) with $\zeta=1$, and
    according to the model,
    the circled nodes are connected to all the nodes lying within
    their disks, as indicated by the thick links. In particular, the
    two circled nodes lie within each other's disks.
    The peculiar shape of these disks shows that the hyperbolic distance between
    any two points other than the origin is \emph{not} equal to the Euclidean distance between them.
    In particular, the farther away from the origin are the two
    nodes, located at the same Euclidean distance in the tangential direction,
    the longer is the hyperbolic distance between them, which explains
    why peripheral nodes are not connected to each other, and why
    a majority of links appear radially oriented.
    \label{fig:sample_network}
    }
\end{figure}

The $\mathbb{H}^2$ model described so far has thus only two
parameters, $\alpha/\zeta\geq1/2$ and $\nu>0$, controlling the
degree distribution shape and average degree. The model produces
scale-free networks with any power-law degree distribution exponent
$\gamma=2\alpha/\zeta+1\geq2$. The uniform node density in the
hyperbolic space corresponds to $\alpha=\zeta$, and results in
$\gamma=3$, i.e., the same exponent as in the original preferential
attachment model~\cite{BarAlb99}. Since
$\xi=(\alpha/\zeta)/(\alpha/\zeta-1/2)=(\gamma-1)/(\gamma-2)$,
the average degree of nodes at distance $r$ from the
origin~(\ref{eq:k(r)-alpha-zeta}), and the total average degree in
the network $\bar{k}=2\nu\xi^2/\pi$ are
\begin{eqnarray}
\bar{k}(r)&=&\bar{k}\frac{\gamma-2}{\gamma-1}e^{\zeta(R-r)/2},\label{eq:k(r)-gamma-zeta}\\
\bar{k}&=&\nu\frac{2}{\pi}\left(\frac{\gamma-1}{\gamma-2}\right)^2.\label{eq:bar-k-step}
\end{eqnarray}
A sample network is visualized in Fig.~\ref{fig:sample_network}.

\section{Heterogeneous topology implies hyperbolic geometry}\label{sec:model_equivalence}

In the previous section, we have shown that networks constructed over
hyperbolic spaces naturally possess heterogeneous scale-free degree
distributions. In this section we show the converse. Assuming that a
scale-free network has some metric structure underneath, we show
that metric distances can be naturally rescaled such that the
resulting metric space is hyperbolic.

To accomplish this task we use the $\mathbb{S}^1$ model
from~\cite{SeKrBo08} where the underlying metric structure is
abstracted by the simplest possible compact metric space, circle
$\mathbb{S}^1$. This model generates networks as follows. First, $N$
nodes are placed, uniformly distributed, on a circle of radius
$N/(2\pi)$ so that the node density on the circle is fixed to $1$.
Then each node is assigned its expected degree, which is a random variable
$\kappa$ drawn from the continuous power-law distribution
\begin{equation}\label{eq:rho(kappa)}
\rho(\kappa) = \kappa_0^{\gamma-1}(\gamma-1)\kappa^{-\gamma}, \quad \kappa \ge \kappa_0,
\end{equation}
where $\gamma>2$ is the target degree distribution exponent, and $\kappa_0$
is the minimum expected degree. Finally, each node pair with
expected degrees $(\kappa,\kappa')$ and angular coordinates
$(\theta,\theta')$ located at distance $d=N\Delta\theta/(2\pi)$ over
the circle ($\Delta\theta=\pi-|\pi-|\theta-\theta'||$) is connected
with probability $\tilde{p}(\chi)$, which can be any integrable
function of
\begin{equation}\label{eq:chi}
\chi = \frac{d}{\mu\kappa\kappa'},
\end{equation}
where $\mu>0$ is the parameter controlling the average degree in the
network. This form of the argument of the connection probability
function is the only requirement to ensure that the average degree
$\bar{k}(\kappa)$ of nodes with expected degree $\kappa$ in the
constructed network is indeed proportional to $\kappa$, specifically
$\bar{k}(\kappa)/\bar{k}=\kappa/\bar{\kappa}$, where $\bar{k}$ is
the average degree in the network as before, and
\begin{equation}\label{eq:bar-kappa}
\bar{\kappa}=\int_{\kappa_0}^\infty\kappa\rho(\kappa)d\kappa=\kappa_0\frac{\gamma-1}{\gamma-2}.
\end{equation}
Due to this proportionality, the degree distribution in the network
is indeed power-law distributed with exponent $\gamma$.

To see that condition~(\ref{eq:chi}) ensures
$\bar{k}(\kappa)/\bar{k}=\kappa/\bar{\kappa}$, set $\theta=0$
without loss of generality, let
$I = \int_0^\infty\tilde{p}(\chi)\,d\chi$,
and observe that~\cite{BoPa03}
\begin{eqnarray}\label{eq:k(kappa)-general}
\bar{k}(\kappa) &=& \frac{N}{2\pi}\iint\rho(\kappa')\tilde{p}(\chi)d\kappa'd\theta'\nonumber\\
&=&2\mu\kappa\int_{\kappa_0}^\infty \kappa'\rho(\kappa')d\kappa'\int_0^{N/(2\mu\kappa\kappa')}\tilde{p}(\chi)d\chi\nonumber\\
&=&2\mu I\bar{\kappa}\kappa.
\end{eqnarray}
Since
\begin{equation}\label{eq:bar-k-general}
\bar{k} = \int\bar{k}(\kappa)\rho(\kappa)d\kappa=2\mu I\bar{\kappa}^2,
\end{equation}
we conclude that $\bar{k}(\kappa)=\kappa\bar{k}/\bar{\kappa}$, and
confirm that $\mu$ controls the average degree in the network. We
also note that $\kappa_0$ is a dumb parameter, which can be set to
$\kappa_0=\bar{k}(\gamma-2)/(\gamma-1)$ leading to
$\bar{k}(\kappa)=\kappa$.

We now establish the equivalence between this $\mathbb{S}^1$ model
and the $\mathbb{H}^2$ model described in the previous section.
To do so, we need to find a change of variables from
$\kappa$, expected degree of a node, to $r$, its radial coordinate
on a disk of radius $R$, such that if variable $\kappa$ is
power-law distributed according to~(\ref{eq:rho(kappa)}), then
after this $\kappa$-to-$r$ change of variables, variable $r$ is
exponentially distributed according to~(\ref{eq:rho(r)-alpha}).
The change of variables that accomplishes this task is
\begin{equation}\label{eq:kappa(r)}
\kappa = \kappa_0 e^{\zeta(R-r)/2},
\end{equation}
where $\zeta>0$ is a parameter defining $\alpha$ in~(\ref{eq:rho(r)-alpha})
after this change of variables. The resulting value of $\alpha$ is
$\alpha=\zeta(\gamma-1)/2$, which is the same relationship
among $\alpha$, $\zeta$, and $\gamma$ as
in~(\ref{eq:P(k)-H2-final}). In other words, after the $\kappa$-to-$r$
mapping above, the nodes get distributed
on the disk as in the $\mathbb{H}^2$ model, suggesting that
parameter $\zeta$ is actually the space curvature.

To check if it is indeed the case, and if the two models are indeed
equivalent, we have to verify that the pairs of nodes connected or
disconnected in the $\mathbb{S}^1$ model with expected degree
$\kappa$ mapped to radial coordinate $r$ correspond to,
respectively, connected or disconnected nodes in the native
$\mathbb{H}^2$ model. That is, we have to demonstrate that the
connection probabilities in the two models are consistent,
$p(x)=\tilde{p}(\chi)$. To show this we first fix the disk
radius $R$ to its value in the $\mathbb{H}^2$
model~(\ref{eq:N.vs.R,zeta}), and then observe that if we set
\begin{equation}\label{eq:chi.vs.x}
\nu=\pi\mu\kappa_0^2, \quad\text{yielding
$\bar{k}=\nu I\frac{2}{\pi}\left(\frac{\gamma-1}{\gamma-2}\right)^2$},
\end{equation}
then the change of variables~(\ref{eq:kappa(r)}) maps the argument
$\chi$ of the connection probability in the $\mathbb{S}^1$
model~(\ref{eq:chi}) to
\begin{equation}\label{eq:chi(x)}
\chi = e^{\zeta(x-R)/2},
\end{equation}
where $x$ is equal to the second approximation of the hyperbolic
distance in~(\ref{eq:x-approx}). Therefore, the connection
probability $p(x)$ in the $\mathbb{H}^2$ model is approximately
equal to the connection probability
$\tilde{p}\left(e^{\zeta(x-R)/2}\right)$ in the $\mathbb{S}^1$
model. In particular, the step function connection
probability~(\ref{eq:p(x)-step}) in the $\mathbb{H}^2$ model
corresponds to
\begin{equation}
\tilde{p}(\chi)=\Theta(1-\chi)
\end{equation}
in the $\mathbb{S}^1$ model. The integral $I$ of this connection
probability is obviously $1$, so that the $\bar{k}$ vs $\mu$
relationship~(\ref{eq:bar-k-general}) in the $\SM$ model becomes
$\bar{k}=2\mu\left[\kappa_0(\gamma-1)/(\gamma-2)\right]^2$, which is
consistent with the condition $\nu=\pi\mu\kappa_0^2$~(\ref{eq:chi.vs.x})
given the $\bar{k}$.vs.$\nu$ relationship in
the $\mathbb{H}^2$ model~(\ref{eq:bar-k-step}). As the final
consistency check, we observe that the substitution of the
$\kappa$-to-$r$ mapping~$(\ref{eq:kappa(r)})$ into the
proportionality $\bar{k}(\kappa)=\kappa\bar{k}/\bar{\kappa}$ in the
$\SM$ model yields~(\ref{eq:k(r)-gamma-zeta}) in the $\HM$
model. That is, the average degrees $\bar{k}(r)$ of nodes with
radial coordinate $r$ in the $\mathbb{S}^1$ and $\mathbb{H}^2$ models
are the same.

The two models are thus equivalent, and with the appropriate choice
of parameters, generate statistically the same ensembles of
networks, which one can confirm in simulations. In this section the
network metric structure has been modeled the simplest way, by
circle $\SM=\partial\HM$, which by no means is the only possibility
for the hyperbolic space boundary $\partial X$, see
Section~\ref{sec:primer}. Therefore, the established equivalence
between the $\SM$ and $\HM$ models suggests that as soon as a
heterogeneous network has some metric structure induced by distances
$d$ on $\partial X$, this metric structure can be rescaled by node
degrees $\kappa$ to become hyperbolic, using appropriate
modifications of~(\ref{eq:chi},\ref{eq:chi(x)}). The heterogeneous
degree distribution effectively adds an additional dimension to
$\partial X$ (the radial dimension in the $\SM=\partial\HM$ case),
such that the resulting space $X$ ($\HM$ in the considered case) is
hyperbolic, a mechanism conceptually similar to how time in special
relativity, or gravity in~\cite{Maldacena99,GuKl98,Witten98} makes
the higher-dimensional (time)space hyperbolic. In other words,
hyperbolic geometry naturally emerges from network heterogeneity,
the same way as network heterogeneity emerges from hyperbolic
geometry in the previous section.

\section{Hyperbolic geometry versus statistical mechanics}\label{sec:fermi}

In this section we relax the final constraint in the model that the
connection probability is a step function, and provide a statistical
mechanics interpretation of the resulting network ensemble.

Since $\tilde{p}(\chi)$ can be any integrable function in the
$\mathbb{S}^1$ version of the model, $p(x)$ can be any function in
the $\mathbb{H}^2$ version. Given this freedom, we consider the
following family of connection probability functions,
\begin{equation}\label{eq:p(x)-fermi}
p(x)=\frac{1}{e^{\beta(\zeta/2)(x-R)}+1}=\frac{1}{\chi^{\beta}+1}=\tilde{p}(\chi),
\end{equation}
parameterized by $\beta>0$. The $\tilde{p}(\chi)$ function is
integrable for any $\beta>1$,
\begin{equation}\label{eq:I(beta)}
I=\int_0^\infty\tilde{p}(\chi)\,d\chi=\left(\frac{\beta}{\pi}\sin\frac{\pi}{\beta}\right)^{-1}.
\end{equation}
However, we will not restrict $\beta>1$, and will also consider
$\beta\in[0,1)$.

The main motivation for the connection probability
choice~(\ref{eq:p(x)-fermi}) is that it casts the ensemble of graphs
in the model to exponential random
graphs~\cite{DoMeSa03b,PaNe04,GaLo09,AnBi09,Bianconi09}. Exponential
random graphs are maximally random graphs subjected to specific
constraints, each constraint associated with an auxiliary field or
Lagrangian multiplier in the standard entropy maximization approach,
commonly used in statistical mechanics. Each graph $G$ in the
ensemble has probability weight $P(G)=e^{-H(G)}/Z$, where $H(G)$ is
the graph Hamiltonian, and $Z=\sum_Ge^{-H(G)}$ is the partition
function. For example, the ensemble of graphs in the configuration
model, i.e., graphs with a given degree sequence $\{k_i\}$, is
defined by Hamiltonian
$H(G)=\sum_i\omega_ik_i=\sum_{ij}\omega_ia_{ij}=\sum_{i<j}(\omega_i+\omega_j)a_{ij}$,
where $\omega_i$ are the auxiliary fields coupled to nodes $i$, and
$\{a_{ij}\}$ is $G$'s adjacency matrix. A natural generalization of
this ensemble~\cite{PaNe04} is given by the Hamiltonian $H(G) =
\sum_{i<j} \omega_{ij}a_{ij}$ in which the auxiliary fields are
coupled not to nodes $i$ but to links $ij$. The partition function
is then
\begin{equation}\label{eq:Z}
Z=\prod_{i<j}\left(1+e^{-\omega_{ij}}\right),
\end{equation}
and the probability of link existence between nodes $i$ and $j$ is
given by~\cite{PaNe04}
\begin{equation}\label{eq:p_ij}
p_{ij}=-\frac{\partial\ln Z}
{\partial\omega_{ij}}=\frac{1}{e^{\omega_{ij}}+1}.
\end{equation}
The connection probability~(\ref{eq:p(x)-fermi}) thus interprets the
auxiliary fields $\omega_{ij}$ in this ensemble as a linear function
of hyperbolic distances $x_{ij}$ between nodes in the ensemble of
graphs generated by our model,
\begin{equation}\label{eq:omega_ij}
\omega_{ij}=\beta\frac{\zeta}{2}(x_{ij}-R),
\end{equation}
which makes the two ensembles identical.

The connection probability~(\ref{eq:p(x)-fermi}) is nothing
but the Fermi-Dirac distribution. It appears because we allow only
one link between a pair of nodes. If we allowed multiple links, or
if we considered weighted networks, the resulting link statistics
would be Bose-Einstein~\cite{PaNe04,GaLo09}. Hyperbolic distances
$x$ in~(\ref{eq:p(x)-fermi}) can now be interpreted as energies
of fermionic links, whereas hyperbolic disk radius $R$ is the chemical
potential, $2/\zeta$ is the Boltzmann constant, and $\beta=1/T$ is
the inverse temperature. The ensemble is grand canonical with the number
of particles or links $M$ fixed on average. The standard definition of
the chemical potential is then
\begin{equation}\label{eq:R-defined}
M={N\choose2}\int_0^{2R} g(x)p(x)\,dx,
\end{equation}
where $g(x)$ is the degeneracy of energy level $x$. In our case,
$g(x)$ is the probability that two nodes are located at distance
$x$ from each other. We can compute this probability to yield
\begin{equation}
g(x)=\frac{\zeta}{\pi}\left(\frac{\gamma-1}{\gamma-2}\right)^2
e^{\zeta(x-2R)/2}+\left(a+bx\right)e^{\alpha(x-2R)},
\end{equation}
where $a,b$ are some constants, and $\gamma=2\alpha/\zeta+1$.
Substituting this $g(x)$ in definition~(\ref{eq:R-defined}),
using $M=\bar{k}N/2$ there, and keeping the leading terms, we get
\begin{equation}\label{eq:bar-k-stat-mech}
\bar{k}=
N\left[I\frac{2}{\pi}\left(\frac{\gamma-1}{\gamma-2}\right)^2e^{-\zeta R/2}+
\frac{e^{-\beta\zeta R/2}}{(1-\beta)c}\right],
\end{equation}
where $c$ is another constant which we determine in the next
section. If $\beta>1$, we neglect the second term above, and observe
that the standard definition of the chemical potential in
statistical mechanics~(\ref{eq:R-defined}) yields the same result
as~(\ref{eq:N.vs.R,zeta}), obtained using purely geometric
arguments. The same observation applies for the parameter
$\nu=Ne^{-\zeta R/2}$ that we get from~(\ref{eq:bar-k-stat-mech}): it
is the same as in~(\ref{eq:bar-k-general}) with
$\mu=\nu/(\pi\kappa_0^2)$ and $\bar{\kappa}$ in~(\ref{eq:bar-kappa}),
or as in~(\ref{eq:bar-k-step}) if temperature $T=0$, so that $I=1$.

At $T=0$ the system is in the ground, most degenerate state, and all
$M$ links occupy the lowest energy levels until all of them are
filled. In this ground state, Fermi distribution~(\ref{eq:p(x)-fermi}) converges to
the step function~(\ref{eq:p(x)-step}), which {\it a posteriori}
justifies our choice there. At higher temperatures the fermionic
particles start populating higher energy states, and at $T=1$ we
have a phase transition caused by the divergence of
$\tilde{p}(\chi)$ leading to a discontinuity of the partition
function~(\ref{eq:Z}). This discontinuity is due to the
discontinuity of the chemical potential $R$. We see
from~(\ref{eq:bar-k-stat-mech},\ref{eq:I(beta)}) that $R$ diverges
as $\sim-\ln(\beta-1)$ at $\beta\to1_+$. If $\beta<1$, then the
second term in~(\ref{eq:bar-k-stat-mech}) is the leading term, and
instead of~(\ref{eq:N.vs.R,zeta}) we have
\begin{equation}\label{eq:N.vs.R,zeta,beta,incomplete}
N=\bar{k}(1-\beta)c\,e^{\beta\zeta R/2},
\end{equation}
so that at $\beta\to1_-$, the chemical potential $R$ diverges as
$\sim-\ln(1-\beta)$. We investigate what effect this phase
transition has on network topology in the next two sections.

\section{Degree distribution at non-zero temperature}\label{sec:degree_distribution}

\subsection{$\beta>1$}

Since the connection probability $\tilde{p}(\chi)$
in~(\ref{eq:p(x)-fermi}) is integrable in this cold regime, we
immediately conclude that the degree distribution is the same power
law as at the zero temperature, while the average degree is
$\bar{k}=2\mu I\bar{\kappa}^2$~(\ref{eq:bar-k-general}) with $I$
in~(\ref{eq:I(beta)}). In view of the equivalence between the $\SM$
and $\HM$ models established in Section~\ref{sec:model_equivalence},
the power-law exponent $\gamma>2$ is related to the $\HM$ model
parameters $\zeta>0$ and $\alpha>\zeta/2$ via
$\gamma=2\alpha/\zeta+1$, as at $T=0$. The chemical potential is
$R=(2/\zeta)\ln(N/\nu)$ with
$\nu=\pi\mu\kappa_0^2$~(\ref{eq:N.vs.R,zeta},\ref{eq:chi.vs.x}).

\subsection{$\beta<1$}

In this hot regime, the connection probability $\tilde{p}(\chi)$
diverges, and we have to renormalize its integral
$I=\int\tilde{p}(\chi)d\chi$. Specifically, instead of integrating
to infinity as in~(\ref{eq:k(kappa)-general}), we have to explicitly
cut off the integration at the maximum value of
$\chi_{\max}=N/(2\mu\kappa\kappa')$. The exact value of
$\int_0^{\chi_{\max}}\tilde{p}(\chi)d\chi$ with $\tilde{p}(\chi)$
in~(\ref{eq:p(x)-fermi}) is
$_2H_1(1,\beta^{-1};1+\beta^{-1};-\chi_{\max}^\beta)\chi_{\max}$,
where $_2H_1$ is the Gauss hypergeometric function. The leading term
of this product for large $\chi_{\max}$ and $\beta\in[0,1)$ is
$\chi_{\max}^{1-\beta}/(1-\beta)$, substituting which into the
expression for the average degree in the $\SM$
model~(\ref{eq:k(kappa)-general}) we get
\begin{eqnarray}
\frac{\bar{k}(\kappa)}{\langle k \rangle}&=&\frac{\kappa^\beta}{\langle\kappa^\beta\rangle},\label{eq:k(kappa)-beta<1}\\
\langle k \rangle\equiv\bar{k}&=&(2\mu)^\beta\langle\kappa^\beta\rangle^2\frac{N^{1-\beta}}{1-\beta},\label{eq:bar-k-beta<1}\\
\langle\kappa^\beta\rangle&=&\int_{\kappa_0}^\infty\kappa^\beta\rho(\kappa)d\kappa=\kappa_0^\beta\frac{\tilde{\gamma}-1}{\tilde{\gamma}-\beta-1},\label{eq:bar-kappa-beta<1}
\end{eqnarray}
where $\tilde{\gamma}$ is the input value of the $\gamma$-parameter in the
$\SM$ model, i.e., the distribution of the hidden variable $\kappa$
is
$\rho(\kappa)=\kappa_0^{\tilde{\gamma}-1}(\tilde{\gamma}-1)\kappa^{-\tilde{\gamma}}$.
We introduce a new notation for this parameter to differentiate it
from the value of power-law exponent $\gamma$ in generated networks,
which is different from $\tilde{\gamma}$ in this hot regime. Indeed, since
the average degree $\bar{k}(\kappa)$ of nodes with hidden variable $\kappa$ is no
longer proportional to $\kappa$ but to $\kappa^\beta$~(\ref{eq:k(kappa)-beta<1}), the degree
distribution in the modeled networks is
\begin{equation}\label{eq:gamma-beta<1}
P(k) \sim k^{-\gamma},
\quad\text{with}\;\gamma=(\tilde{\gamma}-1)T+1.
\end{equation}

The mapping to the $\HM$ model is achieved via the same change of
variables~(\ref{eq:kappa(r)}), and by requiring that
$\chi=e^{\zeta(x-R)/2}$. Performing this change of variables, and
noticing that
$(\tilde{\gamma}-1)/(\tilde{\gamma}-\beta-1)=(\gamma-1)/(\gamma-2)$, we obtain
the following key relationships in the $\HM$ model:
\begin{eqnarray}
\gamma&=&2\frac{\alpha}{\zeta}T+1, \quad \tilde{\gamma} = 2\frac{\alpha}{\zeta}+1,\label{eq:gamma-beta<1-H2}\\
\bar{k}(r)&=&\bar{k}\frac{\gamma-2}{\gamma-1}e^{\beta\zeta(R-r)/2},\label{eq:k(r)-beta<1}\\
\bar{k}&=&\frac{\nu}{1-\beta}\left(\frac{2}{\pi}\right)^\beta\left(\frac{\gamma-1}{\gamma-2}\right)^2,\label{eq:bar-k-beta<1-H2}\\
N&=&\nu\,e^{\beta\zeta R/2},\quad \nu=\left(\pi\mu\kappa_0^2\right)^\beta N^{1-\beta}\label{eq:N.vs.R-beta<1-H2}.
\end{eqnarray}
The last two equations fill in the $c$ coefficient in the expression
for the chemical potential~(\ref{eq:N.vs.R,zeta,beta,incomplete}).

Finally, we note that in the hot regime the admissible range of
input parameters controlling the degree distribution exponent
$\gamma$ is $\tilde{\gamma}>\beta+1$ ($\SM$) or $\alpha>\beta\zeta/2$
($\HM$), both yielding $\gamma>2$.

\section{Clustering as a function of temperature}\label{sec:clustering}

\subsection{$\beta>1$}

\begin{figure}
\includegraphics[width=3in]{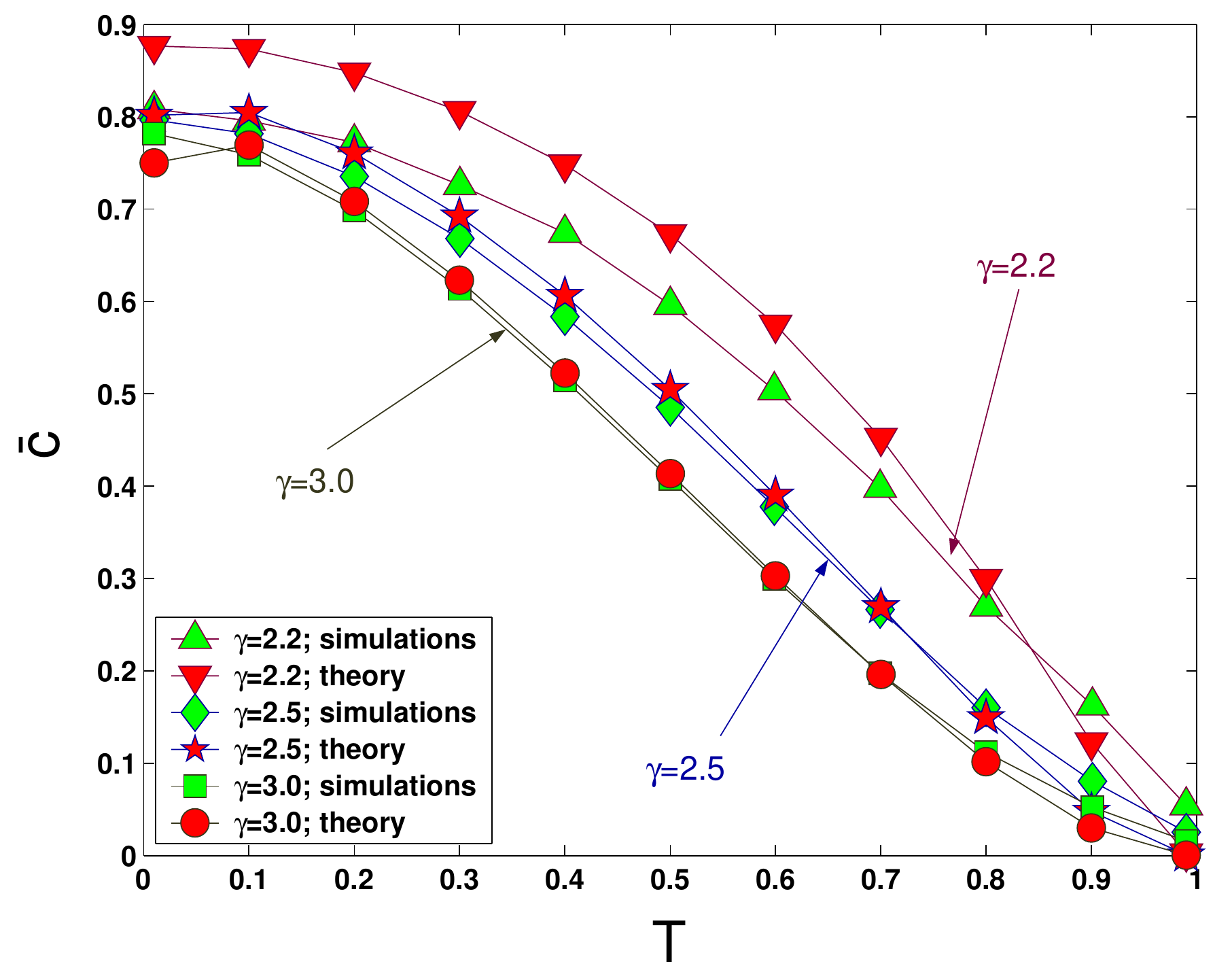}
\caption{(Color online) Average clustering $\bar{c}$ as a function of temperature
$T=1/\beta\in[0,1)$. The simulation results are averaged across
$100$ networks with average degree $\bar{k}=6$ and $N=10^5$ nodes
each. The average clustering is calculated excluding nodes of
degree~$1$. The theoretical results are obtained via the numerical
integration of
$\bar{c}=\int_{\kappa_0}^\infty\bar{c}(\kappa)\rho(\kappa)d\kappa$
with $\bar{c}(\kappa)$ given by Eq.~(\ref{eq:c(kappa)}). The
stronger disagreement between simulations and theory for smaller
values of $\gamma$ is due to the increasingly pronounced finite-size
effects~\cite{BoPaVe04}. \label{fig:c(T)}}
\end{figure}

In the cold regime, the average clustering $\bar{c}$ is a decreasing
function of temperature, see Fig.~\ref{fig:c(T)}. Clustering is
maximized at $T=0$, and it gradually, almost linearly, decreases to
zero at the phase transition point $T=1$.

Unfortunately, $\bar{c}$ cannot be computed analytically, but some
estimates for specific values of $\beta$ are possible. The average
clustering $\bar{c}(\kappa)$ of nodes with expected degree $\kappa$
in the $\SM$ model is the probability that two nodes with expected
degrees and angular coordinates $(\kappa',\theta')$ and
$(\kappa'',\theta'')$, both connected to node with $(\kappa,0)$ (we
set $\theta=0$ without loss of generality), are connected to each
other. Introducing notations for the three rescaled distances
$\chi'=N\theta'/(2\pi\mu\kappa\kappa')$,
$\chi''=N\theta''/(2\pi\mu\kappa\kappa'')$, and
$\chi=N\Delta\theta/(2\pi\mu\kappa'\kappa'')$, where
$\Delta\theta=|\theta'-\theta''|$, this probability is given
by~\cite{BoPa03}
\begin{eqnarray}\label{eq:c(kappa)-def}
\bar{c}(\kappa) &=& \left(\frac{N}{\bar{k}(\kappa)}\right)^2
\iint_{\kappa_0}^\infty d\kappa'd\kappa''\rho(\kappa')\rho(\kappa'')\nonumber\\
&\times&\iint_{-\pi}^\pi d\theta'd\theta''\tilde{p}(\chi')\tilde{p}(\chi'')\tilde{p}(\chi).
\end{eqnarray}
Changing the integration variables from $\theta'$ and $\theta''$ to
$\chi'$ and $\chi''$ in the second integral, extending the
integration limits to infinity, and using the expression for the
average degree in the model~(\ref{eq:k(kappa)-general}), yield
\begin{eqnarray}\label{eq:c(kappa)}
\bar{c}(\kappa) &=&
\frac{1}{\left(2I\bar{\kappa}\right)^2}
\iint_{\kappa_0}^\infty d\kappa'd\kappa''\kappa'\rho(\kappa')\kappa''\rho(\kappa'')\\
&\times&\iint_{-\infty}^\infty
d\chi'd\chi''\tilde{p}(|\chi'|)\tilde{p}(|\chi''|)\tilde{p}\left(\kappa\left|\frac{\chi'}{\kappa''}-\frac{\chi''}{\kappa'}\right|\right).\nonumber
\end{eqnarray}

\begin{figure}
\centerline{\includegraphics[width=2.5in]{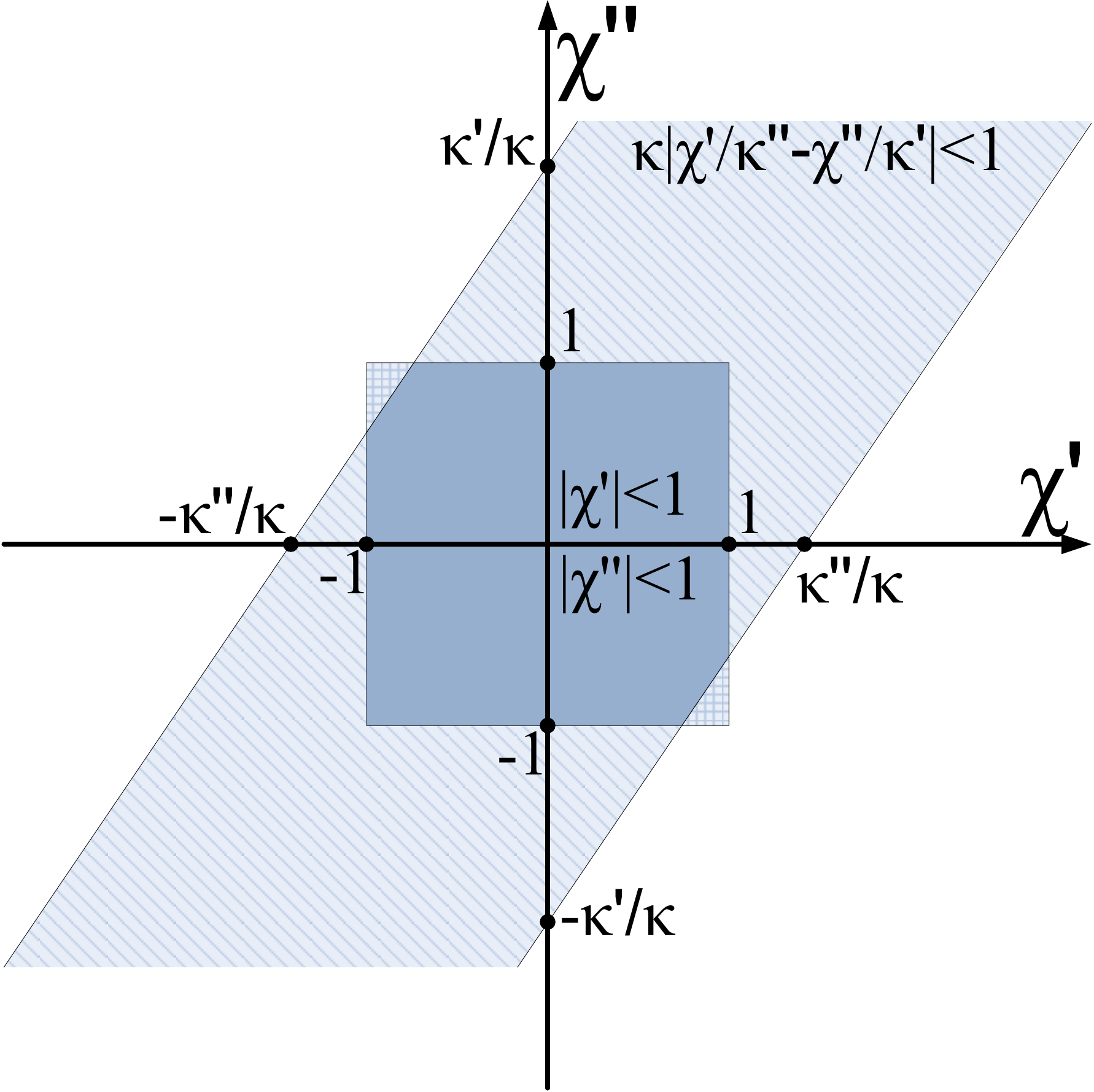}}
\caption{(Color online) The inner integral in~(\ref{eq:c(kappa)}) at the zero
temperature is the dark shaded area in the center.
\label{fig:chichi}}
\end{figure}

At $T=0$, $I=1$, while $\tilde{p}(\chi)\to\Theta(1-\chi)$. Therefore,
the inner integral in the last expression reduces to the area of the
intersection of the square defined in the $(\chi',\chi'')$
coordinates by $\{|\chi'|<1;|\chi''|<1\}$, and the stripe
$\kappa|\chi'/\kappa''-\chi''/\kappa'|<1$, see
Fig.~\ref{fig:chichi}. For small $\kappa$, the stripe is so wide for
almost any combination of $(\kappa',\kappa'')$ that it fully
contains the square, whose area is $4$, so that
$\bar{c}(\kappa_0)\approx1$ proving that clustering is maximized at
the zero temperature. Recall that clustering cannot be $1$ for all
node degrees because of structural constraints~\cite{SeBo05}. For
arbitrary values of $\kappa$, the exact expression for the
intersection area involves cumbersome combinatorial conditions for
the mutual relationship among $\kappa$, $\kappa'$, and $\kappa''$,
which make taking the outer integral in~(\ref{eq:c(kappa)})
problematic. However one can check that for large $\kappa$,
$\bar{c}(\kappa)=g(\gamma)\kappa_0/\kappa$, where $g(\gamma)$ is a
decreasing function of $\gamma$.

For any other values of $T\in(0,1)$, the inner integral
in~(\ref{eq:c(kappa)}) can be taken by residues, but the number of
poles depends on $\beta=1/T$. At $\beta=2$, for example, the inner
integral is
$\pi^2\kappa'\kappa''/(\kappa\kappa'+\kappa\kappa''+\kappa'\kappa'')$,
so that for $\gamma=3$ we have the exact expression for
$\bar{c}(\kappa)$
\begin{eqnarray}
\bar{c}(\kappa) &=&
\kappa_0\{(2\kappa+\kappa_0)\ln(2\kappa+\kappa_0)-2(\kappa+\kappa_0)\ln(\kappa+\kappa_0)\nonumber\\
&+&\kappa_0\ln\kappa_0\}/\kappa^2,
\end{eqnarray}
and $\bar{c}(\kappa_0)=\ln(27/16)=0.52$, while
$\bar{c}(\kappa)=(\ln4)\kappa_0/\kappa$ for large $\kappa$. For
other values of $\gamma$, one can show that
$\bar{c}(\kappa)=\tilde{g}(\gamma)\kappa_0/\kappa$, where
$\tilde{g}(\gamma)$ is also a decreasing function of $\gamma$.

In other words, the degree-dependent clustering $\bar{c}(\kappa)$
decays with $\kappa$ as $\sim\kappa^{-1}$, an effect that was
considered as a signature of the hierarchical organization of
complex networks~\cite{RaSoMoOlBa02,RaBa03}.

\subsection{$\beta<1$}

In the hot regime, temperature has no effect on clustering, which is
always zero for large networks. This effect can be confirmed in
simulations, and seen analytically. Indeed, observe that in view
of~(\ref{eq:k(kappa)-beta<1}-\ref{eq:bar-kappa-beta<1}), the
$\theta$-to-$\chi$ change of variables,
turning~(\ref{eq:c(kappa)-def}) into~(\ref{eq:c(kappa)}), now yields
the pre-factor in the latter equal to
$\left[\left(\mu\kappa\right)^{1-\beta}\left(1-\beta\right)/\left(2^\beta\langle\kappa^\beta\rangle
N^{1-\beta}\right)\right]^2$ instead of $1/(2I\bar{\kappa})^2$. This
new pre-factor is obviously zero in the thermodynamic limit.

\section{Connection to the configuration model and classical random graphs}~\label{sec:random_graphs}

Since clustering does not depend on temperature in the hot regime,
while the power-law exponent~(\ref{eq:gamma-beta<1-H2}) depends on
temperature via the ratio $T/\zeta$, we can let $T\to\infty$ and
$\zeta\to\infty$, but fix their ratio to be a new parameter
$\eta=\zeta/T$. With this parameter the key
equations~(\ref{eq:gamma-beta<1-H2},\ref{eq:bar-k-beta<1-H2},\ref{eq:N.vs.R-beta<1-H2})
in the $\HM$ model become
\begin{equation}\label{eq:alpha',zeta',c'}
\gamma=2\frac{\alpha}{\eta}+1,\quad
\bar{k}=\nu\left(\frac{\gamma-1}{\gamma-2}\right)^2,\quad N=\nu\,e^{\eta R/2}.
\end{equation}
But since curvature $\zeta=\infty$, the last
$\Delta\theta$-dependent term in the expression for the hyperbolic
distance~(\ref{eq:x-approx}) is zero. Since this term reflects the
presence of the metric structure in the network, its disappearance
effectively destroys this structure. More formally, the network
metric structure becomes degenerate, because the hyperbolic distance
$x_{ij}$ between a pair of nodes $i$ and $j$ reduces to the sum of
their radial coordinates, $x_{ij}=r_i+r_j$, as a result of which the
auxiliary fields~(\ref{eq:omega_ij}) decouple,
$\omega_{ij}=\omega_i+\omega_j$, where $\omega_i=\eta(r_i-R/2)/2$.
Therefore, the probability $p_{ij}$ of the existence of link $ij$
in~(\ref{eq:p_ij}) depends now only on the product of $i,j$'s
expected degrees $\bar{k}(r_i),\bar{k}(r_j)$ given
by~(\ref{eq:k(r)-beta<1}),
$p_{ij}=[\bar{k}(r_i)\bar{k}(r_j)]/[\bar{k}N]$, so that the network
ensemble becomes the ensemble of networks in the configuration
model, i.e., the ensemble of graphs with given expected
degrees~\cite{ChLu02b}.

Alternatively, we can keep both $\alpha$ and $\zeta$ finite while
heating the networks up by increasing $T\to\infty$. In this case,
Eqs.~(\ref{eq:gamma-beta<1-H2}-\ref{eq:N.vs.R-beta<1-H2}) converge
to $\gamma\to\infty$, $\bar{k}(r)\to\bar{k}$, $\bar{k}\to\nu$, and
$R\to\infty$, while the Fermi-Dirac connection
probability~(\ref{eq:p(x)-fermi}) becomes uniform $p(x)\to
p=\bar{k}/N$. That is, all nodes get uniformly distributed on the
boundary at infinity $\partial\HM$, and each pair of nodes is
connected with the same probability $p$, independent of their
distances. We note that the distance between two points
$i,j\in\partial\HM$ with angular coordinates $\theta_i,\theta_j$ is
$x_{ij}=\sin(\Delta\theta_{ij}/2)$~\cite{BuSch07-book}---compare
with~(\ref{eq:x-approx}) in $\HM$ and with $x_{ij}=r_i+r_j$ in the
other limiting case, the configuration model. The limiting degree
distribution is Poissonian ($\gamma\to\infty$), and the network
ensemble converges to the ensemble of classical random graphs ${\cal
G}_{N,p}$ with a given average degree $\bar{k}=pN$~\cite{ErRe59}.
The network in this case loses not only its metric structure, but
also its hierarchical heterogeneous organization.

\begin{table}
\begin{centering}
\caption{Network properties in the model---average degree~$\bar{k}$,
power-law exponent~$\gamma$, and average clustering~$\bar{c}$---and
the model parameters controlling these properties, with references
to the corresponding equations. \label{table:parameters}} {\small
\begin{tabular}{|c|c|c|c|c|}\hline
\multirow{2}{*}{Property} & \multicolumn{2}{|c|}{Cold regime} & \multicolumn{2}{|c|}{Hot regime} \\\cline{2-5}
& $\SM$ & $\HM$ & $\SM$ & $\HM$ \\ \hline
$\bar{k}$ & $\mu$~(\ref{eq:bar-k-general}) &
    $\nu$~(\ref{eq:chi.vs.x}) & $\mu$~(\ref{eq:bar-k-beta<1}) & $\nu$~(\ref{eq:bar-k-beta<1-H2}) \\\hline
$\gamma$  & $\gamma$~(\ref{eq:rho(kappa)}) & $\alpha/\zeta$~(\ref{eq:P(k)-H2-final}) &
    $\tilde{\gamma}$~(\ref{eq:gamma-beta<1})& $\alpha/\zeta$~(\ref{eq:gamma-beta<1-H2}) \\\hline
$\bar{c}$ & \multicolumn{2}{|c|}{$\beta=1/T$~(Fig.~\ref{fig:c(T)})} & \multicolumn{2}{|c|}{$0$} \\\hline
\end{tabular}
}
\end{centering}
\end{table}

\begin{figure*}
\centerline{
\includegraphics[width=2.3in]{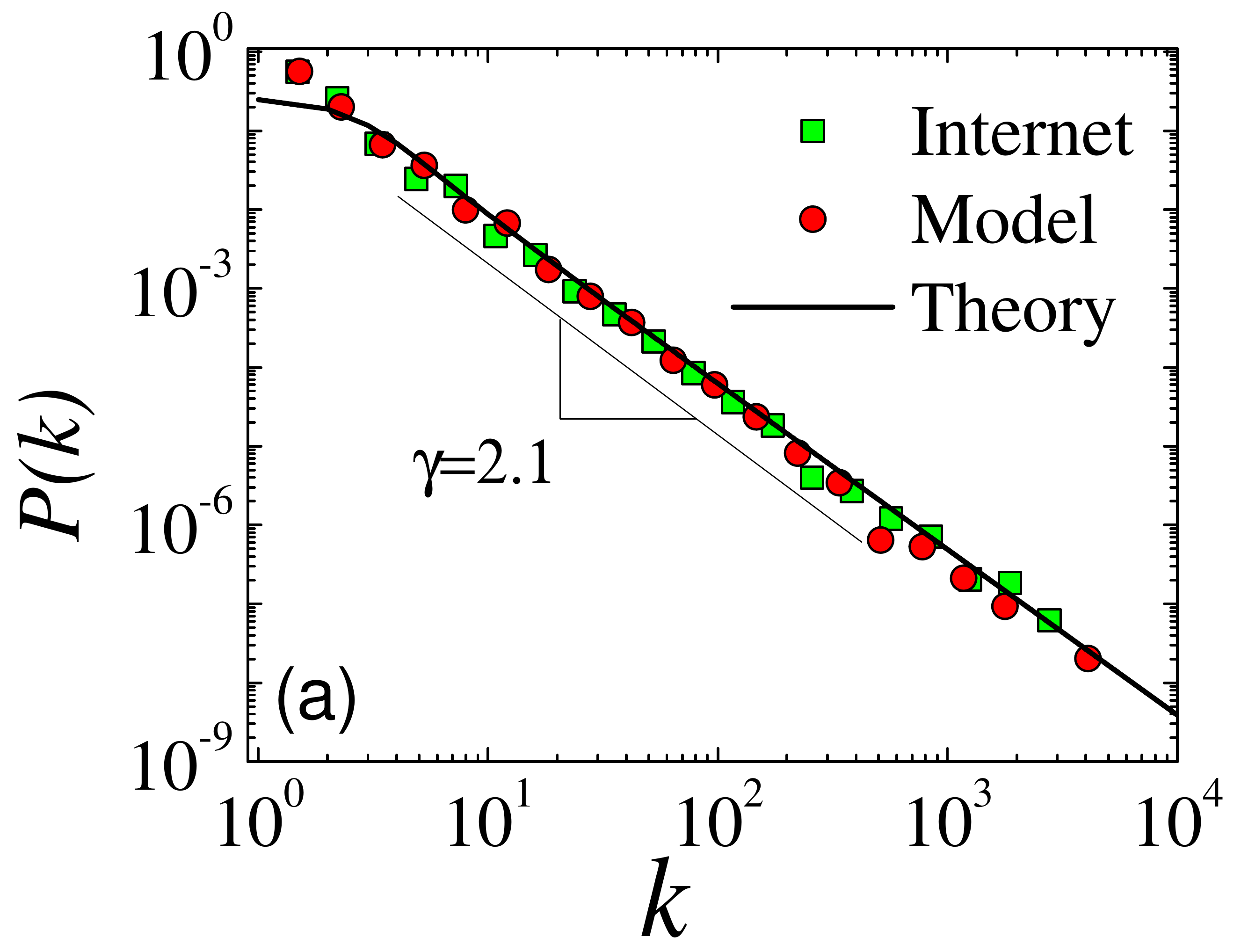}
\includegraphics[width=2.3in]{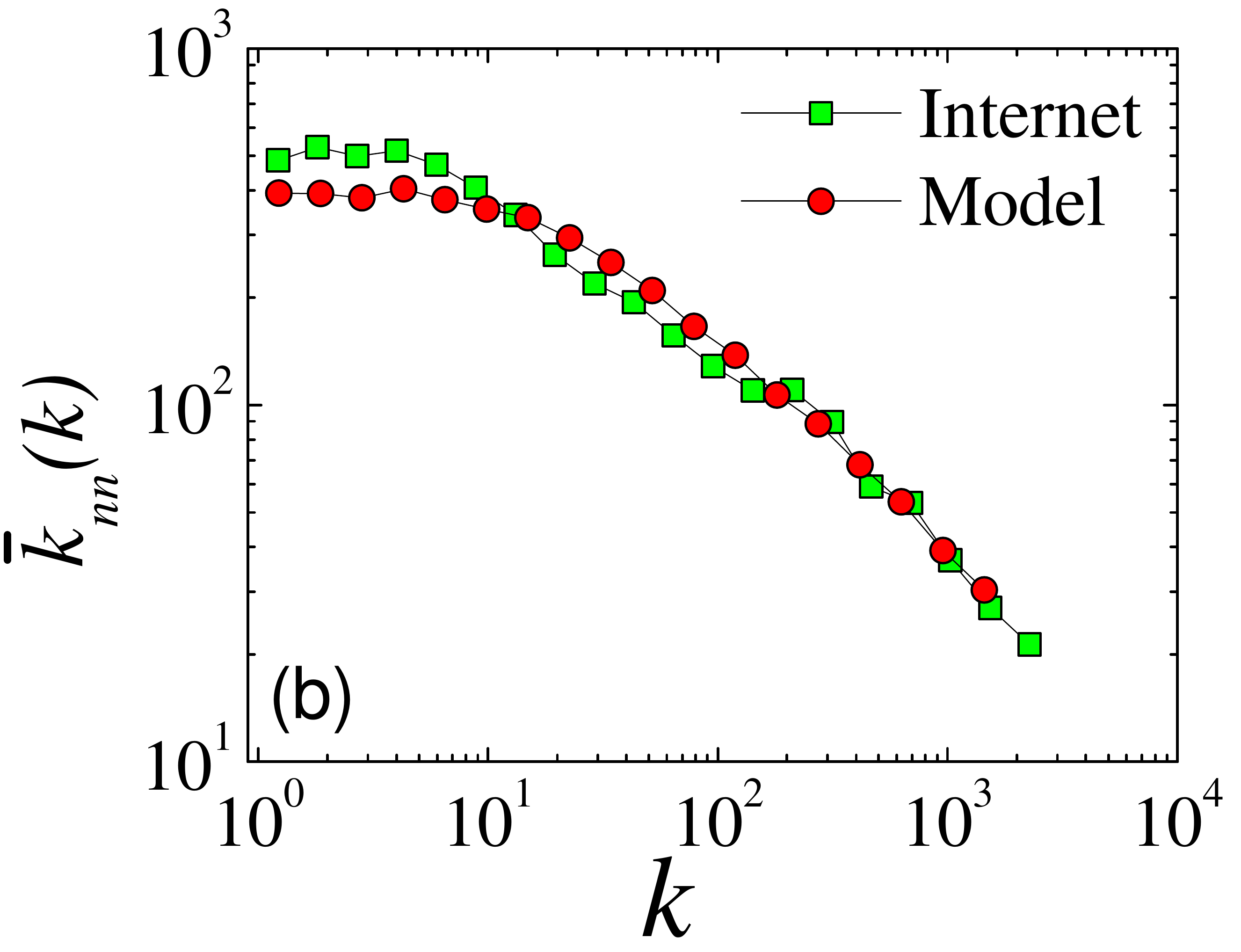}
\includegraphics[width=2.3in]{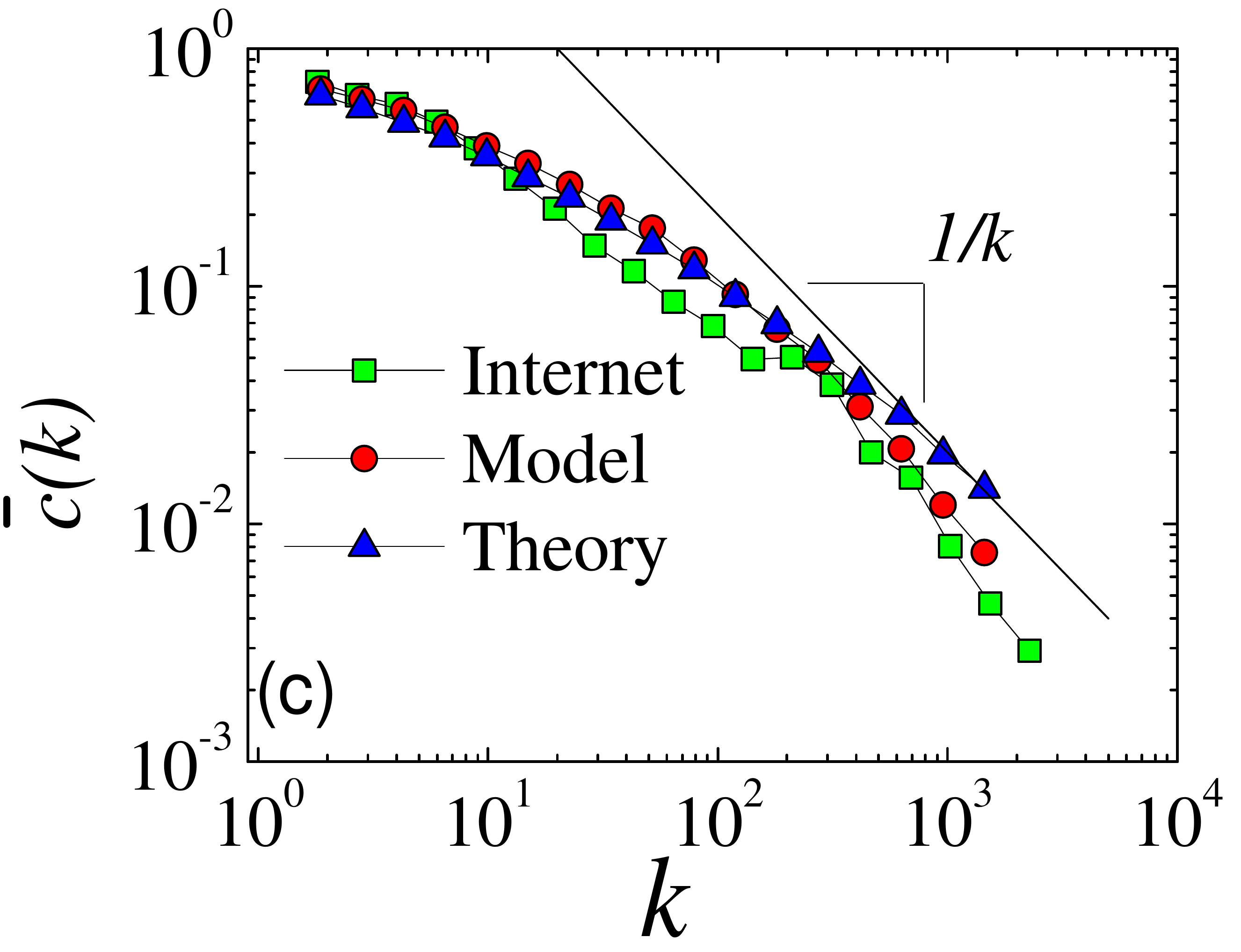}}
\caption{(Color online) The Internet as seen by the CAIDA's Archipelago Measurement
Infrastructure~\cite{ClHy09} vs.\ a network in the $\HM$ model with
$\alpha=0.55$, $\zeta=1$, and $\beta=2$. {\bf (a)}~The degree
distributions $P(k)$ in both networks are power laws with exponent
$\gamma=2.1$. The theoretical curve is given
by~(\ref{eq:P(k)-H2-final-exact}). {\bf (b)}~The average nearest
neighbor degrees~$\bar{k}_{nn}(k)$. {\bf (c)}~The degree-dependent
clustering. The theoretical curve is obtained by a numerical estimate
of the outer integral in~(\ref{eq:c(kappa)}). The inner integral is
$\pi^2\kappa'\kappa''/(\kappa\kappa'+\kappa\kappa''+\kappa'\kappa'')$
at $\beta=2$. The numerical integration is performed by summation over
the node degrees $k$ in the modeled network, i.e., $\int
d\kappa\,\rho(\kappa)\to\sum_kP(k)$, and by mapping $\kappa$'s to
$k$'s via $\kappa=k\bar{\kappa}/\bar{k}$. Random graphs capturing
the three metrics in (a-c) reproduce also many other important structural
properties of the Internet~\cite{MaKrFaVa06-phys}.
\label{fig:model.vs.internet}}
\end{figure*}

Here, we finish the description and analysis of our geometric model
of complex networks. To summarize, the model can produce scale-free
networks with any average degree $\bar{k}$, power-law exponent
$\gamma>2$, and average clustering $\bar{c}$, controlled,
respectively, by parameters $(\gamma,\mu,\beta)$ and
$(\alpha/\zeta,\nu,\beta)$ in the $\SM$ and $\HM$ formulations of
the model, see Table~\ref{table:parameters}. In
Fig.~\ref{fig:model.vs.internet} we observe a good match between the
basic topological properties of the real Internet and a synthetic
network generated by the $\HM$ model with an appropriate choice of
parameters in the cold regime. In the hot regime the model subsumes
the standard configuration model and classical random graphs as two
different limiting cases with degenerate geometric structures.

\section{Efficiency of greedy navigation in modeled networks}\label{sec:simulations}

In this section we shift our attention from the analysis of the
structure of complex networks in our model to the analysis of their
function. Specifically, we are interested in their navigation
efficiency.

One important function that many real networks perform is to
transport information or other media. Examples include the Internet,
brain, or signaling, regulatory, and metabolic networks. The
information transport in these networks is not akin to diffusion.
Instead information must be delivered to specific destinations, such
as specific hosts in the Internet, neuron groups in the brain, or
genes and proteins in regulatory networks. In the latter case, for
example, the network reacts to an increased concentration of some
sugar by expressing not all but very specific proteins, the ones
responsible for digesting this sugar. At the same time the nodes in
the network are not aware of the global network structure, so that
the questions we face are if paths to specific destinations in the
network can be found without such global topology knowledge, and how
optimal these paths can be.

The salient feature of our model is that it allows one to study the
efficiency of such path finding without global knowledge, because
our networks have underlying geometry which enables greedy
forwarding~(GF). Since each node in the network has its address,
i.e., coordinates in the underlying hyperbolic space, a node can
compute the distances between each of its neighbors in the network,
and the destination whose coordinates are written in the information
packet, or encoded in the signal. GF then accounts to forwarding the
information to the node's neighbor closest to the destination in the
hyperbolic space. Since each node knows only its own address, the
addresses of its neighbors, and the destination address of the
packet, no node has any global knowledge of the network structure.

We report simulation results for two forms of GF, {\it original} GF
(OGF) and {\it modified} GF (MGF). The OGF algorithm drops the packet
if the current node is a {\em local minimum}, meaning that it does
not have any neighbor closer to the destination than itself. The MGF
algorithm excludes the current node from any distance comparisons,
and finds the neighbor closest to the destination. The packet is
dropped only if this neighbor is the same as the packet's previously
visited node.

These GF processes can be very inefficient. They can often get stuck
at local minima, or even if they succeed reaching the destination,
they can travel along paths much longer than the optimal shortest
paths available in the network. Furthermore, even if they are
efficient in static networks, their efficiencies can quickly
deteriorate in the presence of network topology dynamics, e.g., they can
be vulnerable with respect to network damage.

To estimate the GF efficiency in static networks, we compute the
following metrics: (i)~the percentage of successful paths, $p_s$,
which is the proportion of paths that reach their destinations;
(ii)~the average hop-length $\bar{h}$ of successful paths; and
(iii)~the average and maximum stretch of successful paths. We
consider three types of stretch. The first stretch is the standard
hop stretch defined as the ratio between the hop-lengths of greedy
paths and the corresponding shortest paths in the graph. We denote
its average and maximum by $s_1$ and $\max(s_1)$. The optimal paths
have stretch equal to $1$. The other two stretches are {\it
hyperbolic}. They measure the deviation of the hyperbolic length,
traveled by a packet along either the greedy or shortest path, from
the hyperbolic distance between the source and destination.
Formally, let $(s,t)$ be a source-destination pair and let
$s=h_0,h_1,...,h_\tau=t$ be the greedy or shortest path between $s$
and $t$, and $\tau$ its hop length. Further, let $x_i, i=1...\tau$,
be the hyperbolic distance between $h_i$ and $h_{i-1}$. The
hyperbolic stretch is the ratio $\sum_i x_i/x_{st}$, where $x_{st}$
is the hyperbolic distance between $s$ and $t$. For greedy paths, we
denote the average and maximum of this stretch by $s_2$ and
$\max(s_2)$; for shortest paths those are denotes by $s_3$ and $\max(s_3)$. The
lower these two stretches, the closer the greedy and shortest paths
stay to the hyperbolic geodesics, and the more congruent we say the
network topology is with the underlying geometry.

We first focus on static networks, where the network topology does
not change, and then emulate the network topology dynamics by
randomly removing one or more links from the topology. For each
generated network instance, we extract the giant connected component
(GCC), and perform GF between $10^4$ random source-destination pairs
belonging to the GCC. All the metrics converge after approximately
$10^3$ source-destination pairs, but we evaluate an order of
magnitude more combinations for more reliable results. The network
size is $N=10^4$, and the average degree is $\bar{k}=6.5$ in all the
experiments, while temperature $T=0$
only in the following two subsections.

\subsection{Static networks}\label{sec:static_networks}

\begin{figure}
    \centerline{
        \subfigure[Success ratio]{\includegraphics[width=1.8in]{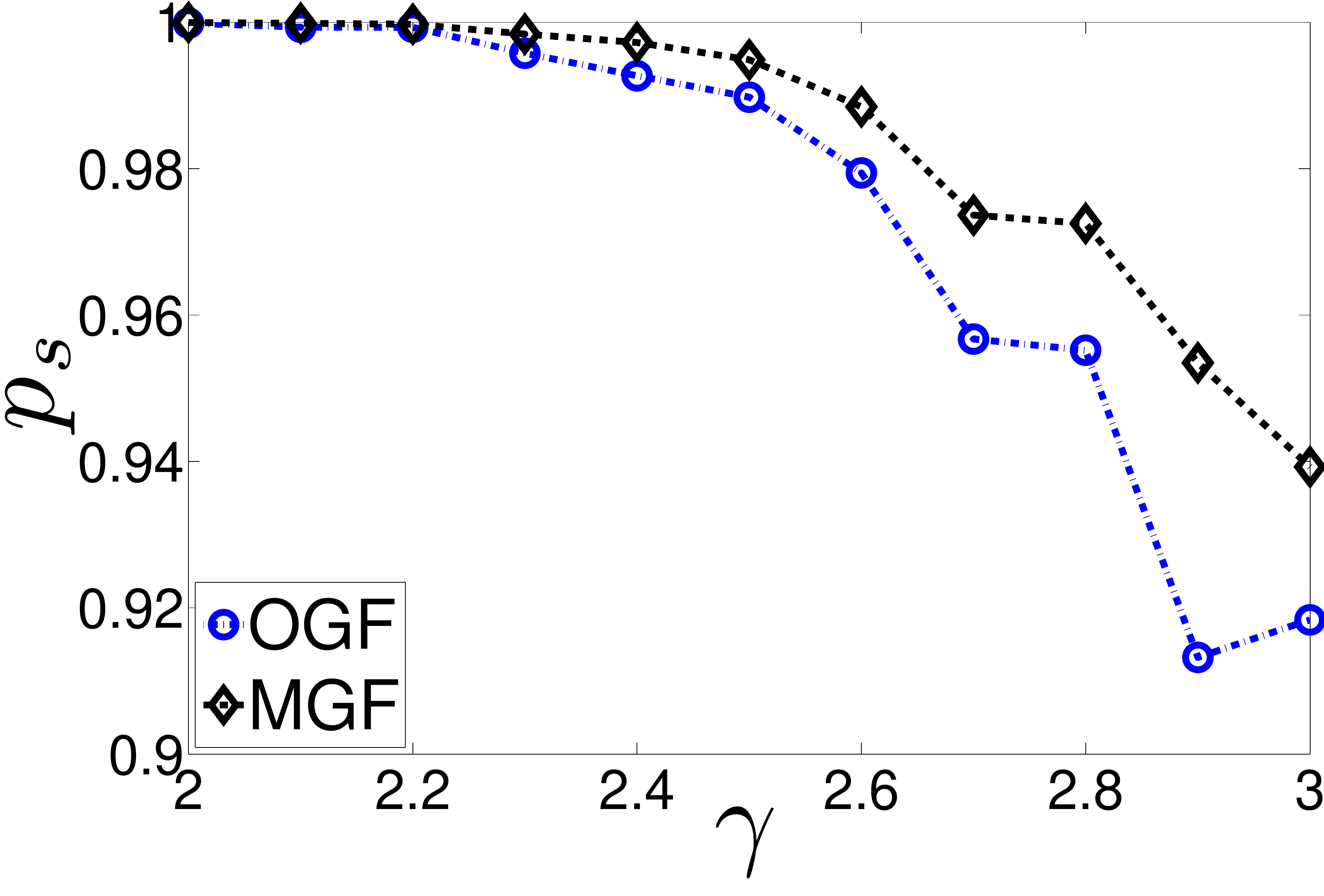}}
        \subfigure[Average hop-length]{\includegraphics[width=1.8in]{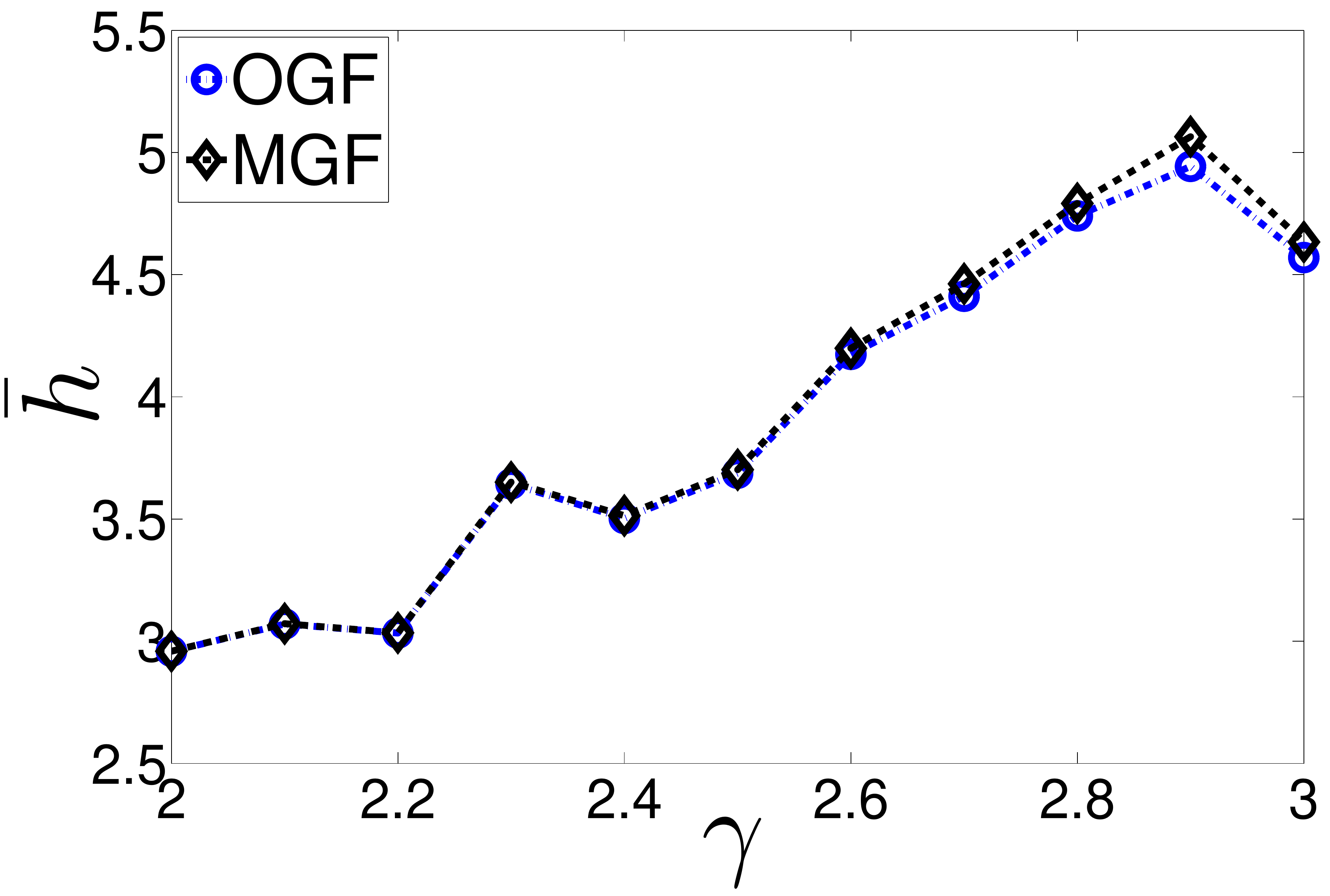}}
    }
    \centerline{
        \subfigure[Average stretch]{\includegraphics[width=1.8in]{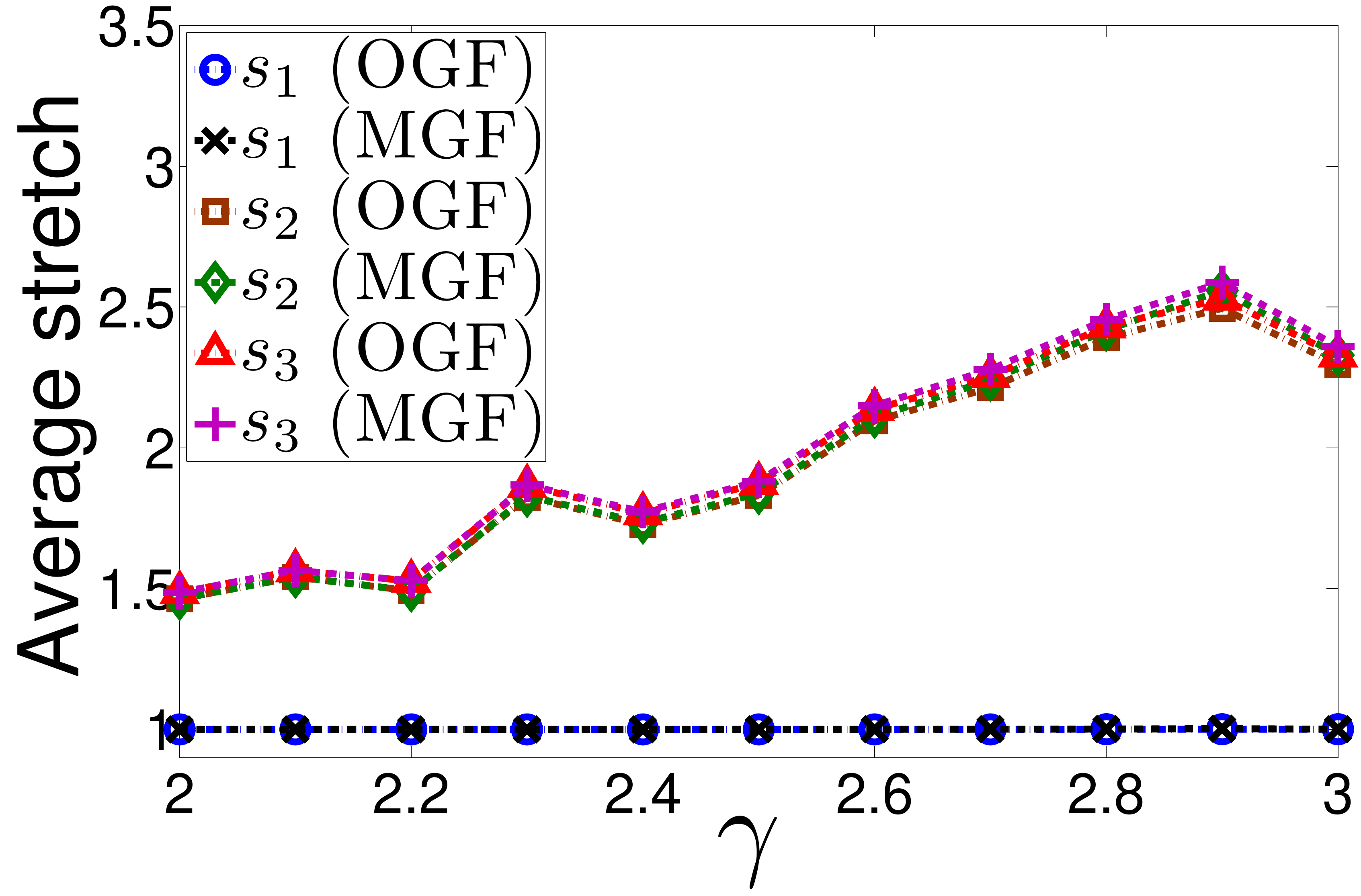}}
        \subfigure[Maximum stretch]{\includegraphics[width=1.8in]{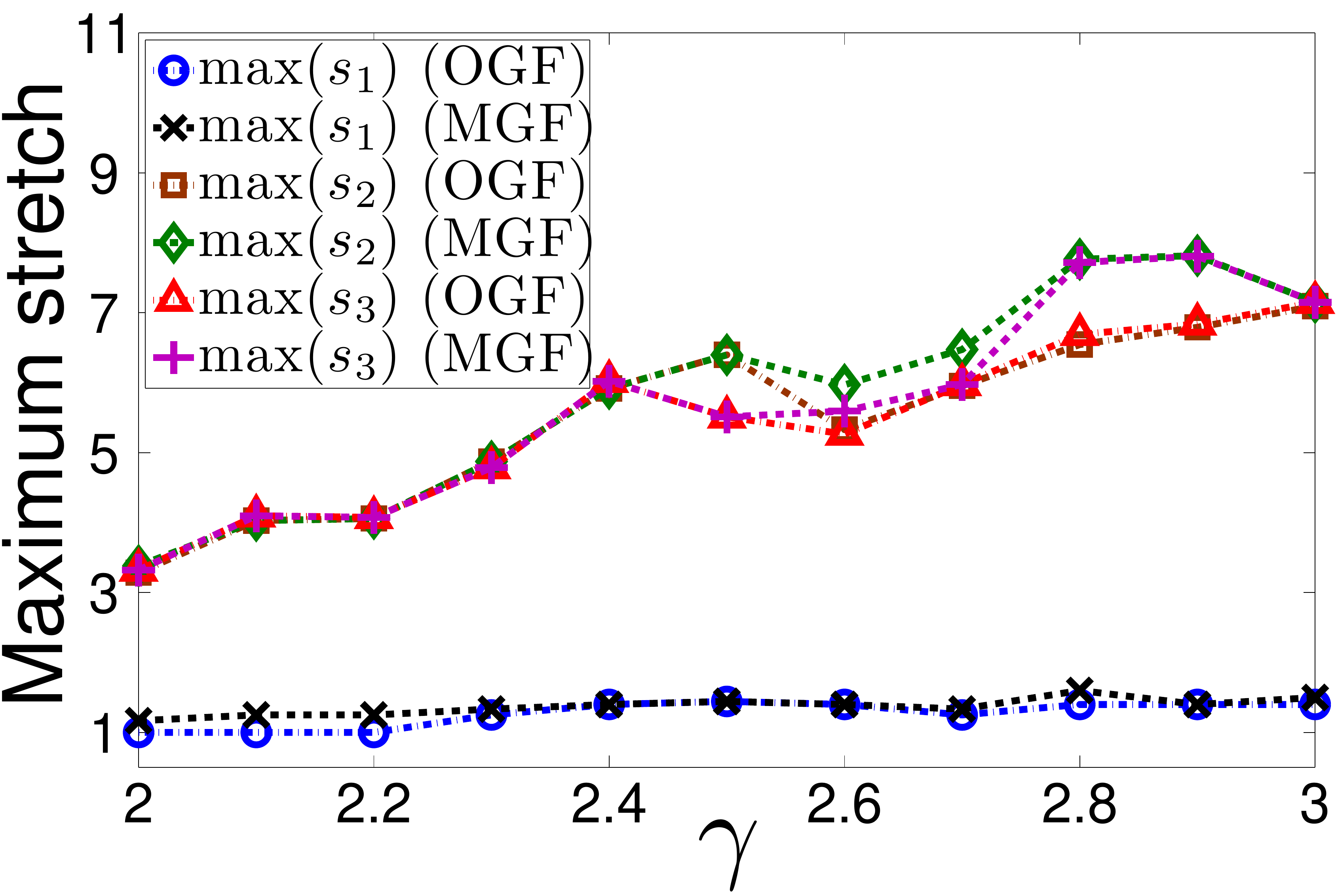}}
    }
    \caption{(Color online) Greedy forwarding in static networks.
    \label{fig:original_graph}
    }
\end{figure}

Figure~\ref{fig:original_graph} shows the results for static networks,
averaged, for each $\gamma$, over five network instances. We see that
the success ratio $p_s$ increases and path length $\bar{h}$ and all
the stretches decrease as we decrease $\gamma$ to $2$. Remarkably,
for $\gamma=2.1$, e.g., equal to the $\gamma$ observed in the
Internet, OGF and MGF yield $p_s=99.92\%$ and $p_s=99.99\%$, with
the OGF's maximum stretch of $1$, meaning that {\em all greedy paths
are shortest paths}. Interestingly, the hyperbolic stretch of
shortest paths ($s_3$ and $\max(s_3)$) is slightly worse (larger)
than of greedy paths ($s_2$ and $\max(s_2)$), which allows us to
informally say that for small $\gamma$'s, {\em greedy paths are
``shorter than shortest''} as they are shortest hop-wise, but GF
tends to select among many shortest paths those of least hyperbolic
stretch.

In summary, GF is exceptionally efficient in static networks,
especially for the small $\gamma$'s observed in the vast majority of
complex networks, including the Internet. The GF efficiency is
maximized in this case, and the two algorithms exhibit almost the
best possible performance, with reachability reaching almost
$100\%$, and all greedy paths being optimal (shortest).

\subsection{Dynamic networks}

We next look at the GF performance in dynamic networks with link
failures. For each $\gamma$, we randomly select a network instance
from above and remove one or more random links in it. We consider
the following two link-failure scenarios. In \underline{Scenario 1}
we remove a percentage $p_r$, ranging from $0\%$ to $30\%$, of all
links in the network, compute the new GCC, and for all
source-destination pairs remaining in it, we recompute the new
success ratio $p^{new}_{s}$, and the average and maximum stretch
$s^{new}_{1}$ and $\max(s^{new}_{1})$. In \underline{Scenario 2} we
provide a finer-grain view focusing on paths that used a removed
link. We remove one link from the network, compute the new GCC, and
for the source-destination pairs that are still in it, we find the
percentage $p^{l}_{s}$ of successful paths, only among those
previously successful paths that traversed the removed link.  For
these still-successful paths, we also compute the new average and
maximum stretch $s^{l}_{1}$ and $\max(s^{l}_{1})$. We then repeat
the procedure for $1000$ random links, and report the average values
for $p^{l}_{s}$ and $s^{l}_{1}$, and the maximum value for
$\max(s^{l}_{1})$.

\begin{figure}
    \centerline{
        \subfigure[Success ratio in Scenario 1]{\includegraphics[width=1.8in]{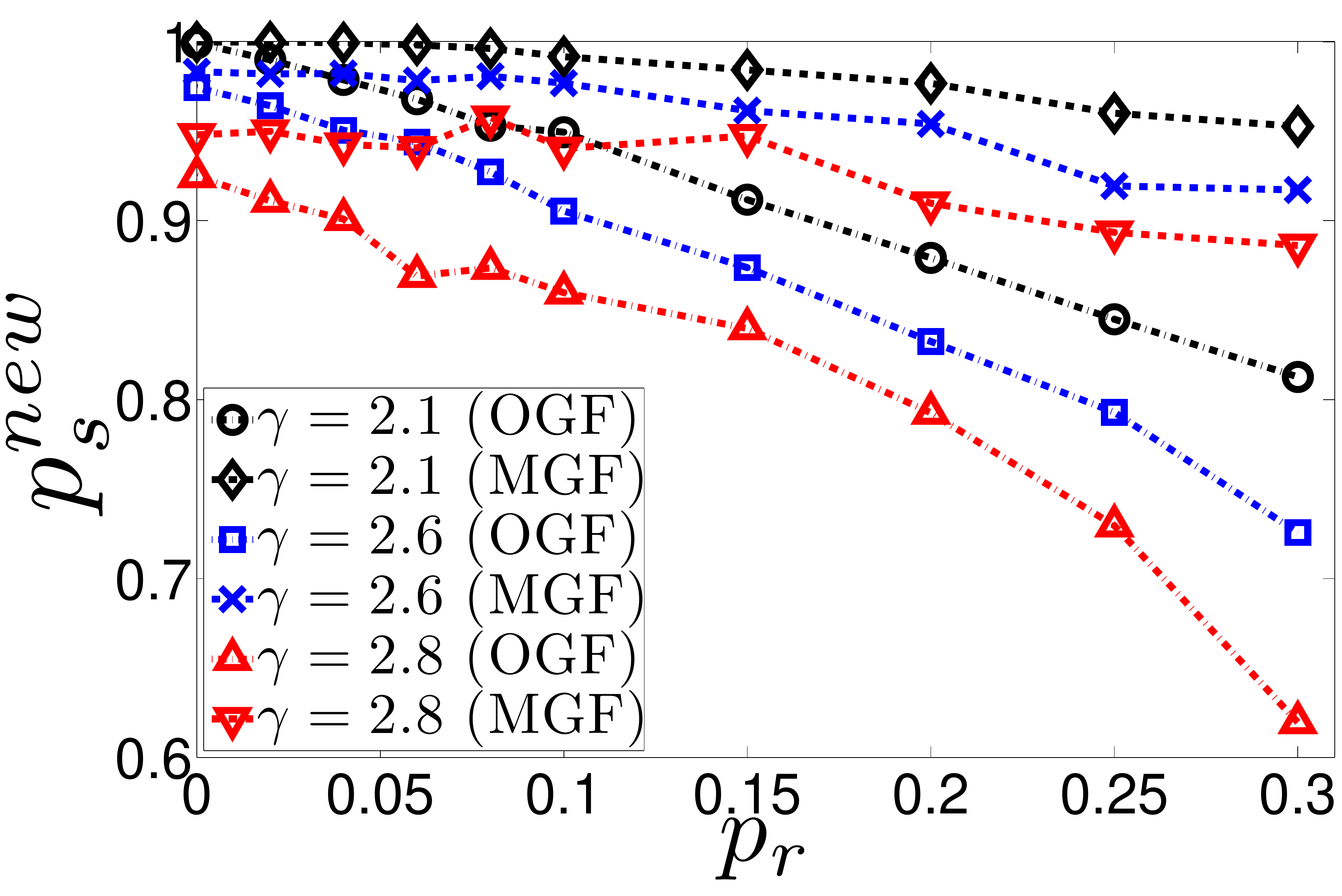}}
        \subfigure[Stretch in Scenario 1]{\includegraphics[width=1.8in]{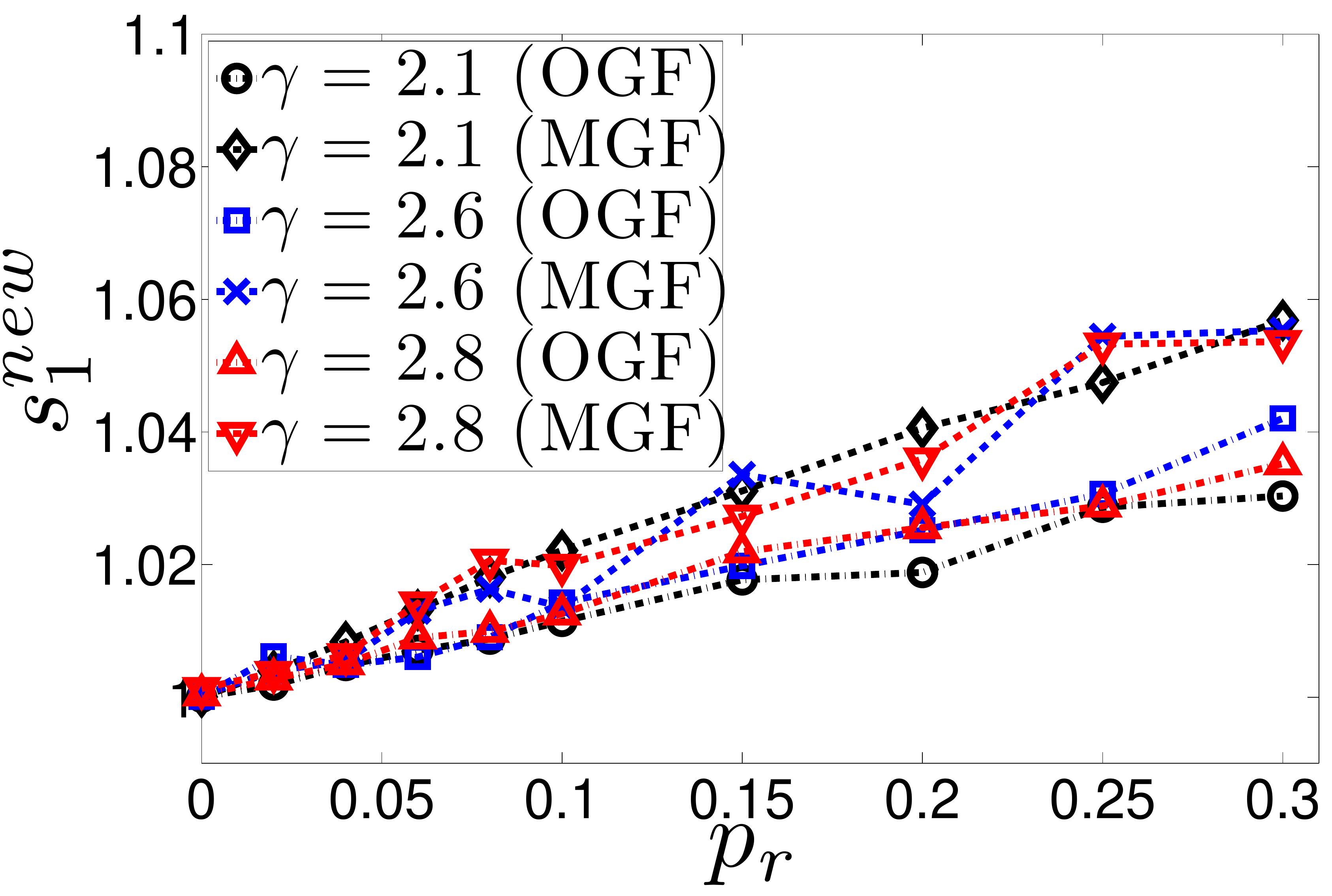}}
    }
    \centerline{
        \subfigure[Success ratio in Scenario 2]{\includegraphics[width=1.8in]{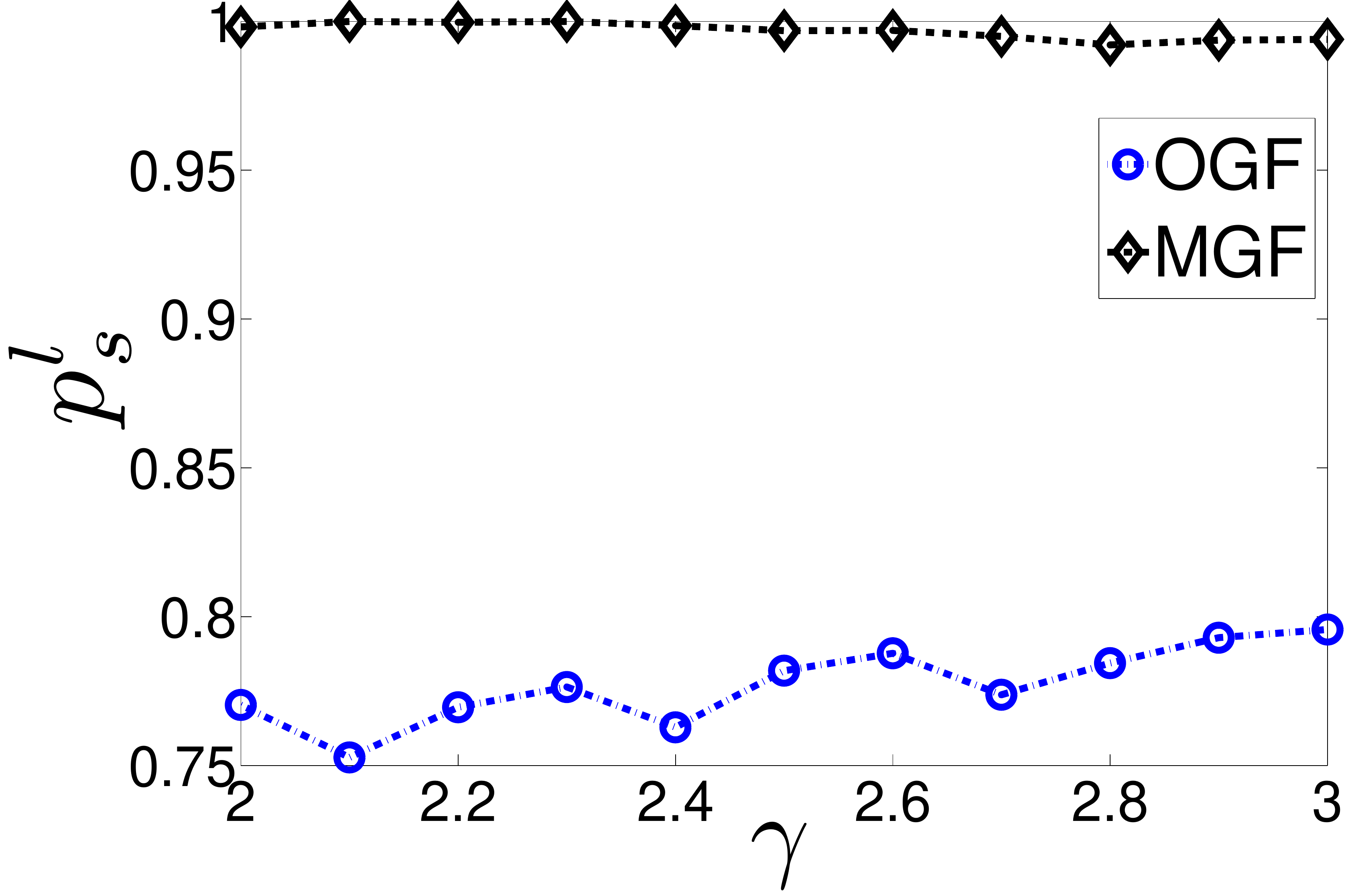}}
        \subfigure[Stretch in Scenario 2]{\includegraphics[width=1.8in]{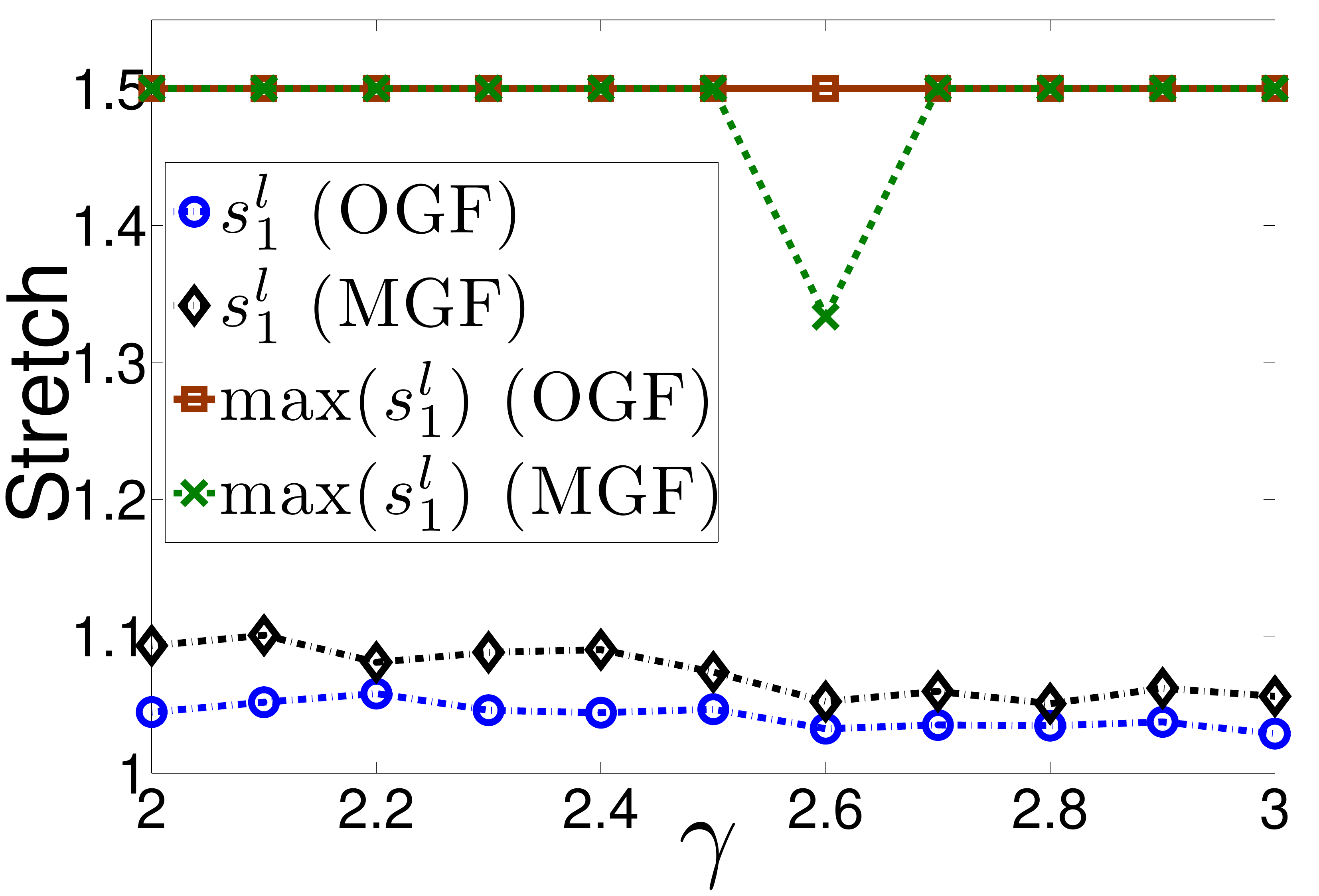}}
    }
    \caption{(Color online) Greedy forwarding in dynamic networks.
    \label{fig:link_rem}
    }
\end{figure}

Figure~\ref{fig:link_rem} presents the results. We see that for small
$\gamma$'s, the success ratio $p^{new}_{s}$ remains remarkably high,
for all meaningful values of $p_{r}$. For example, MGF on networks
with $\gamma=2.1$ and $p_{r} \leq 10\%$, yields $p^{new}_{s} >
99\%$. The simultaneous failure of $10\%$ of the links in a network
such as the Internet is a rare catastrophe, but even after such a
catastrophe the success ratio in our synthetic networks is above
$99\%$. The average stretch $s^{new}_{1}$ slightly increases as we
increase $p_{r}$, but remains quite low. We do not show
$\max(s^{new}_{1})$ to avoid clutter. For $\gamma=2.1$,
$\max(s^{new}_{1}) \leq 2$. The percentage $p^{l}_{s}$ of MGF paths
that used a removed link and that found a by-pass after its removal
is also remarkably close to $100\%$ for small $\gamma$'s. The
average stretch $s^{l}_{1}$ in Scenario~2 also remains low, below
$1.1$, and the maximum stretch $\max(s^{l}_{1})$ never exceeds
$1.5$.

In summary, GF is not only efficient in static networks, but its
efficiency is also robust in the presence of network topology
dynamics. In particular, for small $\gamma$'s matching those found
in real networks such as the Internet, GF maintains remarkably high
reachability and low stretch, even after catastrophic damages to the
network.

\subsection{Role of clustering}

\begin{figure}
    \centerline{
        \subfigure[Success ratio]{\includegraphics[width=1.8in]{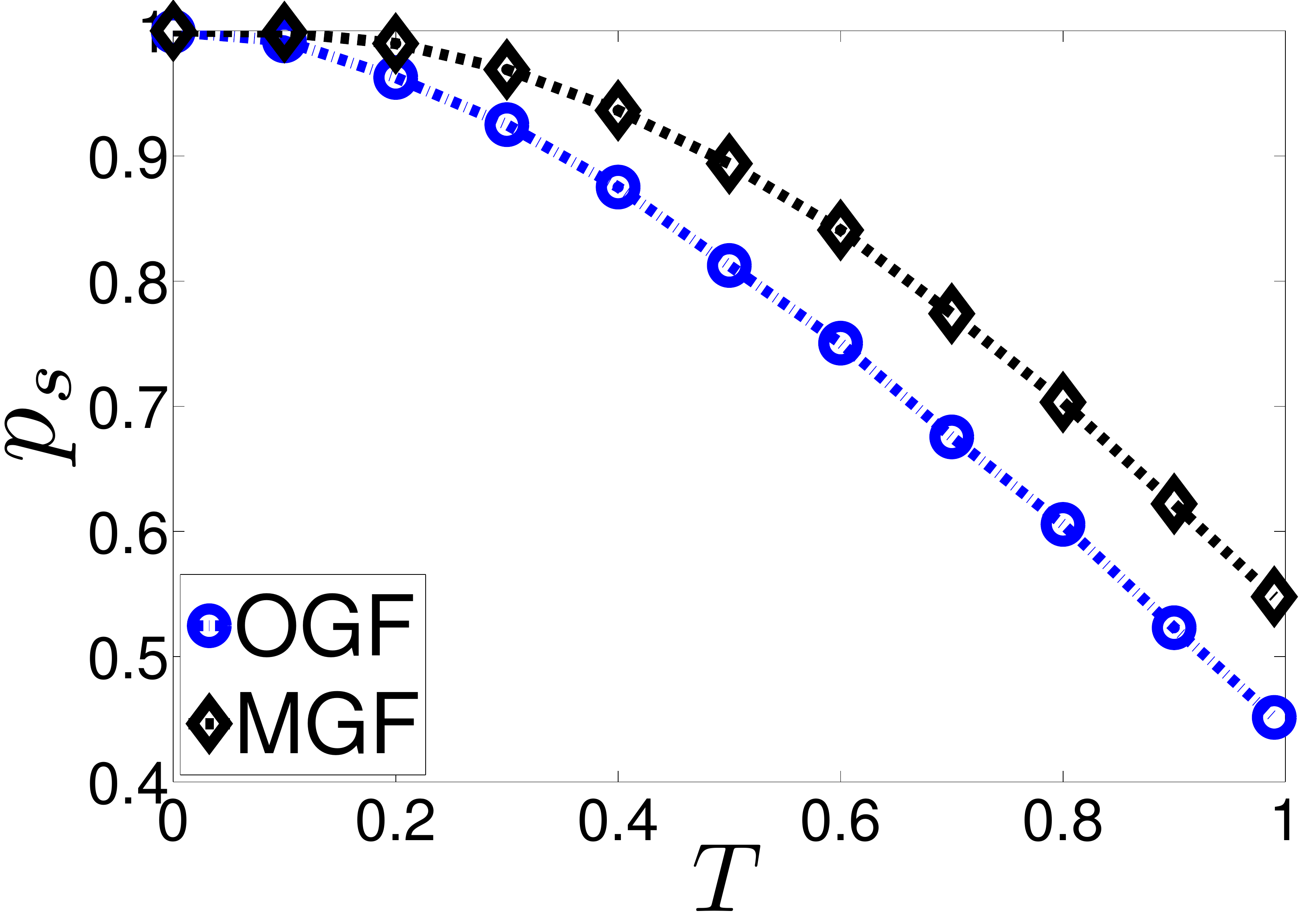}}
        \subfigure[Average hop-length]{\includegraphics[width=1.8in]{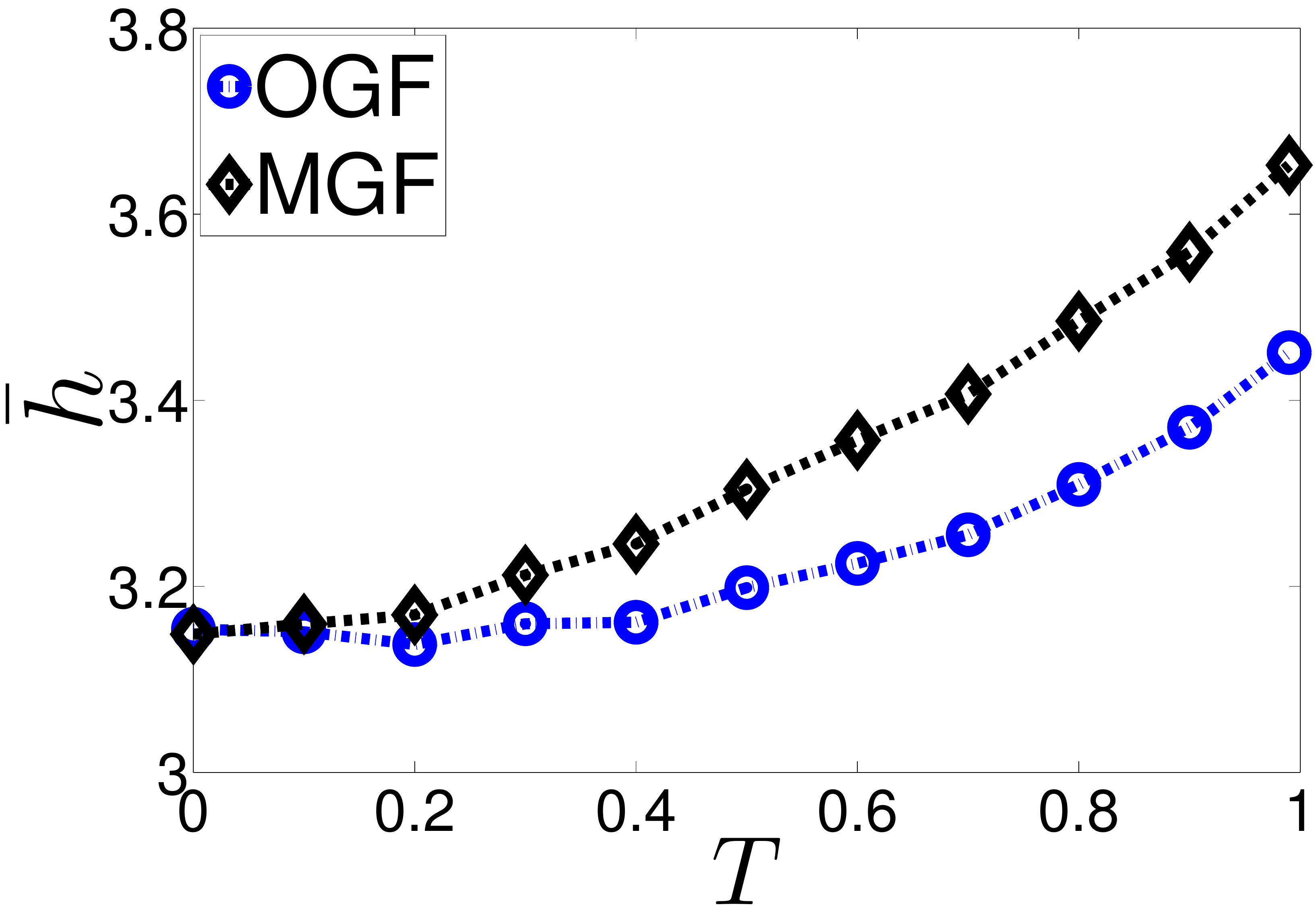}}
    }
    \centerline{
        \subfigure[Average stretch]{\includegraphics[width=1.95in]{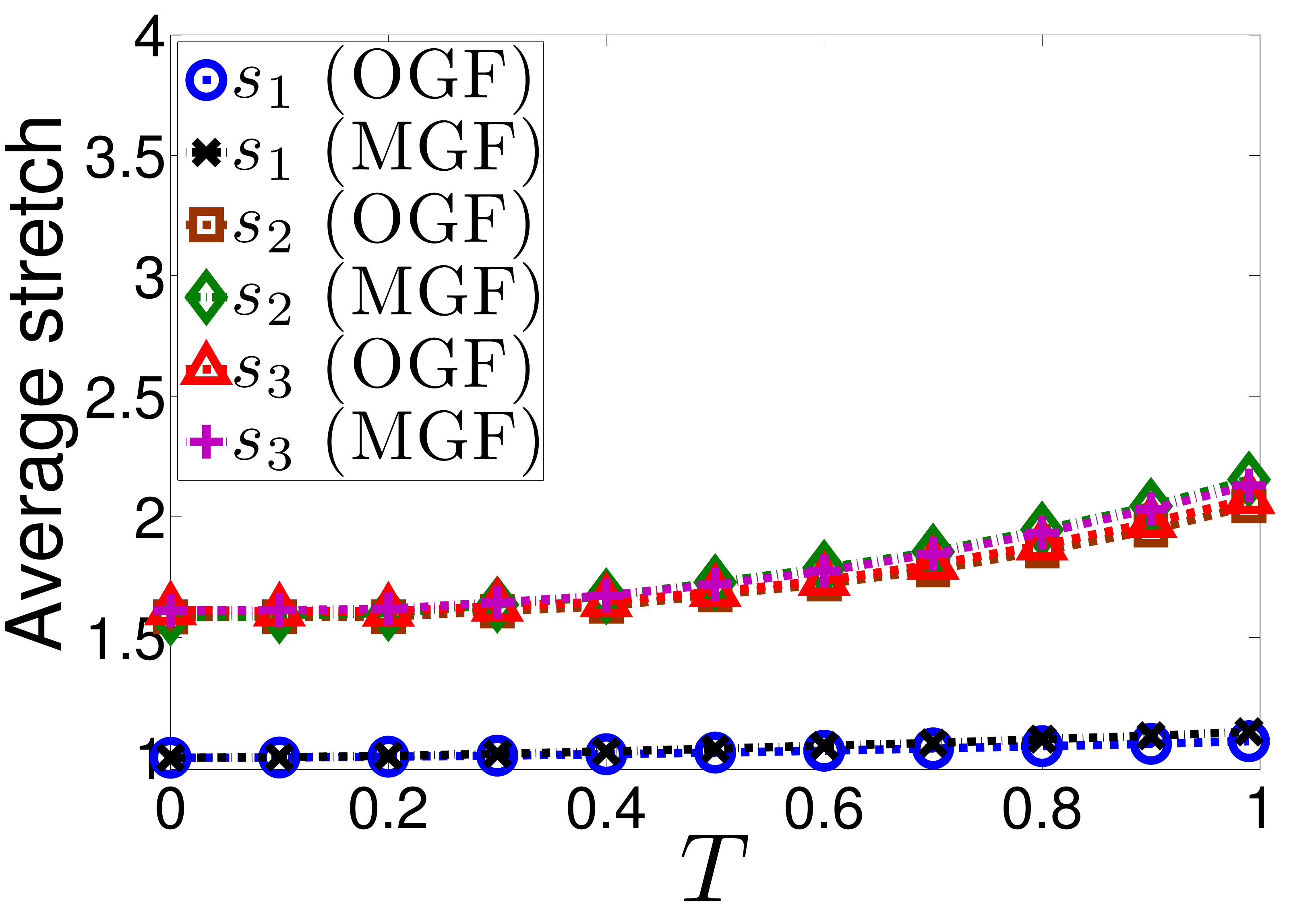}}
        \subfigure[Maximum stretch]{\includegraphics[width=1.8in]{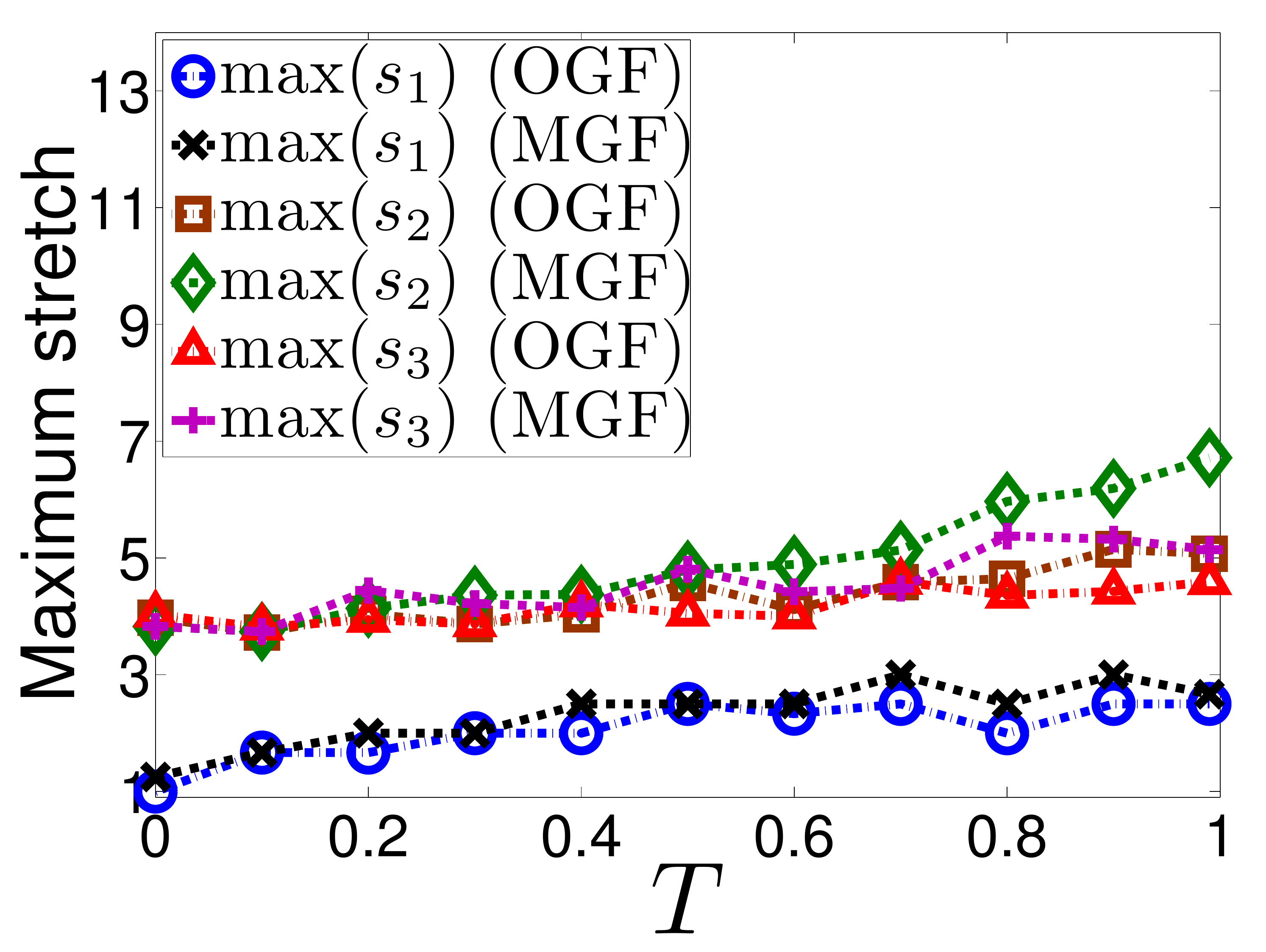}}
    }
    \caption{(Color online) Greedy forwarding as a function of temperature $T$.
    \label{fig:gr_vs_T}
    }
\end{figure}

Next we fix $\gamma=2.1$, and investigate the GF performance as a
function of temperature in Fig.~\ref{fig:gr_vs_T}. The picture is
qualitatively similar to Fig.~\ref{fig:original_graph}. The GF
efficiency is the better, the smaller is the temperature, i.e., the
stronger the clustering, see Fig.~\ref{fig:c(T)}. At zero
temperature where clustering is maximized, GF demonstrates the best
possible performance, as discussed in
Section~\ref{sec:static_networks}.

\subsection{Random graphs}

Finally, we look at the GF performance in the configuration model
and classical random graphs, which are two different degenerate
cases with zero clustering in our geometric network ensemble, see
Section~\ref{sec:random_graphs}.

\begin{figure}
    \centerline{
        \subfigure[Success ratio]{\includegraphics[width=1.8in]{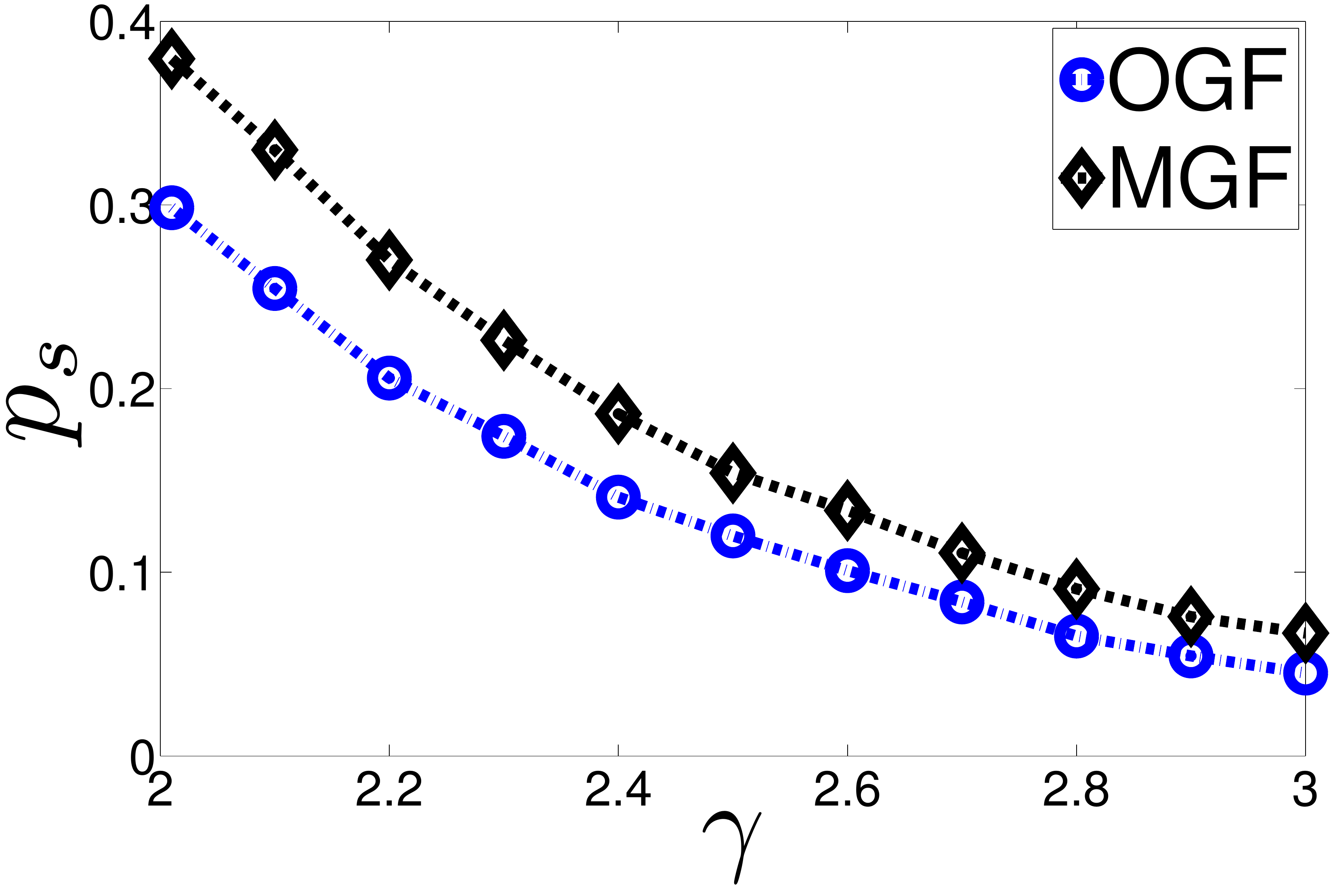}}
        \subfigure[Average hop-length]{\includegraphics[width=1.8in]{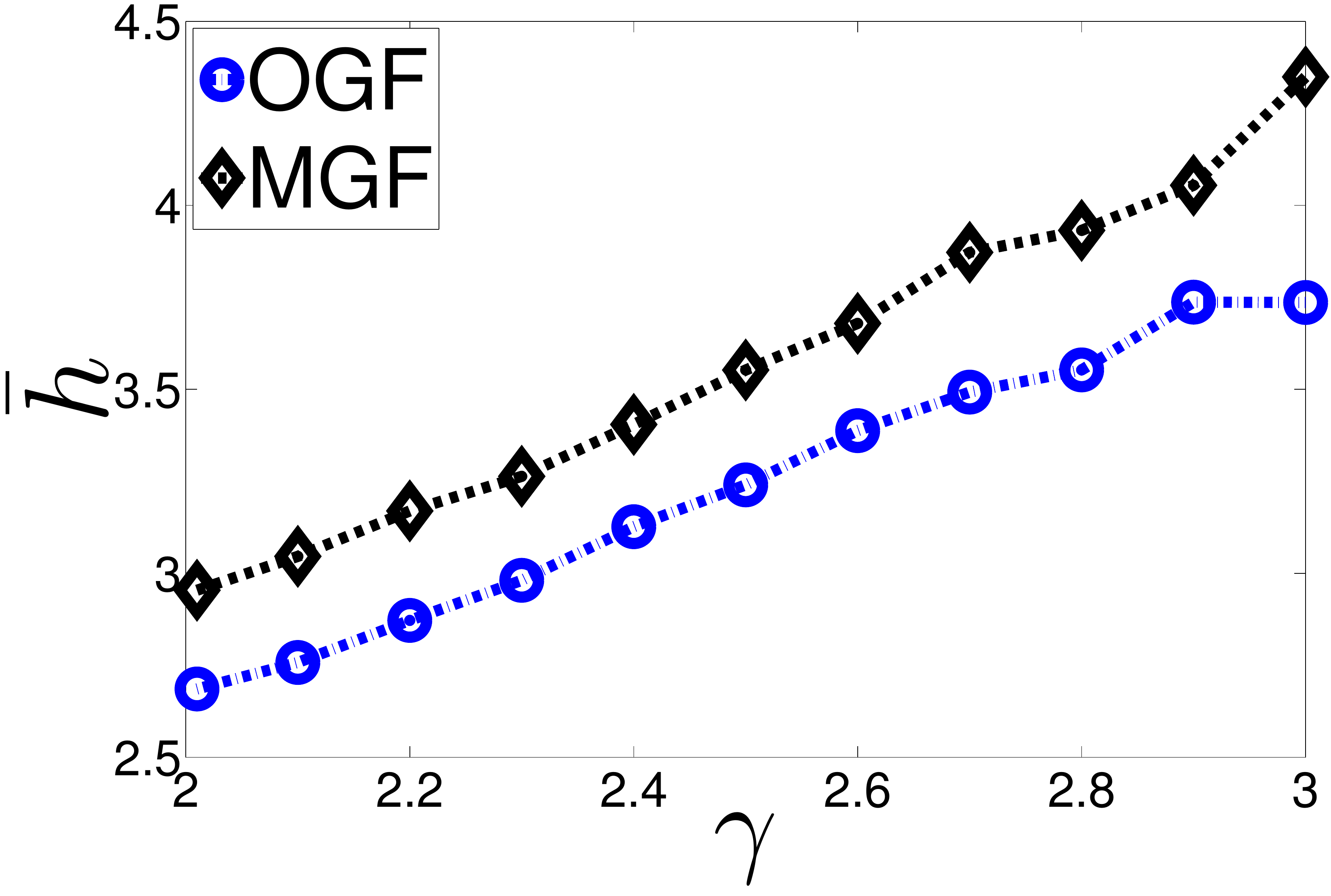}}
    }
    \centerline{
        \subfigure[Average stretch]{\includegraphics[width=1.85in]{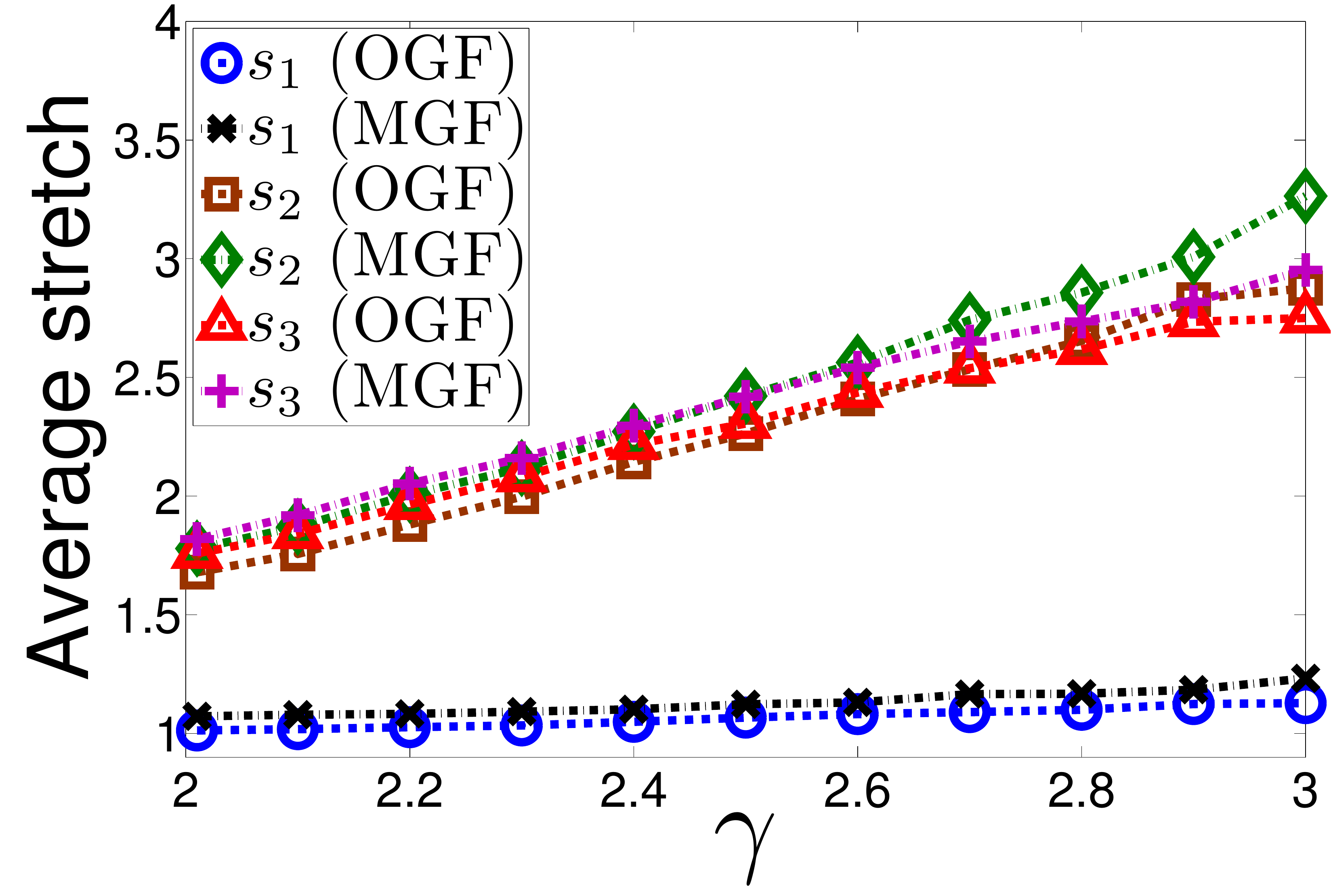}}
        \subfigure[Maximum stretch]{\includegraphics[width=1.8in]{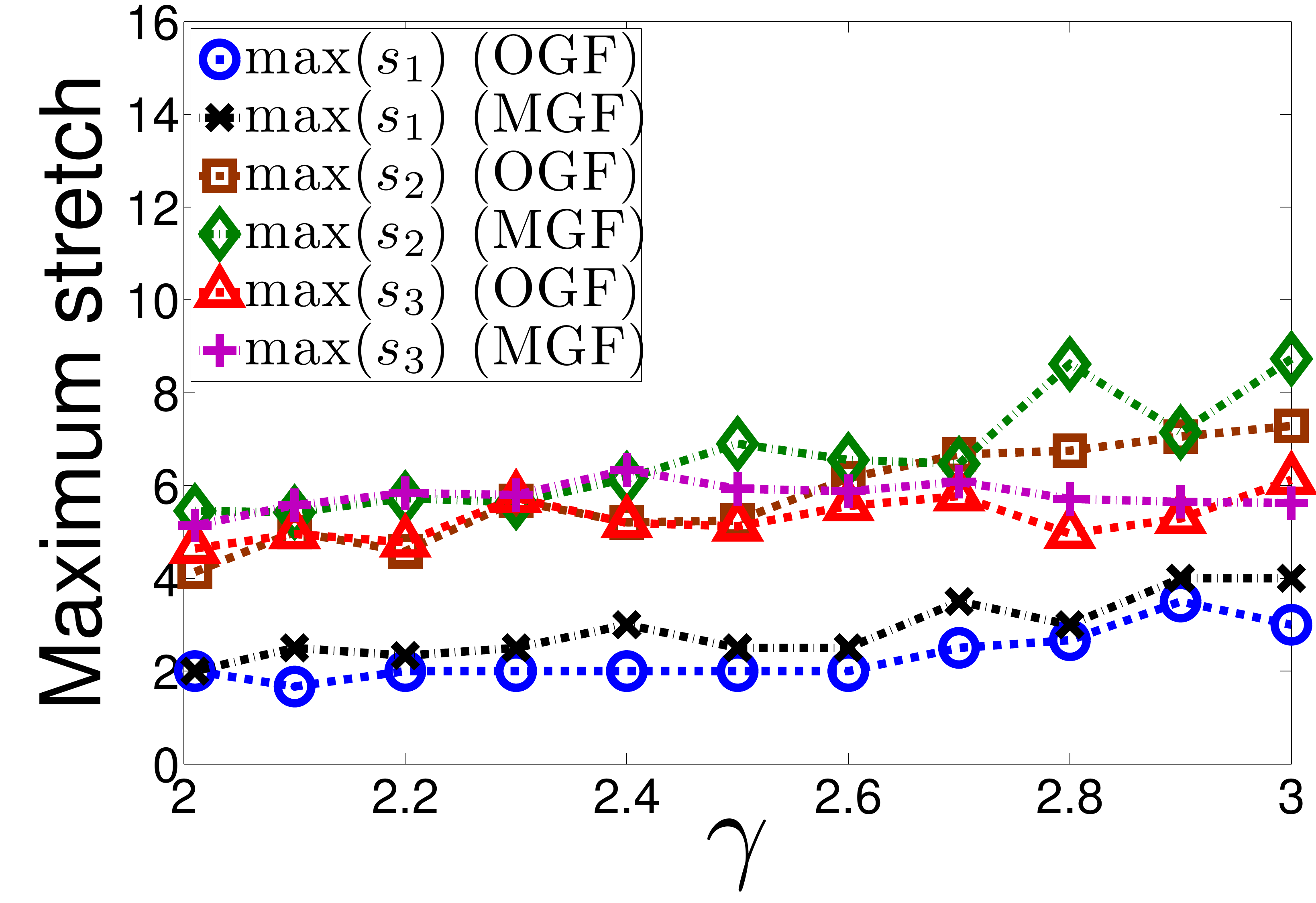}}
    }
    \caption{(Color online) Greedy forwarding in the configuration model.
    \label{fig:gr_config}
    }
\end{figure}

To test the configuration model, we fix $\alpha=1/2$, so that
$\gamma=1/\eta+1$, compute distances between nodes $i,j$ according
to $x_{ij}=r_i+r_j$, and show the results in
Fig.~\ref{fig:gr_config}. We observe that the GF efficiency is poor
in this case. The success ratio $p_s$ never exceeds $40\%$, and
drops to below $10\%$ for large $\gamma$'s. This poor performance is
expected. Indeed, since $x_{ij}=r_i+r_j$, GF reduces to following
the node degree gradient. Each node just forwards the packet to its
highest-degree neighbor $h$ since this neighbor has the smallest
radial coordinate $r_h$, thus minimizing the distance to the
destination. If during this process the packet reaches the
highest-degree hub in the network core, without visiting a node
directly connected to the destination, then it gets stuck at this
hub, because no angular coordinates instructing in what direction to
exit the core are any longer available---a problem, which does not
admit a simple and efficient solution~\cite{AdLu01}.

To test classical random graphs, we assign to nodes their angular
coordinates $\theta$ uniformly distributed on $[0,2\pi]$, connect
each node pair with the same probability $p=\bar{k}N$, and compute
distances according to $x_{ij}=\sin(\Delta\theta_{ij}/2)$. Greedy
forwarding is extremely inefficient in this case. The OGF and MGF
average success ratios $p_s$ are $0.17\%$ and $0.21\%$.

In summary, the hierarchical organization (heterogeneous degree
distribution) and metric structure (strong clustering) in the
network are both critically important for network navigability.

\subsection{Why hierarchical structure and strong clustering
ensure efficient navigation}

We have seen that more heterogeneous networks (smaller $\gamma$)
with stronger clustering (smaller $T$) are more navigable. Here, we
explain why it is the case.

\begin{figure*}
    \centerline{
        \subfigure[]
            {\includegraphics[width=2.3in]{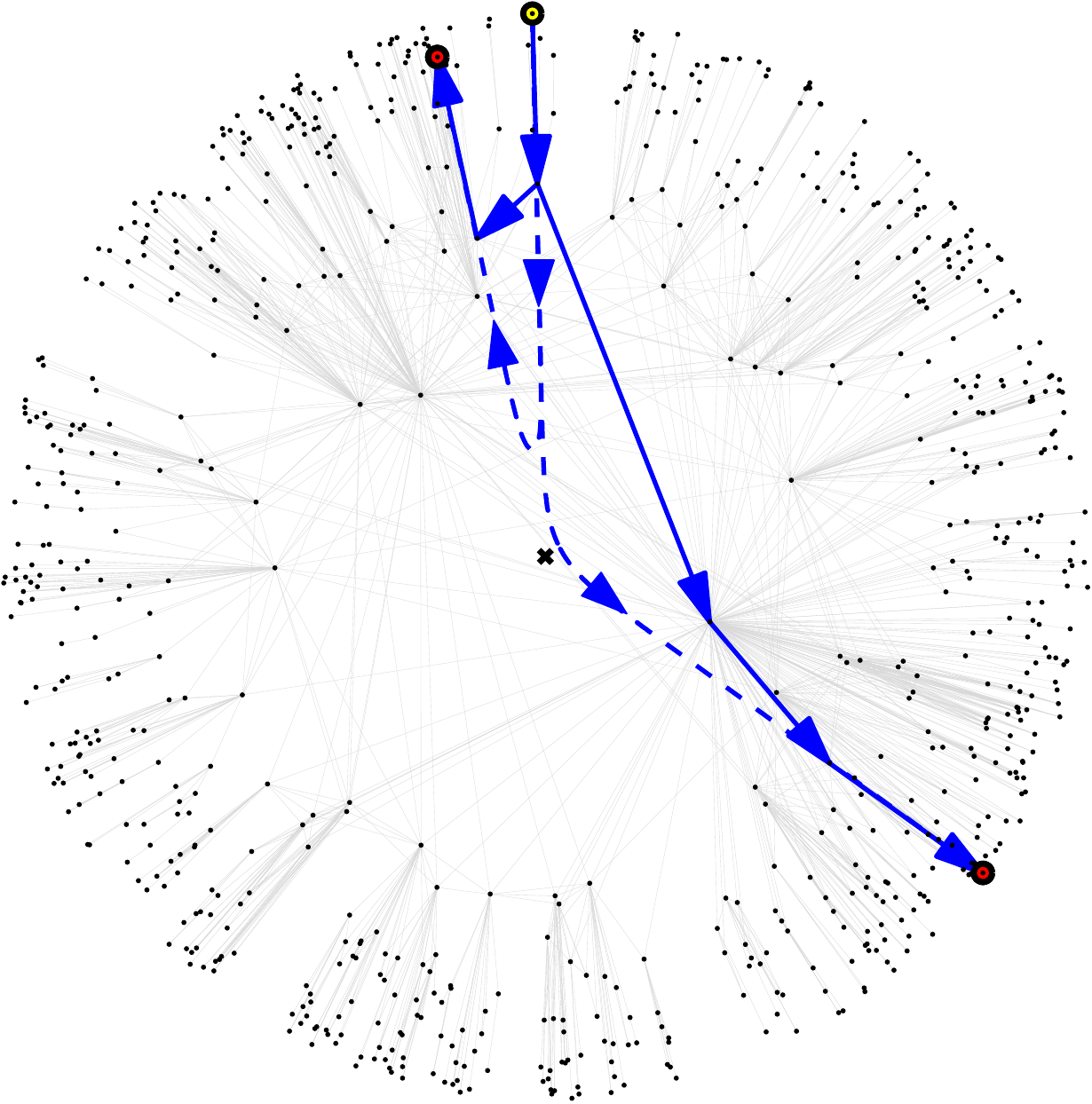}}
        \hfill
        \subfigure[]
            {\includegraphics[width=2.3in]{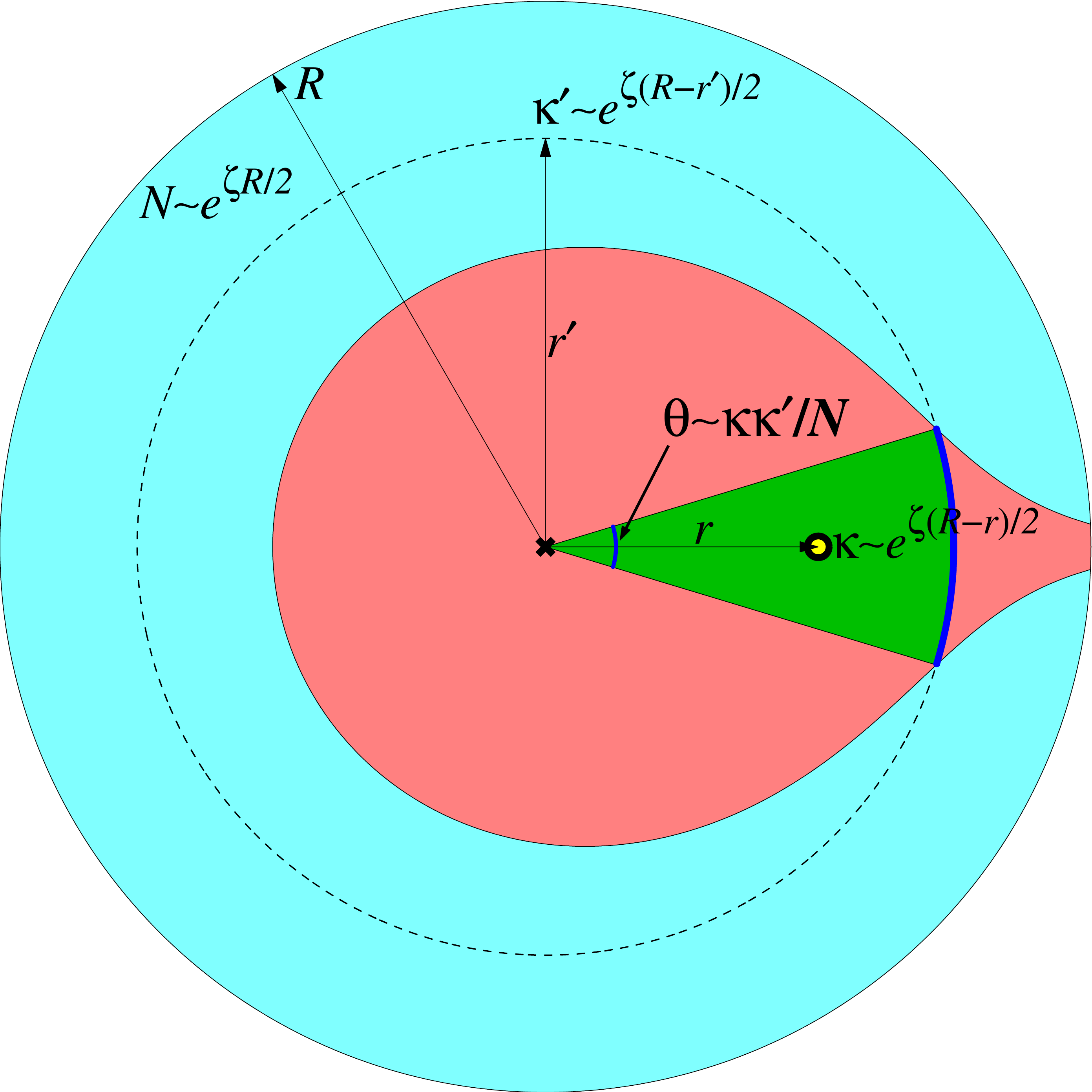}}
        \hfill
        \subfigure[]
            {\includegraphics[width=2.3in]{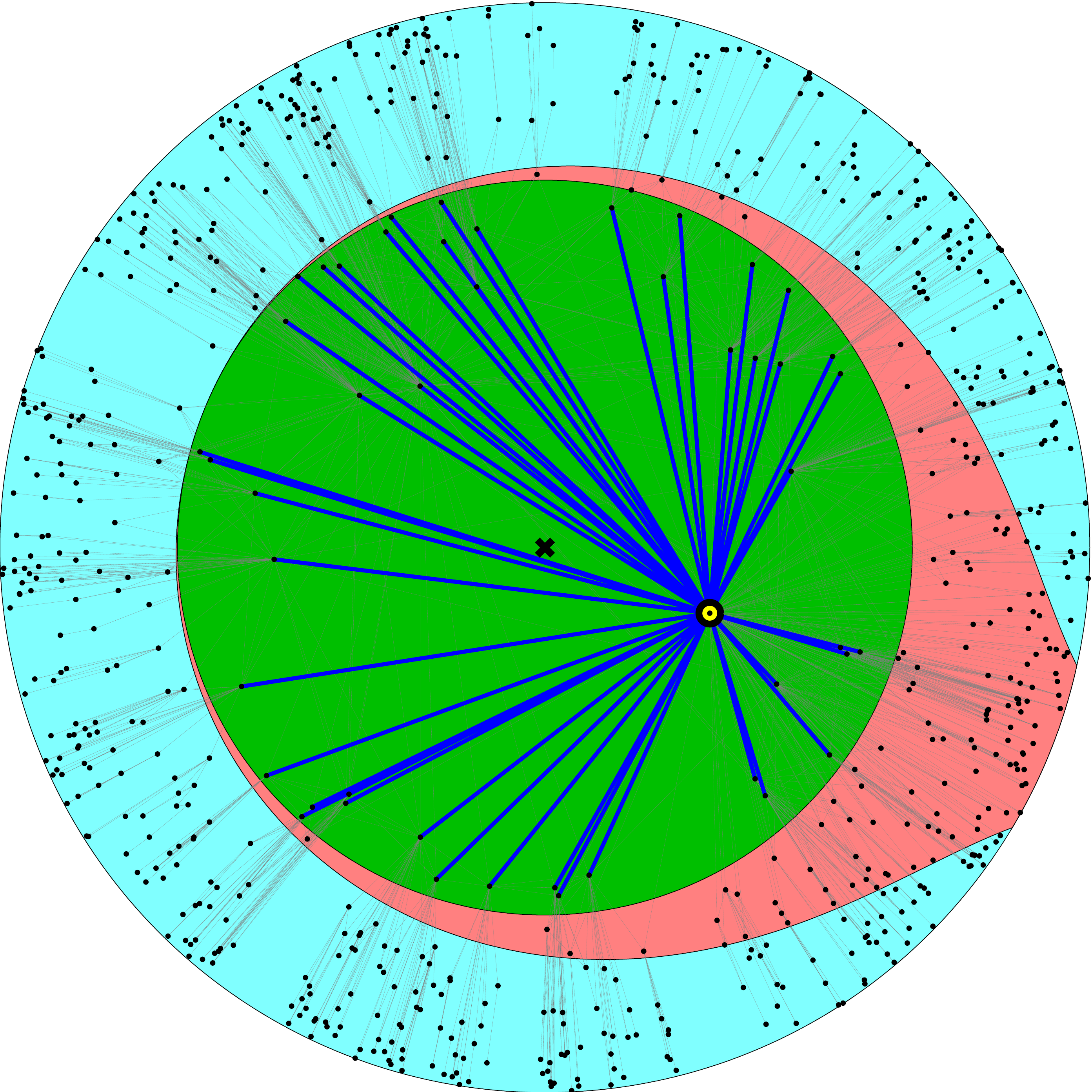}}
    }
    \caption{(Color online) {\bf(a)} Two greedy paths, which are also
    shortest paths ($s_1=1$), from the source at the top to two
    destinations are shown by the solid arrows. The dashed curves
    are the hyperbolic geodesics between the same source and destinations.
    The hyperbolic stretches $s_2=s_3$ of the left
    and right paths are $1.51$ and $1.68$.
    {\bf(b)} The inner triangular shape (green) shows the angular sector $\theta$
    that the outer shape (red), which is the hyperbolic disk of radius $R$
    centered at the circled point located at distance $r$ from the crossed origin, cuts out off the
    dashed circle of radius $r'$ centered at the origin. The
    expected node degrees at $r$ and $r'$ are $\kappa$ and
    $\kappa'$. {\bf(c)} The circled node is an example of a bridge node.
    It is connected to all nodes in its hyperbolic disk of radius $R$
    (the outer shape (red)), including {\em all\/} nodes with expected degrees
    exceeding a certain threshold, or, equivalently, to all nodes with radial
    coordinates below a certain threshold, shown by the innermost
    disk (green) whose radius is $R-r$, where $r$ is the radial coordinate
    of the circled node.
    \label{fig:hierarchy}
    }
\end{figure*}

We first recall that the congruency, measured by the hyperbolic
stretch, between the network topology and hyperbolic geometry is the
stronger, the smaller are the $\gamma$ and $T$, see
Figs.~\ref{fig:original_graph},\ref{fig:gr_vs_T}. To visualize this
effect, we draw in Fig.~\ref{fig:hierarchy}(a) a couple of GF paths
and their corresponding hyperbolic geodesics. We see that the
lengths of the latter are indeed dominated by the sums of the radial
coordinates of the source and destination, minus some
$\Delta\theta$-dependent corrections~(\ref{eq:x-approx}). This
domination of the radial direction shapes the following hierarchical
path pattern of the hyperbolic geodesics, as well as of the
corresponding GF paths: (i)~zoom-out from the network periphery to
the core, moving to increasingly higher-degree nodes, that is, nodes covering
increasingly wider areas by their connections, see
Fig.~\ref{fig:sample_network}; (ii)~turn in the core to the
direction of the destination; and finally (iii)~zoom-in onto it,
moving to lower-degree nodes.
This path pattern is exactly the pattern of {\em hierarchical
paths\/} in~\cite{TruMa04}. A path is called {\em hierarchical\/} in~\cite{TruMa04} if it
consists of two segments: first, a segment of nodes with increasing
degrees, and then a segment of nodes with decreasing degrees. As shown
in~\cite{TruMa04} (see Fig.~2(a) there), the percentage of shortest
paths that are also hierarchical approaches $100\%$ with $\gamma\to2$.
Remarkably, this hieratical path pattern also characterizes the
policy-compliant paths followed by information packets in the Internet~\cite{Gao01,DiKrFo06}.
Since the GF paths, also the shortest paths in the network, follow
the shortest geodesic paths in the hyperbolic space, the resulting
hyperbolic stretch is small. Thanks to strong clustering, the
network has many partially disjoint paths between the same source
and destination, which all follow the same hierarchical pattern. Therefore, even
if some paths are damaged by link failures, other congruent paths
remain, and GF can still find them using the same hyperbolic
geodesic direction, which explains the high robustness of network
navigability with respect to network damage. As clustering weakens,
not only the path diversity in the network decreases, but also the
network metric structure deteriorates, since the edge existence
probability~(\ref{eq:p(x)-fermi}) depends less and less on the
hyperbolic distance between nodes. In the extreme case of classical
random graphs, for example, the connection probability does not
depend on this distance at all. As a result, the congruency between
network topology and underlying geometry evaporates.

Heterogeneity is another key element responsible for high
navigability. This heterogeneity is nothing but a reflection of the
hierarchical, tree-like structure of the underlying hyperbolic
space. Indeed, its hierarchical structure manifests itself in the
hierarchy of node degrees, and in the degree-dependent amount of
space that nodes cover by their connections. As
Fig.~\ref{fig:sample_network} shows, nodes of higher degrees, closer
to the top of the hierarchy, cover wider areas with their
connections. To quantify, at $T=0$ the angular sector
$\theta(\kappa,\kappa')$ that nodes with expected degree $\kappa$
cover by their connections to nodes with expected degree $\kappa'$,
see Fig.~\ref{fig:hierarchy}(b), is
$\theta(\kappa,\kappa')=4\pi\mu\kappa\kappa'/N$. This
degree-dependent hierarchy of space coverage makes the hierarchical
zooming-out/zooming-in path pattern possible and successful.

Finally, the stronger the heterogeneity, the more bridges are in the
network, where by {\it bridges\/} we mean nodes that connect to {\em
all\/} nodes with expected degrees exceeding a certain threshold, an
example is shown in Fig.~\ref{fig:hierarchy}(c). This threshold is
given by the equation $\theta(\kappa,\kappa')=2\pi$, yielding that a
node with expected degree $\kappa$ is connected to all nodes with
expected degrees $\kappa'>N/(2\mu\kappa)$. However such
$\kappa'$-degree nodes may not exist in the network, as the required
$\kappa'$ may exceed the maximum expected degree
$\kappa_{\max}=\kappa_0N^{1/(\gamma-1)}$~\cite{BoPaVe04}. Requiring
$\kappa'<\kappa_{\max}$ leads to
$\kappa>N^{(\gamma-2)/(\gamma-1)}/(2\mu\kappa_0)$. That is, only
such $\kappa$-degree nodes are expected to be bridges. The equation
for the expected bridge existence is then $\kappa<\kappa_{\max}$,
yielding $N^{(\gamma-3)/(\gamma-1)}/(2\mu\kappa_0^2)<1$. That is,
bridges exist in any sufficiently large network with
$\gamma<3$---the smaller the $\gamma$ is, the more bridges and the
longer they are---while networks with $\gamma>3$ have no bridges.
The role of bridges in the network core is straightforward: as soon
as GF reaches a bridge, it can cross the entire network, in any
direction, at one hop~\cite{BoKr09}. Without bridges, GF is doomed
to wander along the network periphery, endangered by getting lost
there at any hop. The GF success ratio in networks with $\gamma>3$
deteriorates to zero in the thermodynamic limit~\cite{BoKrKc08}.

\section{Conclusion}\label{sec:conclusion}

We have developed a framework to study the structure and function of
complex networks in purely geometric terms. In this framework, two
common properties of complex network topologies, strong
heterogeneity and clustering, turn out to be simple reflections of
the basic properties of an underlying hyperbolic geometry.
Heterogeneity, measured in terms of the power-law degree
distribution exponent, is a function of the negative curvature of
the hyperbolic space, while clustering reflects its metric property.

Conversely, a heterogeneous network with a metric structure has an
effective hyperbolic geometry underneath. This finding sheds
light on self-similarity in complex networks~\cite{SeKrBo08}. The
network renormalization procedure considered
in~\cite{SeKrBo08}---throwing out nodes of degrees exceeding a
certain threshold---is equivalent to contracting the radius of the
hyperbolic disk where all nodes reside. This
contraction is a homothety along the radial direction, which is a
symmetry transformation of the hyperbolic space, and self-similarity
of hyperbolic spaces with respect to such homothetic transformations
has been formally defined and studied~\cite{BuSch07-book}.
Self-similarity of complex networks thus appears as a reflection of
self-similarity of hyperbolic geometry, or as the invariance with
respect to symmetry transformations in the underlying space.

The developed framework establishes a clear connection between
statistical mechanics and hyperbolic geometry of complex networks.
The collection of edges in a network, for example, can be treated as a
system of non-interacting fermions whose energies are the hyperbolic
distances between nodes. This geometric interpretation may lead to
further developments applying the standard tools of statistical
mechanics to network analysis.

The network ensemble in our framework subsumes the standard
configuration model and classical random graphs as two limiting
cases with degenerate geometric structures. The hyperbolic distance
between two nodes~(\ref{eq:x-approx}) delicately combines their
radial and angular coordinates. In the configuration model, the
distance degenerates to the sum of radial coordinates only,
destroying the network metric structure. In classical random graphs,
on the contrary, there is no radial distance dependence. The
connection probability between nodes does not depend on any
distances at all. As a result, not only the metric structure of a
network but also its hierarchical heterogeneity gets completely
destroyed.

We have shown that both these properties, strong clustering and
hierarchical heterogeneous organization, are critically important
for navigability, which is the network efficiency with respect to
targeted transport processes without global knowledge. Such
processes are impossible without auxiliary metric spaces since
global knowledge of network topology would be unavoidable in that
case. The developed framework not only provides a set of tools to
study these processes, but also explains why and how strong
clustering and hierarchical network organization makes them
efficient.

We have observed that the strongest clustering and strongest
heterogeneity, often found in real networks, lead to optimal
navigability. The transport efficiency is the best possible in this
case, according to all efficiency measures. Yet more remarkable is
that this efficiency is extremely robust with respect to even
catastrophic disturbances and damages to the network structure.

Complex networks thus appear to have the optimal structure to route
information or other media through their topological fabric. No
complicated and artificial routing schemes or constructions,
impossible in nature anyway, turn out to be needed to route
information optimally through a complex network. Its geometric
underpinning drastically simplifies the routing function, making
efficient the ``dumb'' strategy of transmitting information in the
right hyperbolic direction toward the destination.

Does signaling in real networks, such as cell signaling pathways or
the brain, follow hyperbolic geodesics, and if it does then what
network perturbations might break this signaling, potentially
causing (lethal) diseases? To answer these questions, one has first
to map a real network to its underlying space, finding the
coordinates for each node. In our recent work to reduce the routing
complexity in the Internet~\cite{BoPa10}, we map the Internet to its
hyperbolic space using statistical inference methods. These methods
work well, but require substantial manual intervention, and do not
scale to large networks. An interesting open problem is thus to find
{\em constructive\/} mapping methods, e.g., deriving the underlying
distances between nodes from their similarity measures based on node
attributes and annotations in a given network.

\begin{acknowledgments}

We thank  A.~Goltsev, S.~Dorogovtsev, A.~Samukhin, F.~Bonahon, E.~Jonckheere,
R.~Pastor-Satorras, A.~Baronchelli, M.~Newman, J.~Kleinberg,
Z.~Toroczkai, F.~Menczer, A.~Clauset, V.~Cerf, D.~Clark, K.~Fall, kc claffy,
B.~Huffaker, Y.~Hyun, A.~Vardy, V.~Astakhov, A.~Aranovich, and
others for many useful discussions and suggestions. This work was
supported by NSF Grants No.\ CNS-0964236, CNS-0722070, CNS-0434996; DHS
Grant No.\ N66001-08-C-2029; DGES Grant No.\ FIS2007-66485-C02-02;
and by Cisco Systems.

\end{acknowledgments}


\begin{thebibliography}{56}
\expandafter\ifx\csname natexlab\endcsname\relax\def\natexlab#1{#1}\fi
\expandafter\ifx\csname bibnamefont\endcsname\relax
  \def\bibnamefont#1{#1}\fi
\expandafter\ifx\csname bibfnamefont\endcsname\relax
  \def\bibfnamefont#1{#1}\fi
\expandafter\ifx\csname citenamefont\endcsname\relax
  \def\citenamefont#1{#1}\fi
\expandafter\ifx\csname url\endcsname\relax
  \def\url#1{\texttt{#1}}\fi
\expandafter\ifx\csname urlprefix\endcsname\relax\def\urlprefix{URL }\fi
\providecommand{\bibinfo}[2]{#2}
\providecommand{\eprint}[2][]{\url{#2}}

\footnotesize

\bibitem[{\citenamefont{Horne}()}]{poincare}
\bibinfo{author}{\bibfnamefont{B.}~\bibnamefont{Horne}},
  \emph{\bibinfo{title}{Poincar\'e}},
  \bibinfo{note}{\url{http://poincare.sourceforge.net/}}.

\bibitem[{\citenamefont{Goemans and Williamson}(1995)}]{GoWi95}
\bibinfo{author}{\bibfnamefont{M.~X.} \bibnamefont{Goemans}} \bibnamefont{and}
  \bibinfo{author}{\bibfnamefont{D.~P.} \bibnamefont{Williamson}},
  \bibinfo{journal}{J {ACM}} \textbf{\bibinfo{volume}{42}},
  \bibinfo{pages}{1115} (\bibinfo{year}{1995}).

\bibitem[{\citenamefont{Dimitropoulos et~al.}(2005)\citenamefont{Dimitropoulos,
  Krioukov, Huffaker, kc~claffy, and Riley}}]{DiKrHuClRi05}
\bibinfo{author}{\bibfnamefont{X.}~\bibnamefont{Dimitropoulos}},
  \bibinfo{author}{\bibfnamefont{D.}~\bibnamefont{Krioukov}},
  \bibinfo{author}{\bibfnamefont{B.}~\bibnamefont{Huffaker}},
  \bibinfo{author}{\bibnamefont{kc~claffy}}, \bibnamefont{and}
  \bibinfo{author}{\bibfnamefont{G.}~\bibnamefont{Riley}}, in
  \emph{\bibinfo{booktitle}{Proceedings of the 4th International Workshop on
  Experimental and Efficient Algorithms (WEA 2005), Santorini Island, Greece,
  May 10-13, 2005}}, edited by \bibinfo{editor}{\bibfnamefont{S.~E.}
  \bibnamefont{Nikoletseas}} (\bibinfo{publisher}{Springer},
  \bibinfo{year}{2005}), vol. \bibinfo{volume}{3503} of
  \emph{\bibinfo{series}{Lecture Notes in Computer Science}}, pp.
  \bibinfo{pages}{113--125}, ISBN \bibinfo{isbn}{3-540-25920-1}.

\bibitem[{\citenamefont{Palmer}(2009)}]{Palmer09}
\bibinfo{author}{\bibfnamefont{T.~N.} \bibnamefont{Palmer}},
  \bibinfo{journal}{P R Soc A} \textbf{\bibinfo{volume}{465}},
  \bibinfo{pages}{3165} (\bibinfo{year}{2009}).

\bibitem[{\citenamefont{Krioukov et~al.}(2009)\citenamefont{Krioukov,
  Papadopoulos, Vahdat, and Bogu{\~{n}}\'{a}}}]{KrPa09}
\bibinfo{author}{\bibfnamefont{D.}~\bibnamefont{Krioukov}},
  \bibinfo{author}{\bibfnamefont{F.}~\bibnamefont{Papadopoulos}},
  \bibinfo{author}{\bibfnamefont{A.}~\bibnamefont{Vahdat}}, \bibnamefont{and}
  \bibinfo{author}{\bibfnamefont{M.}~\bibnamefont{Bogu{\~{n}}\'{a}}},
  \bibinfo{journal}{Phys Rev E} \textbf{\bibinfo{volume}{80}},
  \bibinfo{pages}{035101(R)} (\bibinfo{year}{2009}).

\bibitem[{\citenamefont{Newman}(2010)}]{Newman10-book}
\bibinfo{author}{\bibfnamefont{M.~E.~J.} \bibnamefont{Newman}},
  \emph{\bibinfo{title}{Networks: An Introduction}} (\bibinfo{publisher}{Oxford
  University Press}, \bibinfo{address}{Oxford}, \bibinfo{year}{2010}).

\bibitem[{\citenamefont{Dorogovtsev}(2010)}]{Dorogovtsev10-book}
\bibinfo{author}{\bibfnamefont{S.~N.} \bibnamefont{Dorogovtsev}},
  \emph{\bibinfo{title}{Lectures on Complex Networks}}
  (\bibinfo{publisher}{Oxford University Press}, \bibinfo{address}{Oxford},
  \bibinfo{year}{2010}).

\bibitem[{\citenamefont{Maldacena}(1999)}]{Maldacena99}
\bibinfo{author}{\bibfnamefont{J.}~\bibnamefont{Maldacena}},
  \bibinfo{journal}{Int J Theor Phys} \textbf{\bibinfo{volume}{38}},
  \bibinfo{pages}{1113} (\bibinfo{year}{1999}).

\bibitem[{\citenamefont{Gubser et~al.}(1998)\citenamefont{Gubser, Klebanov, and
  Polyakov}}]{GuKl98}
\bibinfo{author}{\bibfnamefont{S.~S.} \bibnamefont{Gubser}},
  \bibinfo{author}{\bibfnamefont{I.~R.} \bibnamefont{Klebanov}},
  \bibnamefont{and} \bibinfo{author}{\bibfnamefont{A.~M.}
  \bibnamefont{Polyakov}}, \bibinfo{journal}{Phys Lett B}
  \textbf{\bibinfo{volume}{428}}, \bibinfo{pages}{105} (\bibinfo{year}{1998}).

\bibitem[{\citenamefont{Witten}(1998)}]{Witten98}
\bibinfo{author}{\bibfnamefont{E.}~\bibnamefont{Witten}}, \bibinfo{journal}{Adv
  Theor Math Phys} \textbf{\bibinfo{volume}{2}}, \bibinfo{pages}{253}
  (\bibinfo{year}{1998}).

\bibitem[{\citenamefont{Clauset et~al.}(2008)\citenamefont{Clauset, Moore, and
  Newman}}]{ClMo08}
\bibinfo{author}{\bibfnamefont{A.}~\bibnamefont{Clauset}},
  \bibinfo{author}{\bibfnamefont{C.}~\bibnamefont{Moore}}, \bibnamefont{and}
  \bibinfo{author}{\bibfnamefont{M.~E.~J.} \bibnamefont{Newman}},
  \bibinfo{journal}{Nature} \textbf{\bibinfo{volume}{453}}, \bibinfo{pages}{98}
  (\bibinfo{year}{2008}).

\bibitem[{\citenamefont{Serrano et~al.}(2008)\citenamefont{Serrano, Krioukov,
  and Bogu{\~{n}}\'{a}}}]{SeKrBo08}
\bibinfo{author}{\bibfnamefont{M.~{\'A}.} \bibnamefont{Serrano}},
  \bibinfo{author}{\bibfnamefont{D.}~\bibnamefont{Krioukov}}, \bibnamefont{and}
  \bibinfo{author}{\bibfnamefont{M.}~\bibnamefont{Bogu{\~{n}}\'{a}}},
  \bibinfo{journal}{Phys Rev Lett} \textbf{\bibinfo{volume}{100}},
  \bibinfo{pages}{078701} (\bibinfo{year}{2008}).

\bibitem[{\citenamefont{Dorogovtsev et~al.}(2003)\citenamefont{Dorogovtsev,
  Mendes, and Samukhin}}]{DoMeSa03b}
\bibinfo{author}{\bibfnamefont{S.~N.} \bibnamefont{Dorogovtsev}},
  \bibinfo{author}{\bibfnamefont{J.~F.~F.} \bibnamefont{Mendes}},
  \bibnamefont{and} \bibinfo{author}{\bibfnamefont{A.~N.}
  \bibnamefont{Samukhin}}, \bibinfo{journal}{Nucl Phys B}
  \textbf{\bibinfo{volume}{666}}, \bibinfo{pages}{396} (\bibinfo{year}{2003}).

\bibitem[{\citenamefont{Park and Newman}(2004)}]{PaNe04}
\bibinfo{author}{\bibfnamefont{J.}~\bibnamefont{Park}} \bibnamefont{and}
  \bibinfo{author}{\bibfnamefont{M.~E.~J.} \bibnamefont{Newman}},
  \bibinfo{journal}{Phys Rev E} \textbf{\bibinfo{volume}{70}},
  \bibinfo{pages}{066117} (\bibinfo{year}{2004}).

\bibitem[{\citenamefont{Garlaschelli and Loffredo}(2009)}]{GaLo09}
\bibinfo{author}{\bibfnamefont{D.}~\bibnamefont{Garlaschelli}}
  \bibnamefont{and} \bibinfo{author}{\bibfnamefont{M.}~\bibnamefont{Loffredo}},
  \bibinfo{journal}{Phys Rev Lett} \textbf{\bibinfo{volume}{102}},
  \bibinfo{pages}{038701} (\bibinfo{year}{2009}).

\bibitem[{\citenamefont{Anand and Bianconi}(2009)}]{AnBi09}
\bibinfo{author}{\bibfnamefont{K.}~\bibnamefont{Anand}} \bibnamefont{and}
  \bibinfo{author}{\bibfnamefont{G.}~\bibnamefont{Bianconi}},
  \bibinfo{journal}{Phys Rev E} \textbf{\bibinfo{volume}{80}},
  \bibinfo{pages}{045102(R)} (\bibinfo{year}{2009}).

\bibitem[{\citenamefont{Bianconi}(2009)}]{Bianconi09}
\bibinfo{author}{\bibfnamefont{G.}~\bibnamefont{Bianconi}},
  \bibinfo{journal}{Phys Rev E} \textbf{\bibinfo{volume}{79}},
  \bibinfo{pages}{036114} (\bibinfo{year}{2009}).

\bibitem[{\citenamefont{Chung and Lu}(2002)}]{ChLu02b}
\bibinfo{author}{\bibfnamefont{F.}~\bibnamefont{Chung}} \bibnamefont{and}
  \bibinfo{author}{\bibfnamefont{L.}~\bibnamefont{Lu}}, \bibinfo{journal}{Proc
  Natl Acad Sci USA} \textbf{\bibinfo{volume}{99}}, \bibinfo{pages}{15879}
  (\bibinfo{year}{2002}).

\bibitem[{\citenamefont{Erd\H{o}s and R\'{e}nyi}(1959)}]{ErRe59}
\bibinfo{author}{\bibfnamefont{P.}~\bibnamefont{Erd\H{o}s}} \bibnamefont{and}
  \bibinfo{author}{\bibfnamefont{A.}~\bibnamefont{R\'{e}nyi}},
  \bibinfo{journal}{Publ Math} \textbf{\bibinfo{volume}{6}},
  \bibinfo{pages}{290} (\bibinfo{year}{1959}).

\bibitem[{\citenamefont{Korman and Peleg}(2006)}]{KoPe06}
\bibinfo{author}{\bibfnamefont{A.}~\bibnamefont{Korman}} \bibnamefont{and}
  \bibinfo{author}{\bibfnamefont{D.}~\bibnamefont{Peleg}}, in
  \emph{\bibinfo{booktitle}{Proceedings of the 33rd International Colloquium on
  Automata, Languages and Programming (ICALP 2006), Venice, Italy, July 10-14,
  2006}}, edited by \bibinfo{editor}{\bibfnamefont{M.}~\bibnamefont{Bugliesi}},
  \bibinfo{editor}{\bibfnamefont{B.}~\bibnamefont{Preneel}},
  \bibinfo{editor}{\bibfnamefont{V.}~\bibnamefont{Sassone}}, \bibnamefont{and}
  \bibinfo{editor}{\bibfnamefont{I.}~\bibnamefont{Wegener}}
  (\bibinfo{publisher}{Springer}, \bibinfo{year}{2006}), vol.
  \bibinfo{volume}{4051} of \emph{\bibinfo{series}{Lecture Notes in Computer
  Science}}, pp. \bibinfo{pages}{619--630}, ISBN \bibinfo{isbn}{3-540-35904-4}.

\bibitem[{\citenamefont{Anderson}(2005)}]{Anderson05-book}
\bibinfo{author}{\bibfnamefont{J.~W.} \bibnamefont{Anderson}},
  \emph{\bibinfo{title}{Hyperbolic Geometry}}
  (\bibinfo{publisher}{Springer-Verlag}, \bibinfo{address}{London},
  \bibinfo{year}{2005}).

\bibitem[{\citenamefont{Cannon et~al.}(1997)\citenamefont{Cannon, Floyd,
  Kenyon, and Parry}}]{CaFlo97}
\bibinfo{author}{\bibfnamefont{J.}~\bibnamefont{Cannon}},
  \bibinfo{author}{\bibfnamefont{W.}~\bibnamefont{Floyd}},
  \bibinfo{author}{\bibfnamefont{R.}~\bibnamefont{Kenyon}}, \bibnamefont{and}
  \bibinfo{author}{\bibfnamefont{W.}~\bibnamefont{Parry}},
  \emph{\bibinfo{title}{Flavors of Geometry}} (\bibinfo{publisher}{MSRI},
  \bibinfo{address}{Berkeley}, \bibinfo{year}{1997}), chap.
  \bibinfo{chapter}{Hyperbolic Geometry}.

\bibitem[{\citenamefont{Burago et~al.}(2001)\citenamefont{Burago, Burago, and
  Ivanov}}]{BuBuIv01-book}
\bibinfo{author}{\bibfnamefont{D.}~\bibnamefont{Burago}},
  \bibinfo{author}{\bibfnamefont{Y.}~\bibnamefont{Burago}}, \bibnamefont{and}
  \bibinfo{author}{\bibfnamefont{S.}~\bibnamefont{Ivanov}},
  \emph{\bibinfo{title}{A Course in Metric Geometry}}
  (\bibinfo{publisher}{AMS}, \bibinfo{address}{Providence},
  \bibinfo{year}{2001}).

\bibitem[{\citenamefont{Ratcliffe}(2006)}]{Ratcliffe06-book}
\bibinfo{author}{\bibfnamefont{J.}~\bibnamefont{Ratcliffe}},
  \emph{\bibinfo{title}{Foundations of Hyperbolic Manifolds}}
  (\bibinfo{publisher}{Springer}, \bibinfo{address}{New York},
  \bibinfo{year}{2006}).

\bibitem[{\citenamefont{Bridson and Haefliger}(1999)}]{BridsonHaefliger99-book}
\bibinfo{author}{\bibfnamefont{M.~R.} \bibnamefont{Bridson}} \bibnamefont{and}
  \bibinfo{author}{\bibfnamefont{A.}~\bibnamefont{Haefliger}},
  \emph{\bibinfo{title}{Metric Spaces of Non-Positive Curvature}}
  (\bibinfo{publisher}{Springer-Verlag}, \bibinfo{address}{Berlin},
  \bibinfo{year}{1999}).

\bibitem[{\citenamefont{Buyalo and Schroeder}(2007)}]{BuSch07-book}
\bibinfo{author}{\bibfnamefont{S.}~\bibnamefont{Buyalo}} \bibnamefont{and}
  \bibinfo{author}{\bibfnamefont{V.}~\bibnamefont{Schroeder}},
  \emph{\bibinfo{title}{Elements of Asymptotic Geometry}}
  (\bibinfo{publisher}{European Mathematical Society},
  \bibinfo{address}{Z{\"u}rich}, \bibinfo{year}{2007}).

\bibitem[{\citenamefont{Gromov}(2007)}]{Gromov07-book}
\bibinfo{author}{\bibfnamefont{M.}~\bibnamefont{Gromov}},
  \emph{\bibinfo{title}{Metric Structures for Riemannian and Non-Riemannian
  Spaces}} (\bibinfo{publisher}{Birkh{\"a}user}, \bibinfo{address}{Boston},
  \bibinfo{year}{2007}).

\bibitem[{\citenamefont{Girvan and Newman}(2002)}]{GiNe02}
\bibinfo{author}{\bibfnamefont{M.}~\bibnamefont{Girvan}} \bibnamefont{and}
  \bibinfo{author}{\bibfnamefont{M.~E.~J.} \bibnamefont{Newman}},
  \bibinfo{journal}{Proc Natl Acad Sci USA} \textbf{\bibinfo{volume}{99}},
  \bibinfo{pages}{7821} (\bibinfo{year}{2002}).

\bibitem[{\citenamefont{Bogu{\~{n}}\'{a}
  et~al.}(2004{\natexlab{a}})\citenamefont{Bogu{\~{n}}\'{a}, Pastor-Satorras,
  D\'{\i}az-Guilera, and Arenas}}]{BoPa04}
\bibinfo{author}{\bibfnamefont{M.}~\bibnamefont{Bogu{\~{n}}\'{a}}},
  \bibinfo{author}{\bibfnamefont{R.}~\bibnamefont{Pastor-Satorras}},
  \bibinfo{author}{\bibfnamefont{A.}~\bibnamefont{D\'{\i}az-Guilera}},
  \bibnamefont{and} \bibinfo{author}{\bibfnamefont{A.}~\bibnamefont{Arenas}},
  \bibinfo{journal}{Phys Rev E} \textbf{\bibinfo{volume}{70}},
  \bibinfo{pages}{056122} (\bibinfo{year}{2004}{\natexlab{a}}).

\bibitem[{\citenamefont{Watts et~al.}(2002)\citenamefont{Watts, Dodds, and
  Newman}}]{WatDoNew02}
\bibinfo{author}{\bibfnamefont{D.~J.} \bibnamefont{Watts}},
  \bibinfo{author}{\bibfnamefont{P.~S.} \bibnamefont{Dodds}}, \bibnamefont{and}
  \bibinfo{author}{\bibfnamefont{M.~E.~J.} \bibnamefont{Newman}},
  \bibinfo{journal}{Science} \textbf{\bibinfo{volume}{296}},
  \bibinfo{pages}{1302} (\bibinfo{year}{2002}).

\bibitem[{\citenamefont{Redner}(1998)}]{Redner98}
\bibinfo{author}{\bibfnamefont{S.}~\bibnamefont{Redner}}, \bibinfo{journal}{Eur
  Phys J B} \textbf{\bibinfo{volume}{4}}, \bibinfo{pages}{131}
  (\bibinfo{year}{1998}).

\bibitem[{\citenamefont{B{\"{o}}rner et~al.}(2004)\citenamefont{B{\"{o}}rner,
  Maru, and Goldstone}}]{BoMaGo04-pnas}
\bibinfo{author}{\bibfnamefont{K.}~\bibnamefont{B{\"{o}}rner}},
  \bibinfo{author}{\bibfnamefont{J.~T.} \bibnamefont{Maru}}, \bibnamefont{and}
  \bibinfo{author}{\bibfnamefont{R.~L.} \bibnamefont{Goldstone}},
  \bibinfo{journal}{Proc Natl Acad Sci USA} \textbf{\bibinfo{volume}{101}},
  \bibinfo{pages}{5266} (\bibinfo{year}{2004}).

\bibitem[{\citenamefont{Muchnik et~al.}(2007)\citenamefont{Muchnik, Itzhack,
  Solomon, and Louzoun}}]{MuIt07}
\bibinfo{author}{\bibfnamefont{L.}~\bibnamefont{Muchnik}},
  \bibinfo{author}{\bibfnamefont{R.}~\bibnamefont{Itzhack}},
  \bibinfo{author}{\bibfnamefont{S.}~\bibnamefont{Solomon}}, \bibnamefont{and}
  \bibinfo{author}{\bibfnamefont{Y.}~\bibnamefont{Louzoun}},
  \bibinfo{journal}{Phys Rev E} \textbf{\bibinfo{volume}{76}},
  \bibinfo{pages}{016106} (\bibinfo{year}{2007}).

\bibitem[{\citenamefont{Crandall et~al.}(2008)\citenamefont{Crandall, Cosley,
  Huttenlocher, Kleinberg, and Suri}}]{CraCo08}
\bibinfo{author}{\bibfnamefont{D.}~\bibnamefont{Crandall}},
  \bibinfo{author}{\bibfnamefont{D.}~\bibnamefont{Cosley}},
  \bibinfo{author}{\bibfnamefont{D.}~\bibnamefont{Huttenlocher}},
  \bibinfo{author}{\bibfnamefont{J.}~\bibnamefont{Kleinberg}},
  \bibnamefont{and} \bibinfo{author}{\bibfnamefont{S.}~\bibnamefont{Suri}}, in
  \emph{\bibinfo{booktitle}{Proceedings of the 14th ACM SIGKDD International
  Conference on Knowledge Discovery and Data Mining (KDD 2008), Las Vegas,
  Nevada, USA, August 24-27, 2008}}, edited by
  \bibinfo{editor}{\bibfnamefont{Y.}~\bibnamefont{Li}},
  \bibinfo{editor}{\bibfnamefont{B.}~\bibnamefont{Liu}}, \bibnamefont{and}
  \bibinfo{editor}{\bibfnamefont{S.}~\bibnamefont{Sarawagi}}
  (\bibinfo{publisher}{ACM}, \bibinfo{year}{2008}), pp.
  \bibinfo{pages}{160--168}, ISBN \bibinfo{isbn}{978-1-60558-193-4}.

\bibitem[{\citenamefont{Menczer}(2002)}]{menczer02-pnas}
\bibinfo{author}{\bibfnamefont{F.}~\bibnamefont{Menczer}},
  \bibinfo{journal}{Proc Natl Acad Sci USA} \textbf{\bibinfo{volume}{99}},
  \bibinfo{pages}{14014} (\bibinfo{year}{2002}).

\bibitem[{\citenamefont{Nei and Kumar}(2000)}]{Phylogenetic-book00}
\bibinfo{author}{\bibfnamefont{M.}~\bibnamefont{Nei}} \bibnamefont{and}
  \bibinfo{author}{\bibfnamefont{S.}~\bibnamefont{Kumar}},
  \emph{\bibinfo{title}{Molecular Evolution and Phylogenetics}}
  (\bibinfo{publisher}{Oxford University Press}, \bibinfo{address}{Oxford},
  \bibinfo{year}{2000}).

\bibitem[{\citenamefont{Leiner et~al.}(1997)\citenamefont{Leiner, Cerf, Clark,
  Kahn, Kleinrock, Lynch, Postel, Roberts, and Wolff}}]{internet-history}
\bibinfo{author}{\bibfnamefont{B.}~\bibnamefont{Leiner}},
  \bibinfo{author}{\bibfnamefont{V.}~\bibnamefont{Cerf}},
  \bibinfo{author}{\bibfnamefont{D.}~\bibnamefont{Clark}},
  \bibinfo{author}{\bibfnamefont{R.}~\bibnamefont{Kahn}},
  \bibinfo{author}{\bibfnamefont{L.}~\bibnamefont{Kleinrock}},
  \bibinfo{author}{\bibfnamefont{D.}~\bibnamefont{Lynch}},
  \bibinfo{author}{\bibfnamefont{J.}~\bibnamefont{Postel}},
  \bibinfo{author}{\bibfnamefont{L.}~\bibnamefont{Roberts}}, \bibnamefont{and}
  \bibinfo{author}{\bibfnamefont{S.}~\bibnamefont{Wolff}},
  \bibinfo{journal}{Commun ACM} \textbf{\bibinfo{volume}{40}},
  \bibinfo{pages}{102} (\bibinfo{year}{1997}).

\bibitem[{\citenamefont{Dhamdhere and Dovrolis}(2008)}]{DhDo08}
\bibinfo{author}{\bibfnamefont{A.}~\bibnamefont{Dhamdhere}} \bibnamefont{and}
  \bibinfo{author}{\bibfnamefont{K.}~\bibnamefont{Dovrolis}}, in
  \emph{\bibinfo{booktitle}{Proceedings of the 8th ACM SIGCOMM Conference on
  Internet Measurement (IMC 2008), Vouliagmeni, Greece, October 20-22, 2008}},
  edited by \bibinfo{editor}{\bibfnamefont{K.}~\bibnamefont{Papagiannaki}}
  \bibnamefont{and} \bibinfo{editor}{\bibfnamefont{Z.-L.} \bibnamefont{Zhang}}
  (\bibinfo{publisher}{ACM}, \bibinfo{year}{2008}), pp.
  \bibinfo{pages}{183--196}, ISBN \bibinfo{isbn}{978-1-60558-334-1}.

\bibitem[{\citenamefont{Dimitropoulos et~al.}(2006)\citenamefont{Dimitropoulos,
  Krioukov, Riley, and kc~claffy}}]{DiKrRi06}
\bibinfo{author}{\bibfnamefont{X.}~\bibnamefont{Dimitropoulos}},
  \bibinfo{author}{\bibfnamefont{D.}~\bibnamefont{Krioukov}},
  \bibinfo{author}{\bibfnamefont{G.}~\bibnamefont{Riley}}, \bibnamefont{and}
  \bibinfo{author}{\bibnamefont{kc~claffy}}, in
  \emph{\bibinfo{booktitle}{Proceedings of the 7th International Workshop on
  Passive and Active Network Measurement (PAM 2006), Adelaide, Australia, March
  30-31, 2006}}, edited by
  \bibinfo{editor}{\bibfnamefont{M.}~\bibnamefont{Allman}} \bibnamefont{and}
  \bibinfo{editor}{\bibfnamefont{M.}~\bibnamefont{Roughan}}
  (\bibinfo{year}{2006}), pp. \bibinfo{pages}{91--100},
  \urlprefix\url{http://www.pamconf.net/2006/papers/pam06-proceedings.pdf}.

\bibitem[{\citenamefont{Bogu{\~{n}}\'{a} and Pastor-Satorras}(2003)}]{BoPa03}
\bibinfo{author}{\bibfnamefont{M.}~\bibnamefont{Bogu{\~{n}}\'{a}}}
  \bibnamefont{and}
  \bibinfo{author}{\bibfnamefont{R.}~\bibnamefont{Pastor-Satorras}},
  \bibinfo{journal}{Phys Rev E} \textbf{\bibinfo{volume}{68}},
  \bibinfo{pages}{036112} (\bibinfo{year}{2003}).

\bibitem[{\citenamefont{Newman}(2005)}]{newman05}
\bibinfo{author}{\bibfnamefont{M.~E.~J.} \bibnamefont{Newman}},
  \bibinfo{journal}{Contemp Phys} \textbf{\bibinfo{volume}{46}},
  \bibinfo{pages}{323} (\bibinfo{year}{2005}).

\bibitem[{\citenamefont{Krapivsky and Redner}(2005)}]{KraRe05}
\bibinfo{author}{\bibfnamefont{P.~L.} \bibnamefont{Krapivsky}}
  \bibnamefont{and} \bibinfo{author}{\bibfnamefont{S.}~\bibnamefont{Redner}},
  \bibinfo{journal}{Phys Rev E} \textbf{\bibinfo{volume}{71}},
  \bibinfo{pages}{036118} (\bibinfo{year}{2005}).

\bibitem[{\citenamefont{Barab\'{a}si and Albert}(1999)}]{BarAlb99}
\bibinfo{author}{\bibfnamefont{A.-L.} \bibnamefont{Barab\'{a}si}}
  \bibnamefont{and} \bibinfo{author}{\bibfnamefont{R.}~\bibnamefont{Albert}},
  \bibinfo{journal}{Science} \textbf{\bibinfo{volume}{286}},
  \bibinfo{pages}{509} (\bibinfo{year}{1999}).

\bibitem[{\citenamefont{Bogu{\~{n}}\'{a}
  et~al.}(2004{\natexlab{b}})\citenamefont{Bogu{\~{n}}\'{a}, Pastor-Satorras,
  and Vespignani}}]{BoPaVe04}
\bibinfo{author}{\bibfnamefont{M.}~\bibnamefont{Bogu{\~{n}}\'{a}}},
  \bibinfo{author}{\bibfnamefont{R.}~\bibnamefont{Pastor-Satorras}},
  \bibnamefont{and}
  \bibinfo{author}{\bibfnamefont{A.}~\bibnamefont{Vespignani}},
  \bibinfo{journal}{Eur Phys J B} \textbf{\bibinfo{volume}{38}},
  \bibinfo{pages}{205} (\bibinfo{year}{2004}{\natexlab{b}}).

\bibitem[{\citenamefont{Serrano and Bogu{\~{n}}\'{a}}(2005)}]{SeBo05}
\bibinfo{author}{\bibfnamefont{M.~{\'A}.} \bibnamefont{Serrano}}
  \bibnamefont{and}
  \bibinfo{author}{\bibfnamefont{M.}~\bibnamefont{Bogu{\~{n}}\'{a}}},
  \bibinfo{journal}{Phys Rev E} \textbf{\bibinfo{volume}{72}},
  \bibinfo{pages}{036133} (\bibinfo{year}{2005}).

\bibitem[{\citenamefont{Ravasz et~al.}(2002)\citenamefont{Ravasz, Somera,
  Mongru, Oltvai, and Barab\'{a}si}}]{RaSoMoOlBa02}
\bibinfo{author}{\bibfnamefont{E.}~\bibnamefont{Ravasz}},
  \bibinfo{author}{\bibfnamefont{A.~L.} \bibnamefont{Somera}},
  \bibinfo{author}{\bibfnamefont{D.~A.} \bibnamefont{Mongru}},
  \bibinfo{author}{\bibfnamefont{Z.~N.} \bibnamefont{Oltvai}},
  \bibnamefont{and} \bibinfo{author}{\bibfnamefont{A.-L.}
  \bibnamefont{Barab\'{a}si}}, \bibinfo{journal}{Science}
  \textbf{\bibinfo{volume}{297}}, \bibinfo{pages}{1551} (\bibinfo{year}{2002}).

\bibitem[{\citenamefont{Ravasz and Barab\'{a}si}(2003)}]{RaBa03}
\bibinfo{author}{\bibfnamefont{E.}~\bibnamefont{Ravasz}} \bibnamefont{and}
  \bibinfo{author}{\bibfnamefont{A.-L.} \bibnamefont{Barab\'{a}si}},
  \bibinfo{journal}{Phys Rev E} \textbf{\bibinfo{volume}{67}},
  \bibinfo{pages}{026112} (\bibinfo{year}{2003}).

\bibitem[{\citenamefont{Claffy et~al.}(2009)\citenamefont{Claffy, Hyun, Keys,
  Fomenkov, and Krioukov}}]{ClHy09}
\bibinfo{author}{\bibfnamefont{K.}~\bibnamefont{Claffy}},
  \bibinfo{author}{\bibfnamefont{Y.}~\bibnamefont{Hyun}},
  \bibinfo{author}{\bibfnamefont{K.}~\bibnamefont{Keys}},
  \bibinfo{author}{\bibfnamefont{M.}~\bibnamefont{Fomenkov}}, \bibnamefont{and}
  \bibinfo{author}{\bibfnamefont{D.}~\bibnamefont{Krioukov}}, in
  \emph{\bibinfo{booktitle}{Proceedings of the 2009 Cybersecurity Applications
  \& Technology Conference for Homeland Security (CATCH 2009), Washington, DC,
  March 3-4, 2009}} (\bibinfo{publisher}{IEEE Computer Society},
  \bibinfo{year}{2009}), pp. \bibinfo{pages}{205--211}, ISBN
  \bibinfo{isbn}{978-0-7695-3568-5},
  \urlprefix\url{http://www.caida.org/projects/ark/}.

\bibitem[{\citenamefont{Mahadevan et~al.}(2006)\citenamefont{Mahadevan,
  Krioukov, Fall, and Vahdat}}]{MaKrFaVa06-phys}
\bibinfo{author}{\bibfnamefont{P.}~\bibnamefont{Mahadevan}},
  \bibinfo{author}{\bibfnamefont{D.}~\bibnamefont{Krioukov}},
  \bibinfo{author}{\bibfnamefont{K.}~\bibnamefont{Fall}}, \bibnamefont{and}
  \bibinfo{author}{\bibfnamefont{A.}~\bibnamefont{Vahdat}},
  \bibinfo{journal}{Comput Commun Rev} \textbf{\bibinfo{volume}{36}},
  \bibinfo{pages}{135} (\bibinfo{year}{2006}).

\bibitem[{\citenamefont{Adamic et~al.}(2001)\citenamefont{Adamic, Lukose,
  Puniyani, and Huberman}}]{AdLu01}
\bibinfo{author}{\bibfnamefont{L.~A.} \bibnamefont{Adamic}},
  \bibinfo{author}{\bibfnamefont{R.~M.} \bibnamefont{Lukose}},
  \bibinfo{author}{\bibfnamefont{A.~R.} \bibnamefont{Puniyani}},
  \bibnamefont{and} \bibinfo{author}{\bibfnamefont{B.~A.}
  \bibnamefont{Huberman}}, \bibinfo{journal}{Phys Rev E}
  \textbf{\bibinfo{volume}{64}}, \bibinfo{pages}{046135}
  (\bibinfo{year}{2001}).

\bibitem[{\citenamefont{Trusina et~al.}(2004)\citenamefont{Trusina, Maslov,
  Minnhagen, and Sneppen}}]{TruMa04}
\bibinfo{author}{\bibfnamefont{A.}~\bibnamefont{Trusina}},
  \bibinfo{author}{\bibfnamefont{S.}~\bibnamefont{Maslov}},
  \bibinfo{author}{\bibfnamefont{P.}~\bibnamefont{Minnhagen}},
  \bibnamefont{and} \bibinfo{author}{\bibfnamefont{K.}~\bibnamefont{Sneppen}},
  \bibinfo{journal}{Phys Rev Lett} \textbf{\bibinfo{volume}{92}},
  \bibinfo{pages}{178702} (\bibinfo{year}{2004}).

\bibitem[{\citenamefont{Gao}(2001)}]{Gao01}
\bibinfo{author}{\bibfnamefont{L.}~\bibnamefont{Gao}}, \bibinfo{journal}{IEEE
  ACM T Network} \textbf{\bibinfo{volume}{9}}, \bibinfo{pages}{733}
  (\bibinfo{year}{2001}).

\bibitem[{\citenamefont{Dimitropoulos et~al.}(2007)\citenamefont{Dimitropoulos,
  Krioukov, Fomenkov, Huffaker, Hyun, kc~claffy, and Riley}}]{DiKrFo06}
\bibinfo{author}{\bibfnamefont{X.}~\bibnamefont{Dimitropoulos}},
  \bibinfo{author}{\bibfnamefont{D.}~\bibnamefont{Krioukov}},
  \bibinfo{author}{\bibfnamefont{M.}~\bibnamefont{Fomenkov}},
  \bibinfo{author}{\bibfnamefont{B.}~\bibnamefont{Huffaker}},
  \bibinfo{author}{\bibfnamefont{Y.}~\bibnamefont{Hyun}},
  \bibinfo{author}{\bibnamefont{kc~claffy}}, \bibnamefont{and}
  \bibinfo{author}{\bibfnamefont{G.}~\bibnamefont{Riley}},
  \bibinfo{journal}{Comput Commun Rev} \textbf{\bibinfo{volume}{37}},
  \bibinfo{pages}{29} (\bibinfo{year}{2007}).

\bibitem[{\citenamefont{Bogu{\~{n}}\'{a} and Krioukov}(2009)}]{BoKr09}
\bibinfo{author}{\bibfnamefont{M.}~\bibnamefont{Bogu{\~{n}}\'{a}}}
  \bibnamefont{and} \bibinfo{author}{\bibfnamefont{D.}~\bibnamefont{Krioukov}},
  \bibinfo{journal}{Phys Rev Lett} \textbf{\bibinfo{volume}{102}},
  \bibinfo{pages}{058701} (\bibinfo{year}{2009}).

\bibitem[{\citenamefont{Bogu{\~{n}}\'{a}
  et~al.}(2009)\citenamefont{Bogu{\~{n}}\'{a}, Krioukov, and
  kc~claffy}}]{BoKrKc08}
\bibinfo{author}{\bibfnamefont{M.}~\bibnamefont{Bogu{\~{n}}\'{a}}},
  \bibinfo{author}{\bibfnamefont{D.}~\bibnamefont{Krioukov}}, \bibnamefont{and}
  \bibinfo{author}{\bibnamefont{kc~claffy}}, \bibinfo{journal}{Nature Physics}
  \textbf{\bibinfo{volume}{5}}, \bibinfo{pages}{74} (\bibinfo{year}{2009}).

\bibitem[{\citenamefont{Bogu{\~n}\'{a}
  et~al.}(2010)\citenamefont{Bogu{\~n}\'{a}, Papadopoulos, and
  Krioukov}}]{BoPa10}
\bibinfo{author}{\bibfnamefont{M.}~\bibnamefont{Bogu{\~n}\'{a}}},
  \bibinfo{author}{\bibfnamefont{F.}~\bibnamefont{Papadopoulos}},
  \bibnamefont{and} \bibinfo{author}{\bibfnamefont{D.}~\bibnamefont{Krioukov}},
  \bibinfo{journal}{Nature Comms} \textbf{\bibinfo{volume}{1}},
  \bibinfo{pages}{62} (\bibinfo{year}{2010}).

\end{thebibliography}
\end{document}